\documentstyle[11pt,aasms4]{article}
\input epsf

\begin{document}
\title{Reconstructing the spectrum of the pregalactic density field from astronomical data.}

\author{A. Kashlinsky\altaffilmark{}}
\affil{Hughes STX, Code 685, Goddard Space Flight Center, Greenbelt, MD 20771;\\
NORDITA, Blegdamsvej 17, DK-2100 Copenhagen 0, Denmark;\\
Theoretical Astrophysics Center, Juliane Maries Vej 30, 
DK-2100 Copenhagen 0, Denmark}



\def\plotone#1{\centering \leavevmode
\epsfxsize=\columnwidth \epsfbox{#1}}

\def\wisk#1{\ifmmode{#1}\else{$#1$}\fi}

\def\lt     {\wisk{<}}
\def\gt     {\wisk{>}}
\def\le     {\wisk{_<\atop^=}}
\def\ge     {\wisk{_>\atop^=}}
\def\lsim   {\wisk{_<\atop^{\sim}}}
\def\gsim   {\wisk{_>\atop^{\sim}}}
\def\kms    {\wisk{{\rm ~km~s^{-1}}}}
\def\Lsun   {\wisk{{\rm L_\odot}}}
\def\Zsun   {\wisk{{\rm Z_\odot}}}
\def\Msun   {\wisk{{\rm M_\odot}}}
\def\um     {$\mu$m}
\def\sig    {\wisk{\sigma}}
\def\etal   {{\sl et~al.\ }}
\def\eg	    {{\it e.g.\ }}
\def\ie     {{\it i.e.\ }}
\def\bsl    {\wisk{\backslash}}
\def\by     {\wisk{\times}}
\def\half {\wisk{\frac{1}{2}}}
\def\third {\wisk{\frac{1}{3}}}

\begin{abstract}
In this paper we evaluate the spectrum of the pregalactic density field on scales $1h^{-1}{\rm Mpc} <
r < 100h^{-1}$Mpc from a variety of astronomical data. We start with the APM data on the projected
angular correlation function, $w(\theta)$, in six narrow magnitude bins 
and check whether possible evolutionary effects can affect inversion
of the $w(\theta)$ data in terms of the underlying power spectrum.
This is done by normalizing to the angular correlation function on small scales
where the underlying 3-dimensional galaxy correlation function, $\xi(r)$, is known. 
Using the APM data in narrow magnitude bins allows us to test
the various fits to the APM data power spectrum more accurately. We find that for linear scales $r>10h^{-1}$Mpc
the Baugh and Efstathiou (1993) spectrum of galaxy distribution gives the best fit to the data at all depths. 
Fitting power spectra of CDM models to the data at all depths requires $\Omega h=0.2$ 
if the primordial index $n=1$ and $\Omega h=0.3$
if the spectrum is tilted with $n=0.7$. Next we compare the peculiar velocity field predicted by 
the APM spectrum of galaxy (light) distribution with the actual velocity data. 
The two fields are consistent and the comparison suggests that the bias factor is
scale independent with $\Omega^{0.6}/b\simeq$(0.2-0.4). These steps 
enable us to fix the pregalactic mass 
density field on scales between 10 and $\sim 100 h^{-1}$Mpc. The next dataset we use to determine the pregalactic
density field comes from the cluster correlation data. We calculate in detail the amplification of the cluster 
correlation function due to gravitational clustering and use the data on both the slope of the cluster
correlation function and its amplitude-richness dependence. Cluster masses are normalized using the Coma cluster.
We find that no CDM model can fit all the three datasets: APM data on 
$w(\theta)$, the data on cluster correlation function, and the data on the latter's amplitude-richness 
dependence. Next we show that the data on the amplitude-richness dependence can be used directly to obtain the spectrum
of the pregalactic density field. Applying the method to the data, we recover the density field on scales
between 5 and 25$h^{-1}$Mpc whose slope is in good
agreement with the APM data on the same scales. Requiring the two amplitudes 
to coincide, fixes the
value of $\Omega$ to be 0.3 in agreement with observations of the dynamics of the Coma cluster. We then use the
data on high-$z$ objects to constrain the small-scale part, (1-5)$h^{-1}$Mpc 
of the pregalactic density field. We argue that the data at 
high redshifts require more power than given by CDM models normalized to the APM and cluster data. Then we reconstruct
the pregalactic density field out of which modern-day galaxies have formed. We use the data on blue absolute 
luminosities, the fundamental plane relations and the latest X-ray data on the halo velocity dispersion. From this 
we recover the pregalactic density field on comoving scales between 1 and 5$h^{-1}$Mpc which is in reasonable agreement 
with the 
simple power-law extrapolation from the larger scales.
\end{abstract}


\newpage
\section{Introduction}

The origin of structure in the Universe is one of the most outstanding problems in
modern cosmology. 
In the gravitational instability picture it is assumed that structures in the Universe
formed by gravitational growth of small density fluctuations seeded at some early epoch 
of the Universe evolution.
The COBE discovery of the microwave background anisotropies (Smoot et al
1992, Bennett et al 1996) proved
convincingly that density fluctuations were already present at $z \simeq 1000$. It is 
then reasonable to assume that these were indeed the seeds of the density field that 
were to lead to the present day structures. Following the COBE discovery of the
large-scale structure at $z\sim 1000$ it is therefore even more imperative to try to reconstruct 
the pregalactic density field on scales and at epochs inaccessible to the microwave background
measurements.

The spectrum of the density field determines the epoch and the order of galaxy and structure formation. 
Its normalization point is fixed by observations that show that fluctuations in the galaxy counts today 
at $z=0$ have unity amplitude on scale $r_8=8h^{-1}$Mpc. 
Given this amplitude at $r_8$, or mass scale $\sim 10^{15}M_\odot$, and the slope of the
spectrum, one can predict when and on what scales the first objects in the Universe began to collapse.
The converse is also true. Similarly, on scales $>r_8$ where the density field is still in linear regime,
the various properties of galaxy and cluster distribution can be used to determine the spectrum of the
density field on scales currently inaccessible to space-borne microwave background measurements.

On theoretical level, the density field is assumed to have been seeded at some early epoch in the evolution
of the Universe. The most popular of such mechanisms involves inflationary cosmology with cold-dark-matter
contributing most of the mass in the Universe.
Various discussions have also concentrated on the density fluctuations seeded by topological defects,
such as strings and textures, that can be produced naturally during early phase transitions. Another
possibility is the empirical approach, not necessarily motivated by the current ideas from high-energy physics,
whereby one deduces the properties of the early Universe from the present data, instead of 
``predicting" what the present-day Universe should be. 

The density field at some early time can be characterised by both its spatial and Fourier components which
are related via $\delta({\boldmath\mbox{$x$}}) = (2\pi)^{-3} \int \delta_{\boldmath\mbox{$k$}}
\exp(-i{\boldmath\mbox{$k$}} \cdot {\boldmath\mbox{$x$}}) d^3{\boldmath\mbox{$k$}}$. If one assumes the density
field to be a random variable, one can describe its statistical properties 
via the moments of its probability distribution.  Assuming an overall spherical symmetry
the correlation function of the field is defined as 
the generalized scalar product $\xi(r) = \langle
\delta({\boldmath\mbox{$x$}}+{\boldmath\mbox{$r$}})\delta({\boldmath\mbox{$x$}})\rangle$. The power
spectrum is defined as $P(k) =\langle |\delta_{\boldmath\mbox{$k$}}|^2\rangle$, where the average is performed
over all phases. The correlation function and the power spectrum represent a pair of three dimensional
Fourier transforms. In addition the mean square fluctuation over a sphere of volume $V$ containing mass $M$
is related to $\xi(r)$ via: $\Delta^2(M) = \langle \left(\frac{\int 
\delta({\boldmath\mbox{$x$}}) dV}{V}\right)^2 \rangle = \frac{\int \xi(r) dV}{V}$.
If the phases of $\delta_{\boldmath\mbox{$k$}}$ are random, the distribution of the
density field is Gaussian, and the correlation function (or its Fourier 
transform) uniquely describes all 
properties of the density distribution. The central theorem ensures that the Gaussian distribution is reached in 
most inflationary mechanisms for generating energy-density fluctuations. In models with topological defects,
however, the density field is not Gaussian and, in addition to $\xi(r)$, it is determined by higher moments of
the probability distribution.

Enough data have by now been accumulated over a 
large range of scales to strongly constrain the models, on the one hand, and on the other to attempt to
determine the spectrum of the pregalactic density field
independently of theoretical prejudices. This is the aim of this article where
we analyse the various data to show that the data lead to a consistent and unique density field 
in the pregalactic Universe.

When using the data derived from galaxy catalogs one determines the distribution of light,
whereas in order to understand the physics of galaxy formation and early Universe one needs to determine
the distribution of mass. Therefore, one has to introduce the biasing factor which determines how the two 
are related to each other. The most common assumption is that of linear basing, $\delta_{\rm light}\propto
\delta_{\rm mass}$, although more complex and complicated biasing schemes can be considered. Therefore,
before we proceed to the main results of the paper we address the reliability of the general assumption
of light-tracing-mass by comparing the velocity field predicted by the galaxy correlation function
data with peculiar velocities in the Great Attractor region.

The plan of the paper is as follows: In sec.2 we briefly define the 
cold-dark-matter models which we
will test against the data in the paper and discuss the microwave background (COBE DMR) constraints on 
the very large scale part of the density field where it is likely that
the power spectrum has preserved its original slope.
In Sec.3 we analyse the data from the APM survey divided into narrow magnitude bins ($\Delta m \simeq 0.5)$
which contain galaxies at the same evolutionary stages. 
In this way we establish whether significant galaxy evolution effects
could affect determination of the spectrum of the density field from the projected APM data. 
Following this, we discuss the various fits to the APM data and conclude that they all give essentially
the same field of light (galaxy) distribution on linear scales, $r >r_8$. The data in the narrow magnitude
bins, however, set somewhat stronger limits on CDM models than used before. In Sec.4 we elaborate on a method 
to interrelate the data on the galaxy correlation function from both APM and CfA catalogs 
to that on peculiar 
velocities. This establishes if the light traces light and the constraints both datasets suggest
for $\Omega$ and the bias factor. 
Thus we determine the spectrum of the pregalactic density field and its correlation
function on scales (10-100)$h^{-1}$Mpc. In Sec.5 we calculate the amplification in the cluster correlation
function due to gravitational clustering of density fluctuations. We then use the data on both
the spatial slope of the cluster correlation function and on its richness-amplitude dependence 
in order to set further constraints
on the CDM models. We devise a method to invert the richness - correlation amplitude dependence of clusters
of galaxies in order to obtain directly the spectrum of the pregalactic density field on scales (5-20)$h^{-1}$Mpc. 
In Sec.6 we provide comparison between the density field deduced independently 
from the APM catalog and from the cluster correlation data in order
to constrain $\Omega$ by requiring that the two match on the same scales. In sec.7
we discuss the limits on the small-scale part of the pregalactic density field from the data on the existence and
ages of  high-$z$ galaxies and clusters. We also reconstruct the pregalactic density field from the data
on the fundamental plane of modern-day (elliptical) galaxies, their absolute photometry and the dynamics
of their haloes from the recent X-ray observations. At the end of this, we reconstruct in Sec.8 
the {\it pregalactic} density field over two decades in linear scale, $\sim$(1-100)$h^{-1}$Mpc.
Conclusions are summarized in Section 9.

\section{The early Universe anzatz: microwave background and very large scales}

Because of its elegance and simplicity inflation is probably the most popular
of the current theories for the origin of structure in the Universe.
The density field in inflationary 
picture originates from quantum energy fluctuations generated during the slow
roll-over of the inflaton field during this era. Exponential expansion
of the underlying space-time during inflation ensures that the fluctuations are
seeded by normal causal processes on scales well in excess of the current particle
horizon of $\sim 6000 h^{-1}$Mpc. In its simplest and most natural form
inflation leads to gaussian adiabatic density fluctuations with the initial
power spectrum which is scale-free and of the Harrison-Zeldovich slope, $P_i(k)
\propto k^n$ with $n=1$. In conjunction with the COBE observations inflationary models generally
require flat Universe (Kashlinsky, Tkachev and Frieman 1994) while Big Bang
nucleosynthesis implies a small baryon density (Pagel 1997, Walker et al 1991). Thus in order
to reconcile inflationary prejudices with observations one has to postulate the
existence of cold-dark-matter that was not directly coupled to the 
baryon-photon plasma in the prerecombination Universe (e.g. Blumenthal et al
1984 and references cited therein).

Dynamical evolution of cold-dark-matter (CDM) in the early Universe is well understood by 
now (Peebles 1982, Vittorio and Silk 1985, Bond and Efstathiou 1985). In the CDM model one
assumes that evolution during some early (inflationary) epoch resulted
in adiabatic gaussian density fluctuations with {\it initially} scale-invariant
power spectrum. Later the evolution of density fluctuations leads to 
modification of the power spectrum because of the different growth rates of sub- and
super-horizon harmonics during the radiation-dominated era. Thus in CDM models
the power spectrum of the density field at the epoch of recombination is
given by:
\begin{equation}
P(k) \propto k^n T^2(k)
\end{equation}
The transfer function, $T(k)$, depends mainly on the size of the horizon
scale at the matter-radiation equality, $\propto (\Omega h^2)^{-1}$, and is
\begin{equation}
T(k) = \frac{\ln(1+a_0k)}{a_0k} [1+a_1k+(a_2k)^2 + (a_3 k)^3 +(a_4k)^4]^{-1/4}
\end{equation}
where $a_0=0.6a_1=0.14a_2=0.43a_3=0.35a_4=2.34(\Omega h)^{-1} h^{-1}$Mpc 
(Bardeen et al 1986).  On scales greater than the horizon scale at the
matter-radiation equality, $\sim 13 (\Omega h)^{-1} h^{-1}$Mpc, the spectrum
retains its original form and power slope index $n$. On smaller scales it is
modified in a unique, for given $\Omega h$, way.
The values of $n$ that are usually considered are $n\simeq 1$, required by
the simplest CDM model based on inflation. However, a case has been made also
for a tilted CDM model with $n \simeq 0.7$ (Cen et al 1992).

In models invoking topological defects, such as strings, the evolution of 
the initial density field can also be calculated
although with significantly more uncertainty 
because of the possible evolutionary modes of strings when
they enter horizon (Albrecht and Stebbins 1992). In models which involve only baryonic matter, the 
primeval density fluctuations have to be purely isocurvature, and the evolution of the density
field is further complicated by the possible reheating and reionisation effects after recombination
(e.g. Peebles 1987).

In what follows we will distinguish between the ``primordial spectrum" and what we term
the ``pregalactic spectrum" of the density field. By primordial we will mean the spectrum that 
was presumably produced during the very early stages of the Universe's evolution and which is assumed 
to be self-similar and characterised only by its power slope index $n$. ``Pregalactic" in the terminology 
of the paper will refer to the spectrum of the density field after recombination but prior to galaxy
formation, e.g. at redshifts $>100$. In principle, the primordial spectrum can be recovered
from the pregalactic one by assuming a transfer function. We will not do this, since it involves
various uncertain assumptions, such as the nature of the dark matter, the type of density fluctuations,
the values of cosmological parameters, assumptions about the degree and history of reionisation after 
recombination, etc.

In what follows we will aim to reconstruct the pregalactic density field in 
the Universe from a variety of astronomical data, each dataset responsible
for uncovering the spectrum of the density field over a certain range of 
scales. Because of the theoretical appeal of the CDM models
we will compare the results in each range of scales to the anzatz eqs. (1),(2)
predicted by inflationary cold-dark-matter models.

In all models of the early Universe evolution, the density field on the largest scales
which were always outside the horizon during radiation dominated era, $\gg$
a few hundred $h^{-1}$Mpc, is expected to have 
preserved it initial, primordial, form. The spectrum of the density 
distribution on such scales is constrained by the COBE DMR data probing the microwave background
anisotropies on angular scales $> 7^{\rm o}$. Assuming a scale-free primordial power spectrum
with adiabatic initial conditions, COBE DMR data (Smoot et al 1992, Bennett et al 1996),
after correcting for the Galactic cut, imply a roughly power-law spectrum with $n\simeq 1.1\pm 0.3$
at the 68\% confidence level (Gorski et al 1996). Normalization to the COBE data also fixes 
the value of the bias parameter $b$ for a given power spectrum of the density field and
$\Omega$ (Kashlinsky 1992; Efstathiou, Bond and White 1992). CDM models normalized to the COBE DMR
data require $b>1$ (e.g. Stompor et al 1995).

On smaller scales the pregalactic density field can be constrained and, as we show below, determined from
the astronomical data out to $z$ of a few. In
the remainder of this article we recover the pregalactic density field from the various datasets on the
present-day Universe.

\section{Two-point galaxy correlation function on linear scales: constraining the spectrum of
light distribution on scales $> 10h^{-1}$Mpc}

As was mentioned in the Introduction
the two-point galaxy correlation function, $\xi(r)$, and the power spectrum represent a pair
of three-dimensional Fourier transforms and, for Gaussian density fluctuations, 
define all properties of the density field. They are related via:
\begin{equation}
\xi(r) = \frac{1}{2\pi^2} \int_0^\infty P(k) j_0(kr) k^2 dk
\end{equation}
Measurements of the two-point correlation function from galaxy catalogs show that on small scales
$\xi(r)$ is very close to a power law $\xi(r) = (r/r_*)^{-1-\gamma}$, where $r_* = 5.5h^{-1}$Mpc 
and $\gamma=0.7$ (Groth and Peebles 1977). This in turn means that the rms fluctuation in galaxy
counts
\begin{equation}
\langle (\frac{\delta N_{\rm gal}}{N_{\rm gal}})^2 \rangle = 3\int_0^1 \xi(xr)x^2 dx =
\frac{1}{2\pi^2} \int_0^\infty P(k) W_{TH}(kr) k^2 dk
\end{equation}
is unity on
scale $r_8 = (\frac{3}{2-\gamma})^{-\frac{1}{1+\gamma}} r_* \simeq 8h^{-1}$Mpc (Davis and 
Peebles 1983). Here $j_n(x)$ is the $n$-th order spherical Bessel function and
$W_{TH}=[3j_1(x)/x]^2$.
Eq. (4) also defines the connection between the ``counts-in-cells" analysis used
in some galaxy surveys (e.g. Saunders et al 1991) and the underlying power spectrum or
the 2-point correlation function, $\xi(r)$.

The data on the correlation function on larger scales are now available from the APM catalog
measurements (Maddox et al 1990, hereafter MESL). The APM survey contains about 2.5 million galaxies 
in the blue magnitude range $17.5 \leq b_J \leq 20.5$. MESL
measured the \underline{projected} 2-point angular correlation, $w(\theta)$, down to a systematic
error of $w_{pp}\simeq 1.5\times 10^{-3}$, remaining due to plate-to-plate gradients. These data
are consistent with a variety of measurements of $\xi$ from measurements at other bands 
(Picard 1991), catalogs
(Collins, Nichol and Lumdsen 1992) and counts in cells analysis of IRAS galaxies (Saunders et al 1991). These
measurements of $w(\theta)$ probe $\xi(r)$ to considerably larger scales than the non-linear scales
of $r<r_8$.

If galaxy distribution along the line of sight is known one can relate
$w(\theta)$ to the 3-dimensional correlation function, $\xi(r)$,
via the Limber equation (Limber 1953; Peebles 1980). Its relativistic
form is:
\begin{equation}
w(\theta) = \int_0^\infty dz \phi(z) \int_{-\infty}^\infty d\Delta \xi(r_{12};z)
\end{equation}
where $r_{12}^2 = (c\frac{dt}{dz}\Delta)^2 + \frac{x^2(z) \theta^2}{(1+z)^2}$ is the proper length,
$x(z)$ is the co-moving distance to $z$,
$\phi(z) \equiv (dN/dz/N_{tot})^2$ accounts for the fraction of galaxies 
in the redshift interval
$[z;z+dz]$ and $N_{tot}=\int_0^\infty \frac{dN}{dz} dz$. The number of galaxies per $dz$
with apparent magnitude $m_l \leq m \leq m_u$ is given by:
\begin{equation}
\frac{dN}{dz} = \frac{dV}{dz} \int_{L(m_u)}^{L(m_l)} \Phi(L;z) dL 
\end{equation}
where $dV/dz = \frac{ cH_0^{-1} x^2(z) }{ (1+z)^4\sqrt{1+\Omega z} }$ for $\Lambda=0$,
$\Phi(L;z)$ is the galaxy luminosity function at redshift $z$ and $L(m)$ is the absolute luminosity
of galaxy with apparent magnitude $m$ at redshift $z$.

Eq.(5) can be rewritten to relate $w(\theta)$ directly to the power spectrum
(Kashlinsky 1991a; Peacock 1991). Substituting (3) into (5) and using that
$\int_{-\infty}^\infty j_0(\sqrt{x^2+y^2}) dy = \pi J_0(x)$ leads to:
\begin{equation}
w(\theta) = \pi \frac{ \int_0^\infty dz (cdt/dz)^{-1} \phi(z)
\int_0^\infty dk P(k;z) kJ_0\left(\frac{kx\theta}{1+z}\right) }
{ \int_0^\infty P(k;0) W_{TH}(kr_8) k^2 dk }
\end{equation}
where $J_0(x)$ is the zero-order cylindrical Bessel function. Note
that deducing the 3-dimensional power spectrum from the projected
galaxy angular correlation function is independent of the distortions
caused by peculiar velocities 
in the distribution of galaxies mapped in the redshift space (Kaiser 1987).
Throughout the paper we will deal mostly with the shape of the power
spectrum, rather than its amplitude. Hence we chose to normalize eq.(7) 
by introducing in the denominator the explicit expression, eq.(4), for unity galaxy counts 
over a sphere of radius $r_8$. As eqs.(3),(5),(7) show a power law for
$w(\theta)\propto \theta^{-\gamma}$ implies a power law behaviour for
$\xi(r) \propto r^{-1-\gamma}$ or power spectrum $P(k) \propto k^{\gamma-2}$.

The APM survey represents the most accurate and biggest dataset for determining the 
correlation function galaxies. On small scales the projected angular 
correlation behaves like a power law $w(\theta) \propto \theta^{-\gamma}$
with slope $\gamma=0.7$. On larger scales $w(\theta)$ falls off sharply and eventually
becomes lost in the systematic errors dominated by the plate to plate gradients of
the APM catalog, $w_{pp}\simeq 1.5 \times 10^{-3}$. The falloff clearly implies the 
rollover in $P(k)$ at large scales, or small $k$. The APM catalog contains galaxies
down to the limiting blue magnitude of $b_J =20.5$ corresponding to $z \sim 0.2$. Thus in order
to interpret the data on $w(\theta)$ in terms of the power spectrum today (or at any
other coeval epoch) one must eliminate or reduce the possible effects related to 
evolutionary uncertainties of the luminosity function, K-correction, and galaxy
clustering.

Various fits have been suggested for the power spectrum to describe the APM data.
Kashlinsky (1992a; hereafter K92) suggested an empirical fit to the power spectrum
to describe the data: on small scales $P(k) \propto k^{\gamma-2}$ to reproduce
$w(\theta) \propto \theta^{-\gamma}$ at small angles. On large scales, or
$k\leq k_0$, the power spectrum was assumed to go into the Harrison-Zeldovich
regime consistent with the COBE DMR data. K92 used the APM data divided into 
six narrow magnitude of $\Delta m \simeq 0.5$. This way one can reduce effects of 
evolution since such narrow slices are more likely to contain galaxies at similar depths. 
Approximating the selection function like that of the Lick catalog (Groth and
Peebles 1974; Peacock 1991) K92 found
that satisfactory fits to the APM data can be obtained for $k_0^{-1} \sim 40
h^{-1}$Mpc. 

Baugh and Efstathiou (1993 - hereafter BE93) have developed an iterative deprojection
technique using the Lucy method (Lucy 1974). They approximated the galaxy
selection analytically and assumed that the time evolution of the power spectrum
is scale independent. BE93 then applied the method to invert $w(\theta)$ 
from the entire APM dataset of
$17.5 \leq b_J \leq 20.5$. The resultant three dimensional power spectrum with the 
error bars from the method and the APM data uncertainties is plotted in Fig.7 
of BE93. Application of their method to the two-dimensional power spectrum 
from the entire APM dataset led to similar numbers for $P(k)$ (Baugh and
Efstathiou 1994).

In terms of CDM models, the APM data require significantly more large-scale power
than predicted by the standard CDM model with $\Omega=1, h=0.5,n=1$ (MESL).
Various suggestions have been made to account for the observed excess
of large scale power on scales where the present-day density field is still
linear, $r > r_8$. Efstathiou et al (1990) have noted that the required large-scale
power can be reproduced by low-$\Omega$ CDM models in which case a non-vanishing
cosmological constant would be required to keep the Universe flat in accordance with the standard
inflationary scenario. Cen et al (1992) suggested that the excess power can be explained
by the so-called tilted CDM model with $n\simeq$0.7-0.8. Either of these modifications,
decreasing $\Omega$ or introducing $n<1$, would boost the large-scale power of the CDM
models on scales which are in the linear regime today. (At the same time this would, 
however, suppress the small scale power in such models;
implications of this are discussed later in the paper).

We now turn to quantifying how good the above fits and models for $P(k)$ are
viz-a-viz the various evolutionary corrections that can affect the accuracy 
of the various interpretations. The 
evolution of the luminosity function will affect the selection function of the
Limber equation, while evolution of spectral energy distribution will determine
which galaxies appear in the B-band at $z=0$. Finally, the rate of the evolution of 
clustering pattern will introduce another redshift-dependent factor in the integrand
of the numerator of (7). In addition, eq.(7) contains a dependence on $\Omega$.

MESL presented their data in two ways: 1) the entire sample of 2.5 million
galaxies with $b_J$ magnitudes between 17.5 and 20.5, and 2) the data for sub-samples of 
galaxies binned in 6 narrow magnitude slices of $\Delta m \simeq 0.5$. In the first
dataset, which was used in BE93, the galaxies are at very different depths and 
possibly very different evolutionary stages. The second APM dataset used in K92 is more 
immune from effects of evolution since it is more likely to separate 
galaxies at different depths
and any evolution can be more readily uncovered and corrected for. We thus consider the
APM data on $w(\theta)$ for the six slices each $\Delta m \simeq 0.5$ in width.

Fig.1 plots the selection function, $z \phi(z)$, for the six slices with $\Delta m\simeq0.5$ of the APM 
survey from $b_J=17.5$ to $b_J=20.5$. The luminosity function in eq.(6) was adopted
from Loveday et al (1992): i.e. the Schechter (1976) luminosity function with
$M^*_{b_J}=-19.5$ for $h=1$ and $\alpha=-1$. The relativistic $K$ correction was
modelled as $\delta m = K z$ with $K=3$ which adequately describes
galactic spectra in the visible bands (cf. Yoshii and Takahara 1988). Solid lines 
correspond to $\Omega=0.1$ and dotted to $\Omega=1$ and no evolution was assumed.
One can see that there is little dependence on $\Omega$. Similarly, there would
be little change for other plausible values of $0 \leq K \leq 4$. The nearest 
slice of the APM galaxies contains galaxies typically lying at $z\simeq 0.07$ and
the farthest slice contains galaxies at $z \simeq 0.2$, so certain amount
of evolution could have occurred. There is no overlap at the FWHM level between 
galaxies in the nearest and most remote slices, although certain overlap exists
between galaxies in the nearby slices.

On co-moving scales less than $r_8$ the present day density fluctuations are 
non-linear and the clustering pattern has been strongly distorted by gravitational
evolution (e.g. Davis et al 1985). On larger scales the density field is still in the
linear regime and the clustering pattern there is likely to have preserved the original
power spectrum of the density field. We use the small scales, $r \ll r_8$, where the power spectrum
is well known and its evolution with $z$ better understood, to constrain the possible
evolutionary effects that can plague the APM data interpretation.

As $\theta \rightarrow 0$ the APM two-point correlation function is 
approximated well by a pure
power law. At the same time, for small angular separations, the contribution to eq.(5)
from $\xi(r)$ at large separations becomes negligibly small. Thus in the limit of small
angular separations the 3-dimensional two-point correlation function at $z=0$ can be taken
from the Lick catalog to be $\xi(r)=(r/r_*)^{-1-\gamma}$ with $r_*=5.5h^{-1}$Mpc. The time
dependence of $\xi(r)$ can be modelled in the very non-linear regime as $\xi(r;z) =
\xi(r;0) \Psi^2(z)$. Two extremes of clustering evolution are generally considered
for very small scales (Peebles 1980): If clustering is stable in co-moving coordinates 
$\Psi^2(z) = (1+z)^{-3}$; if it is stable in proper coordinates $\Psi^2(z) =(1+z)^{-\gamma}$.
Thus for small scales the evolutionary effects must be constrained 
to match the
APM data for all six slices: $w(\theta) =A_{w,{\rm slice}} \theta^{-\gamma}$. Substituting
$\xi(r;z)=(r/r_*)^{-1-\gamma} \Psi^2(z)$ into (5) leads to:
\begin{equation}
A_w = \frac{ \Gamma(\frac{1}{2}) \Gamma(\frac{\gamma}{2}) }
{ \Gamma(\frac{1+\gamma}{2}) } \left(\frac{r_*}{cH_0^{-1}}\right)^{1+\gamma}
 \int_0^\infty \phi(z) \left[\frac{cH_0^{-1}(1+z)}{x(z)}\right]^\gamma
\Psi^2(z) (1+z)^2 \sqrt{1+\Omega z} dz 
\end{equation}
Comparison between (8) and the amplitude measured in the APM data provides an integral
constraint on the amount of possible evolution.

Fig.2 plots the data from the APM survey (MESL) divided into 6 narrow magnitude bins. The
magnitude limits for each bin are shown on the top of each box. The data is plotted with
open triangles. The power-law fits corresponding to $A_w \theta^{-\gamma}$ are plotted 
with two sets of straight lines. Dashed lines are for $\Omega=1$ and dotted are for 
$\Omega=0.1$ and $\Lambda=0$. The two lines of each type correspond to clustering
pattern stable in co-moving (lower lines) and proper coordinates. The amplitudes $A_w$
are shown for $K=3$. One can see that no-evolution models describe the low-angle behaviour
of $w(\theta)$ extremely well. The dependence on $\Omega$ can be neglected and the dependence
on the precise rate of the evolution of clustering pattern is also small. For clarity of 
the figure we do not show the lines for other values of $K$, but note that the change is very small.
Increasing the value of $K$ would move the lines a little up, decreasing it to $< 3$ would make
them lie exactly on top of the data at small angular scales. Similarly, making the clustering 
pattern evolve a little faster than $\Psi^2(z) =(1+z)^{-3}$, as is suggested by simulations
of non-linear gravitational clustering (Melott 1992) would shift the lines even closer to the data
points. Thus we conclude that the APM data are consistent with no or little evolution out to the
epochs accessible to the APM magnitude limit of $b_J=20.5$. The slight excess in the computed 
amplitude $A_w$ can be easily reduced to zero by adopting a slightly lower value of $K$
or requiring the the
clustering pattern to evolve faster than $\Psi^2(z) = (1+z)^{-3}$. The first of these is quite 
reasonable given the evolution of stellar populations in the relevant bands (e.g. Bruzual 1983).
The second could be required if mergers play significant role in the evolution of clustering on
non-linear scales (Melott 1992); this may even be suggested by the deep galaxy counts data
(Broadhurst et al 1992). Thus we conclude that luminosity evolution plays a minor role out to
$z\simeq 0.2$ for the standard population of blue galaxies with $b_J \leq 20.5$.
The no-evolution models with non-linear clustering 
stable in co-moving coordinates can be used sufficiently accurately in interpreting the APM
data on $w(\theta)$ in terms of the underlying power spectrum of galaxy distribution.

Thin solid lines in Fig.2 show the BE93 fits to the six slices of the APM data assuming no luminosity
evolution and with the luminosity function adopted from Loveday et al (1992). The three lines
correspond to the best determined $P(k)$ of BE93 and to one standard deviation uncertainty. 
The clustering
pattern was assumed to be stable in co-moving coordinates on all scales. The BE93 power spectrum
fits well the data at the various depths. It under-predicts the power on large scales by a small
margin for the most distant of the APM slices. This, however, can be accounted for by assuming 
that the power spectrum on linear scales evolves less rapidly with redshift than the $\Psi^2(z)
=(1+z)^{-3}$ adopted from the non-linear scales. Indeed, linear scales should evolve differently
in accordance with the growth predictions in linear regime. In principle, one can try to
reproduce the consistent evolution of the power spectrum on all scales following the
prescription of Hamilton et al (1991). However, it is not yet clear how well such fits work
for arbitrary cosmologies and (non-power-law) power spectra (cf. Peacock and Dodds
1994). The fit of BE93 power spectrum to
the APM data is good and we adopt it as a reasonable approximation to the power spectrum of galaxy
distribution. The K92 fit is essentially the same as shown in Fig.1 of the K92 paper and for brevity and
clarity we do not show this set of lines in Fig.2, other than to point out that the BE93 power
spectrum fits the APM data much better than the earlier K92 empirical fit.

The K92 and BE93 power spectra fit the clustering hierarchy today over both linear and non-linear
scales. On linear scales the power spectrum today should reflect initial conditions, but on scales
less than $r_8$ the initial power spectrum was distorted by gravitational effects. CDM models 
predict power spectrum only in the linear regime so it makes little sense to plot them in the already 
condensed Fig.2. In order to constrain CDM models from the binned APM data we proceed as follows:
the APM data is accurate for $w(\theta) > w_{pp} \simeq 1.5\times 10^{-3}$ 
and the small angular scales analysis 
suggested that evolutionary effects are small for scales and depths probed 
by the APM survey. The largest angular
scales, where $w(\theta)$ can still be probed, are dominated by contributions 
from linear scales, $r>r_8$, and the data should be 
reproduced by eqs.(1),(2) within the framework of CDM models. 

Thus for a given CDM model specified by $n$ and $\Omega h$ we computed 
the angular scale, $\theta(w)$, on which the predicted angular correlation 
function drops to
the value $w$. Fig.3 plots the resultant
$\theta(w)$ vs the mean magnitude, $b_J$, of the slices in Fig.2 for 
the APM data and for predictions of 
the standard ($n=1$, upper boxes) and tilted ($n=0.7$, lower boxes) CDM models. The left
boxes are for $w=2\times 10^{-3}$, which is just above the systematic error of $w_{pp}=1.5\times
10^{-3}$ induced by the plate-to-plate gradients in the APM survey. The right boxes are for $w=5\times
10^{-3}$ which is significantly larger than $w_{pp}$ and should be quite accurate. The thick plus
signs show the values
of $\theta(w)$ derived from the APM data. CDM models are shown with the other signs: x's correspond
to $\Omega h=0.5$, squares to $\Omega h=0.3$, triangles to $\Omega h=0.2$, rhombs by $\Omega h=0.1$ and 
asterisks to $\Omega h=0.05$. 

As was mentioned the values of $\theta(w=5\times 10^{-3})$ are more reliable, but there is general agreement
between the CDM fits to angular scales at both values of $w$. There is also consistency between the fits for
various slices or depths. 
The fits in Fig.3 show that the linear part of the APM data can be described by 
CDM models which would then require $\Omega h=0.2$ if $n=1$ or $\Omega h=0.3$ if $n=0.7$. Increasing $n$ above
0.7 would require even lower values of $\Omega h$. The limit on $\Omega h$ required by CDM models for $n=1$
is in good agreement with what was claimed before (Efstathiou et al 1990). It is worth noting that the 
limits on the power parameter $\Omega h$ for the CDM models come from the nearby slices. 
The differences in the values of $\theta(w=5\times 10^{-3})$ between CDM models of different
$\Omega h$ are most profound for the nearest (and least affected by any evolution) slices. For the
first three slices the APM data (pluses) would overlap with CDM predictions
only for $\Omega h=0.2$ if $n=1$ and for
$\Omega h=0.3$ if $n=0.7$; CDM models 
with both smaller and larger $\Omega h$ would predict values of the
angular scale $\theta(w)$ where $w=5\times 10^{-3}\simeq 3.3 w_{pp}$ different from what is observed.
The limit on $\Omega h$ (=0.3)
for the tilted model is significantly lower than what was suggested before in Cen et al (1992) who used the
entire APM galaxies lumped together and then scaled to the depth of the Lick catalog. Thus even
if the primordial $n$ were as low as 0.7 one would still require low $\Omega$ to fit the APM
data with the CDM power spectrum.

Strictly speaking, the above discussion applies only to the distribution of light (galaxies). In order
to relate the power spectrum of galaxy distribution to that of mass one has to make assumptions about
biasing or whether and how the light traces mass. The most economical and commonly made assumption to make
is that of linear biasing (Kaiser 1984), i.e. that light traces mass at least to within a scale-independent
constant. We followed this assumption in this section, but it must be remembered that the plots in Figs.2,3
refer to the distribution of luminous matter only. The next section discusses 
this assumption in more detail and there 
we propose an empirical justification for it based on comparing the density
field of the APM survey to that probed by the velocity data.

Assuming that light indeed traces mass, information on 
the initial (pregalactic) power spectrum in present-day galaxy catalogs is preserved only 
on scales where the 
density field is still in the linear regime, $r>r_8$. Thus Fig.4 plots the r.m.s. density fluctuation, 
$\Delta(M)$, over a sphere containing mass $M$ vs the co-moving scale this
mass subtends for the spectra required
by the APM data. It is plotted in units on $\Delta_8$, the r.m.s. fluctuation over a sphere of radius $r_8$,
and the numbers are shown for $r>10h^{-1}$Mpc where the galaxy density field is still in the linear regime.
This ratio is used because in linear regime it is roughly epoch independent and also because for linear
biasing the quantity $\Delta(M)/\Delta_8$ recovered from the APM data is independent of the (constant)
biasing factor $b$. Solid lines are the K92 power spectra with the transition scale $k_0^{-1}=35,45,55
h^{-1}$Mpc from bottom up. Dashed lines correspond to the BE93 power spectrum with three lines showing 
the mean and $\pm$ one standard deviation uncertainty. The CDM models that fit the APM data in six narrow
magnitude slices are shown with thick dashed lines: $n=1, \Omega h=0.2$ and $n=0.7, \Omega h=0.3$. (The
two models would essentially overlap over the range of scales plotted in Fig.4). The various fits to the
data practically overlap on scales $<50h^{-1}$Mpc. On larger scales the BE93 spectrum, which fits the APM
data best, gives the least amount of large-scale power. Within the uncertainty of the methods
the CDM models required by the APM data reproduce roughly the same $\Delta(M)/\Delta_8$ ratio as 
the BE93 spectrum on scales $<100h^{-1}$Mpc.

{\it The quantity $\Delta(M)/\Delta_8$ plotted in Fig.4 is independent of the bias factor provided
the latter is scale-independent. It also measures the spectrum of the density field in linear regime
and is approximately independent of the redshift at which it is evaluated. These are the main
reasons why in this paper we choose to reconstruct this quantity from various datasets.}

To conclude this section: 1) By analysing APM catalog data in narrow magnitude
slices it was demonstrated that evolutionary effects are unlikely to affect interpretation of the APM data. 2) The 
power spectrum derived by BE93 provides a good fit to the APM data at all depths. 3) Assuming
that light traces mass, the APM data can be fit at all depths by the standard CDM models with 
$n=1$ if $\Omega h\simeq 0.2$. 4) Tilted CDM models will require $\Omega h\simeq 0.3$ if $n=0.7$ 
and lower values if $n$ is increased. 5) In any case, all the above fits give practically identical
numbers for $\Delta(M)/\Delta_8$ on linear scales $<50h^{-1}$Mpc. This quantity measures density
field in the linear regime and is approximately redshift independent. If the possible bias factor 
is scale independent the quantity $\Delta(M)/\Delta_8$ determined from the APM data is 
independent of the bias factor.  6) Fig.4 demonstrates consistency of this approach and we conclude that,
provided that light-traces-mass, the pregalactic density field on these scales is close to that drawn
in the figure.

\section{Peculiar velocity field vs galaxy correlation data: constraining $\Omega$ and the bias
factor}

The previous section established the power spectrum of galaxy distribution on scales $10 < r < 
$(60-100)$h^{-1}$Mpc. In order to reliably identify it with the spectrum of the pregalactic
density field one has to establish the relation between the light and mass distributions.
Furthermore, as was discussed in the framework of the CDM models 
APM data would require a low $\Omega$
Universe. An independent test of these findings and assumptions can be provided by inter-comparison
between the observed properties of peculiar velocity and density fields (Kashlinsky 1992b, 1994).
We address these questions in the section.

Measurements of peculiar velocities provide one with direct probe of the mass distribution
in the Universe and thus set strong limits on the large-scale structure models (e.g. Vittorio, 
Juszkiewicz and Davis 1987). Recent advances with the new distance indicators
(Djorgovski and Davis 1987; Dressler et al 1987a) allowed one to measure the
peculiar flows in the local part of the Universe up to $\sim 50-100 h^{-1}$Mpc (see 
Strauss and Willick 1995 for a review). 
The results from using the fundamental plane properties of elliptical galaxies
(Dressler et al 1987b) suggest a large coherence length and amplitude of the local peculiar velocity 
field. The results from analysis of 1,355 spirals using the Tully-Fisher
relation to determine distances (Mathewson et al 1992) would lead to an even larger coherence length and
a similar amplitude. Other samples are also in agreement with the
Great Attractor findings (e.g. Willick 1990). In a more controversial finding Postman and Lauer 
(1995) find from analysis of clusters of galaxies that the peculiar flows can be coherent over a
much larger scale than the Great Attractor findings.
All the results indicate significant deviations from the Hubble flow with a large coherence length.

Since peculiar velocities probe the mass distribution and also depend on the density parameter 
$\Omega$ one can determine the latter by comparing the mass distribution implied by the velocity
field with the observed distribution of galaxies.
The density parameter is thus determined to within the uncertainty of the bias factor $b$
by determining the factor $\Omega^{0.6}/b$. 
There are various ways to use this information to get at $\Omega$. 
Bertschinger and Dekel (1989) developed POTENT method whereby the mass distribution is 
reconstructed by using the analog of the Bernoulli equation for irrotational flows.
They used the method to analyse in great detail and accuracy the peculiar velocity field
out to about 60$h^{-1}$Mpc (Bertschinger et al 1990). In Dekel et al (1993) they
compared the previously determined velocity field with the 
observed distribution of galaxies and concluded that the best fit is 
achieved with $\Omega^{0.6}/b \simeq 1$.

A different method was presented by Kashlinsky
(1992b) and was based on comparing the velocity correlation function with the galaxy correlation 
function from the APM catalog; this led to low values of $\Omega^{0.6}/b \simeq$0.2-0.3. Herebelow 
we present a more detailed analysis based on the latter scheme and show that the results 
are indeed suggestive of low values of the density parameter. We also show that the results are
in agreement with the scale independent bias factor.

We define the dot velocity correlation function $\nu(r) = \langle{\boldmath\mbox{$v$}}(
{\boldmath\mbox{$x$}} +{\boldmath\mbox{$r$}}) \cdot {\boldmath\mbox{$v$}}{(\boldmath\mbox{$x$})}
\rangle$ (Vittorio, Juszkiewicz and Davis 1987, Peebles 1987). 
\begin{equation}
\nu(r) = \frac{1}{2\pi^2} \int_0^\infty |v_k|^2 j_0(kr) k^2 dk
\end{equation}
Since gravity force is conservative, the flow caused by it must be irrotational.
For irrotational flow the $k$-th component of the Fourier transform of the peculiar velocity field is 
${\boldmath\mbox{$v$}_k}=-i{\boldmath\mbox{$k$}} H_0 f(\Omega) \delta_k/k^2$ where $f(\Omega) = \partial
\ln \delta_k/\partial \ln t$. For the case of zero cosmological constant $f(\Omega) = \Omega^{0.6}/b$.
If the cosmological constant, $\Lambda$, is not zero $f(\Omega)$ will be different, but for the most 
interesting case of $\Omega+\Lambda/3H_0^2=1$ the above approximation would work very well (Lahav et
al 1991). Eqs (3),(9),(10) are also valid for the filtered fields in which case both $\nu(r)$ and $\xi(r)$ 
must be replaced with their filtered versions.

Assuming that light traces mass to 
within a constant bias factor, $b$, substituting $v_k$ into (9), and then taking Laplacian operator
of both sides would lead to:
\begin{equation}
 \nabla^2 \nu(r) = - \frac{\Omega^{1.2}}{b^2} H_0^2 \xi(r)
\end{equation}
where it was used $\nabla^2 j_0(kr) = -k^2 j_0(kr)$ along with eq.(3).
This equation was evaluated independently in Kashlinsky (1992b) and Juszkiewicz and Yahil 
(1989) and it can be used to relate the \underline{data} on both peculiar velocity field and
galaxy correlation function to get $\Omega^{0.6}/b$ (Kashlinsky 1992b,1994).
The advantage of using (10) rather than the power spectrum data in conjunction with (9) (e.g.
Kaiser 1983) is that eq.(10) can be integrated to contain only the scales over which the {\it data}
on both $\xi(r)$ and $\nu(r)$ are known, while in the former case assumptions have to be made about the
power spectrum over the wavelengths inaccessible to observations (Vittorio et al 1987). 
(Of course, the data $\xi(r)$ are in effect on integral constraint on $P(k)$ over the infinite
range of wavelengths). Equation (10) can be solved to give the
velocity correlation function in terms of $\xi(r)$:
\begin{equation}
\nu(r) = \nu(0) - \frac{\Omega^{1.2}}{b^2} H_0^2 [J_2(r) - \frac{J_3(r)}{r}]
\end{equation}
where it is defined $J_n(r) \equiv \int_0^r \xi(x) x^{n-1} dx$. Only one of the constants
of integration remains in (11), the other constant of integration vanishes 
because the velocity field must be finite at $r=0$.
The constant of integration remaining in (11) is just the ``central" velocity dispersion, $\nu(0)
= \langle v^2 \rangle$. 

Strictly speaking in order to use eq.(10) the data on  both $\nu(r)$ and $\xi(r)$ must come from the 
same region of the Universe. Since the current velocity data sample galaxies within a radius of 
$\sim 100h^{-1}$Mpc, the numbers for $\xi(r)$ must come from the ``local" surveys such as
the CfA survey (Geller and Huchra 1989). The CfA redshift survey is complete out to $m_B \leq
15.5$ and for the relevant scales, $r\simeq 40h^{-1}$Mpc,
the correlation function in the CfA survey essentially coincides with the one determined from the
APM catalog (Da Costa et al 1994; Vogeley et al 1992). In fact, the numbers 
in this section are in good agreement with each other
when evaluated for $\xi(r)$ from the CfA survey, the power-law model, or the 
BE93 or K92 fits to the APM data on $w(\theta)$.
The galaxy correlation function on the relevant scales from the APM data,
as discussed in the previous section, is plotted in Fig.11; for the CfA survey $\xi(r)$ is plotted
in Fig.3 of Vogeley et al (1992).

The observed correlation function of galaxies would predict large velocities if 
$\Omega^{0.6}/b=1$. For power-law $\xi(r)=(r/r_*)^{-1-\gamma}$ eq.(11) would lead to 
$\sqrt{\nu(0)-\nu(r)} = \frac{H_0r_*}{\sqrt{(1-\gamma)(2-\gamma)} } \frac{\Omega^{0.6}}{b}
(\frac{r}{r_*})^{\half(1-\gamma)}=881\frac{\Omega^{0.6}}{b}(\frac{r}{r_*})^{0.15}$km/sec for
$\gamma=0.7$. The value of $\sqrt{\nu(0)-\nu(r)}$ for $\Omega^{0.6}/b=1$ would be significantly 
larger than the data on peculiar velocities at 40 or $60h^{-1}$Mpc of Bertschinger et al (1990)
(see Table 1).
Fig.5 plots the predicted value of $\sqrt{\nu(0)-\nu(r)}$ 
computed according to eq.(11) for $\Omega^{0.6}/b=1$;
the numbers in Fig.5 scale $\propto \Omega^{0.6}/b$. The data on $\xi(r)$ from which the lines in Fig.5 
were computed are 
for the APM correlation function using the BE93 
deprojection (shown with solid lines) and the correlation function of
the CfA survey from Vogeley et al (1992) (dashed line). The three solid lines correspond to one
standard deviation uncertainty in BE93. 
CfA survey measures $\xi(r)$ out to only $\simeq 40h^{-1}$Mpc; hence the line ends there. 
The lines show that there is good agreement with the dot velocity correlation function values computed 
``locally" (CfA) and ``globally" (APM). There is also good agreement between the shape of the bulk velocity
flows in the Great Attractor region (Bertschinger et al 1990) and Fig.5. 
This is consistent with the assumption of scale independent bias
parameter $b$.

We ask now the following question: given the {\it measurement} of $\nu(r)$ on some scale what
should one expect to find for $\nu(0)$ given the {\it data} on $\xi(r)$? This can be evaluated from
\begin{equation}
\nu(0) = \nu(r) + \frac{\Omega^{1.2}}{b^2} H_0^2 [J_2(r) - \frac{J_3(r)}{r}]
\end{equation}
In order to evaluate $\nu(0)$ from $\nu(r)$ we
use the numbers from the POTENT analysis (Bertschinger et al 1990) who compute the bulk velocity
$V(r)$ field after filtering the data with a Gaussian filter of filtering length $r_f = 12h^{-1}$Mpc. 
The numbers from their analysis of the velocity field are given in Table 1 
which shows the values of $V(r)$,
the amplitude and direction, for three different $r$.

The data with which  $\nu(0)$ ought to be compared are admittedly incomplete, but it is generally
thought that for non-filtered velocity field $\sqrt{\nu(0)} \sim$ (500-600)km/sec
(cf. Peebles 1987). 
This also follows from 
the numbers for velocity dispersions in typical collapsed systems, such as groups of galaxies.
Local measure of  $\sqrt{\nu(0)}$ is probably the dipole velocity which is
observed to be 
630 km/sec with a very small error. (The direction of the microwave background dipole roughly coincides with the
velocity vector from Bertschinger et al (1990) in Table 1.). Since $\nu(r) \rightarrow 0$ as $r\rightarrow \infty$,
eq.(12) shows that the linear perturbative expression for $\nu(0)$ is $\propto J_2(r\rightarrow \infty)\equiv J_{2,\infty}$. 
In this case, because $\xi(r) \propto r^{-1.7}$ at small scales,
a non-negligible contribution to $\nu(0)$ comes from non-linear scales. This suggests the importance of smoothing when
evaluating $\nu(0)$ directly from the data. Peebles (1988) utilized the exact (non-perturbative) Layzer-Irvine cosmic energy
equation to compute $\nu(0)$ using the value of $J_{2,\infty}=164 {\rm e}^{\pm0.15}h^{-2}{\rm Mpc}^2$ evaluated
from the Lick data (Clutton-Brock and Peebles 1981). He obtained $\sqrt{\nu(0)} \simeq$(500-600) km/sec for
$\Omega$ in the range 0.2-0.3 in reasonable agreement with this discussion. 
For comparison,
integrating over $\xi(r)$ from the BE93 deprojection of the APM data gives 
$J_{2,\infty} \simeq 170 \pm 5 h^{-2}{\rm Mpc}^2$, while for the CfA data from Vogeley et al (1992) 
one obtains $J_{2,\infty} \simeq 186 h^{-2}{\rm Mpc}^2$. Both are in agreement with the 
numbers used in Peebles' (1988) analysis.

Table 2 shows the values of $\sqrt{\nu(0)}$ evaluated from eq.(12)
given the velocity data input on $\nu(r)\simeq V^2(r)$ at $r=$40 and 60$h^{-1}$Mpc
from Bertschinger et al (1990). The columns show numbers for $\xi(r)$ on scales $\leq 60h^{-1}$Mpc 
computed according to pure power-law, $\xi(r)=(r/r_*)^{-1-\gamma}$, the K92 and BE93 fits to the APM data;
and the CfA data on $\xi(r)$ from Vogeley (1992) at $r<40h^{-1}$Mpc.
The first column gives the filtering radius in $h^{-1}$Mpc, the 
second the value of $\Omega^{0.6}/b$ used in (3), the next seven columns give the values of $\sqrt{\nu(0)}$
in km/sec computed according to eq.(12) for the various $\xi(r)$. The first four rows show the numbers
with no filtering, the last four rows show the numbers for $r_f=12h^{-1}$Mpc used in Bertschinger et al
(1990) analysis of the velocity field. The data for the CfA measured $\xi(r)$ are shown only for 
$r=40h^{-1}$Mpc, the scale up to which the CfA data can probe $\xi(r)$. 
We have not shown the filtered values for the CfA $\xi(r)$ because the latter was not measured out
to large enough scales to make filtering with $r_f=12h^{-1}$Mpc meaningful. 

The various fits to the galaxy correlation give similar numbers which shows the robustness of 
the various methods. The data on $\xi(r)$ determined from the CfA survey come from the region that includes
galaxies used in the peculiar velocity analyses (Faber et al 1989) and agrees well with $\nu(0)$
predicted by the APM data. This is turn means that the Great Attractor is a typical, rather than
rare, mass concentration in the Universe. There is very good agreement between the numbers 
in Table 2 evaluated at both on 40 and on 60$h^{-1}$Mpc. This is consistent
with the constant bias factor over at least this range of scales.

Low values of $\Omega$ are implied by this analysis. E.g. for the non-filtered field
the predicted value of $\sqrt{\nu(0)}$ should be compared with the data on the dipole velocity
of only 600 km/sec. Also if $\Omega^{0.6}/b$ were as high as unity the typical velocity
dispersion in the collapsed systems would be around 1300km/sec, corresponding to X-ray 
temperature of $>$10 Kev. Observations, however, suggest that typical collapsed structures, such as 
groups and poor clusters of galaxies, have velocity dispersion $\simeq 500$ km/sec and rich
X-ray emitting clusters contain only a small fraction all galaxies. 
For the filtered field with $r_f=12h^{-1}$Mpc the predictions should be compared 
with the value from the Bertschinger et al (1990) analysis that gives 
$457\pm 61$ km/sec. Again the
best agreement between the data and $\nu(0)$ would be achieved for 
$\Omega^{0.6}/b \simeq 0.2-0.3$;
the fact that we reach essentially the same conclusions when comparing the filtered $\nu(r)$ is reassuring.

Thus the above analysis suggests that low-$\Omega$ Universe may not be in conflict with the data on
peculiar velocities and may in fact be even suggested by the latter. Similar conclusions are reached
applying the least action method (Peebles 1989,1990) to the dynamics of the local Group
(Shaya et al 1995). We have not quantified the conclusions from Table 2 in
statistical terms and, hence, take them only as suggestive. Below we propose a further extension of this
discussion that may determine the values of $\Omega^{0.6}/b$ in the new datasets on peculiar flows.

A similar analysis can be applied to the components of the velocity tensor: 
$U_{ij}=\langle { \boldmath\mbox{$v$} }_i({ \boldmath\mbox{$x$} }+{ \boldmath\mbox{$r$} })
\cdot { \boldmath\mbox{$v$} }_j({ \boldmath\mbox{$x$} })\rangle \equiv \Sigma(r) \delta_{ij}
+[\Pi(r)-\Sigma(r)]\frac{{\boldmath\mbox{$r$}}_i \cdot {\boldmath\mbox{$r$}}_j}{r_ir_j}$. 
For irrotational fluid the transverse
velocity correlation function, $\Sigma(r)$,
and the parallel, $\Pi(r)$ are interrelated via:
\begin{equation}
\Pi(r) = \frac{\partial r \Sigma(r)}{\partial r}
\end{equation}
and the dot (total) velocity correlation is given by (e.g. Gorski 1988)
\begin{equation}
\Pi(r)+ 2 \Sigma(r)=\nu(r)
\end{equation}
Eqs(10),(13) and(14) allow one to evaluate both components:
\begin{equation}
\Sigma(r) = \frac{1}{3} \nu(0) - H_0^2 \frac{\Omega^{1.2}}{b^2} \left[ \frac{1}{3}J_2(r)-
\frac{1}{2} \frac{J_3(r)}{r} + \frac{1}{6} \frac{J_5(r)}{r^3}\right]
\end{equation}
\begin{equation}
\Pi(r) = \frac{1}{3} \nu(0) - H_0^2 \frac{\Omega^{1.2}}{b^2} \left[ \frac{1}{3}J_2(r)-
\frac{1}{3} \frac{J_5(r)}{r^3}\right]
\end{equation}

Now note that the quantity $\Sigma(r)-\Pi(r)$ is independent of the (a priori unknown) integration
constant $\nu(0)$ and hence, once measured, it can provide a tool for measuring
$\Omega^{0.6}/b$. It is given by:
\begin{equation}
\Sigma(r)-\Pi(r)=H_0^2 \frac{\Omega^{1.2}}{b^2}\left[ \frac{1}{2} \frac{J_3(r)}{r} 
-\frac{1}{2} \frac{J_5(r)}{r^3}\right]
\end{equation}
Since the logarithmic slope of $\xi(r)$ is $>-2$, the expression on the right-hand-side of (17)
is dominated by the contribution from the integrands in $J_3,J_5$ near $r$. Therefore for this quantity
it is sufficient to choose a large enough separation to ensure the validity of the linear
approximation.

For a power-law $\xi(r)=(r/r_*)^{-1-\gamma}$ one gets $\sqrt{\Sigma(r)-\Pi(r)} = \frac{\Omega^{0.6}}{b}
\frac{H_0r_*}{\sqrt{(2-\gamma)(4-\gamma)}}(\frac{r}{r_*})^{\half(1-\gamma)} \simeq$(400-500)km/sec
for $\gamma=0.7$ if $\Omega^{0.6}/b\simeq 1$.
Fig.6 shows the predicted $\sqrt{\Sigma(r)-\Pi(r)}$ in km/sec for $\Omega^{0.6}/b=1$. The lines were evaluated 
using the CfA and APM data on $\xi(r)$; the notation is the same as in Fig.5. 
Again there is good agreement between the
values computed from the ``locally" determined correlation function (CfA - dashed line) and
the ``global" $\xi(r)$ from the APM data. The amplitude in Fig.6 scales 
$\propto \Omega^{0.6}/b$ and is quite significant if $\Omega^{0.6}/b \simeq 1$. The data
from previous analyses trying to reconstruct the velocity correlation tensor (Gorski et al 1989;
Groth et al 1989) from, by now old, catalogs are too uncertain to use in comparison with
Fig.6. But the magnitude of $\sqrt{\Sigma(r)-\Pi(r)}$ of $\sim$400 km/sec over
the range of (30-60)$h^{-1}$Mpc if $\Omega^{0.6}/b\simeq 1$ suggests that this could be a measurable
task with the new datasets and that in conjunction with the galaxy correlation data this may
further constrain $\Omega^{0.6}/b$.

Thus the analysis and results of this section suggest the following:
1) the spectrum of mass distribution from velocity data is consistent 
with that determined from the galaxy correlation data (cf. Kashlinsky 1992b). 
2) This in turn suggests that the bias factor is consistent with being scale-independent.
3) Low values of $\Omega^{0.6}/b$ may be consistent with the data on peculiar velocities.
4) Thus we interpret the results in Fig.4 as the rms density fluctuation of 
the pregalactic mass density field on scales $r_8 < r < 100h^{-1}$Mpc.

\section{Cluster correlation function and its richness dependence: reconstructing pregalactic 
spectrum on scales 5-25$h^{-1}$Mpc}

The cluster-cluster correlation function is known to have a larger amplitude than the one measured
by galaxies and its amplitude also increases with the cluster richness (Bahcall and Soneira 1983).
Kaiser (1984) suggested that the increase in the amplitude can be explained if clusters formed at the
rare peaks of the initial density field. His analysis was later applied by Bardeen et al (1986) to study
in great detail the properties of peaks in the density field. Kashlinsky (1987,1991b) proposed
that one can combine Kaiser's original suggestion with gravitational clustering theory (Press and
Schechter 1974) to explain both the amplification and its correlation with the cluster richness. He
supposed that structure formation in the Universe forms via a ``natural bias" (cf. Davis et al 1985), i.e.
systems of galaxies (clusters and groups) should be identified with regions that had turn-around time 
time less than the age of the Universe. The model successfully explained the correlation between the
cluster correlation function and richness (mass) with objects that 
turned-around on a larger scale having a greater correlation amplitude. 

Within the gravitational clustering model the properties of the hierarchy would then depend 
uniquely upon the \underline{initial} power spectrum and 
the mass of the objects that formed/turned-around. 
In this section we first evaluate in greater detail the predicted 
properties of the cluster correlation function according to the gravitational clustering picture (Kashlinsky
1987, 1991b) and then apply the results to invert the data on cluster correlation function
to obtain the \underline{pregalactic} spectrum of the density field on scales $5h^{-1}{\rm Mpc} < r< 20 
h^{-1}$Mpc.

We start with the density field at some initial epoch, $z_i \gg 1$, when density fluctuations were linear
on all scales of interest. The final results will be independent of $z_i$. We define with $\delta_M = \int 
\delta({\boldmath\mbox{$x$}})dV/\int dV$ the initial mass over-density over the comoving
volume $V$ that contains mass $M$. On large scales, $r\gg R(M)$, where $R(M)$ is the comoving scale containing
mass $M$, the correlation function of the $\delta_M$ field coincides with $\xi(r)$. At 
zero-lag it equals
the mean square fluctuation over mass $M$: $\Delta_i^2(M) = \langle \delta^2_M \rangle$; the subscript $i$
refers to the values evaluated at $z_i$. Thus the correlation matrix for the $\delta_M$-field is:
\begin{equation}
\{C_{lm} \} = \left( \begin{array}{ll}
	\Delta_i^2(M) \;\;\; \xi_i(r)\\
	\xi_i(r) \;\;\;\;\;\; \Delta_i^2(M)
	\end{array} \right)
\end{equation}

We assume that the \underline{initial} density field was gaussian. Then the probability density
to find two regions containing masses $M_1,M_2$ that had density fluctuations $\delta_1,\delta_2$ at $z_i$
is:
\begin{equation}
p(\delta_1;\delta_2) = \frac{1}{2\pi ||{\boldmath\mbox{$C$}}||}
\exp(-\half { \boldmath\mbox{$\delta \cdot C$}}^{-1} { \boldmath\mbox{$\cdot \delta$} })
\end{equation}
As the clustering evolves deviations from gaussianity will develop due to gravitational effects.
However, since we trace the distribution of the prospective clusters at $z_i$ 
these will not be important for the computation of the cluster correlation function below.
Using properties of the Fourier transform of a multidimensional Gaussian, eq.(19) can be rewritten 
in terms of the direct matrix $\boldmath\mbox{$C$}$ as:
\begin{equation}
p(\delta_1;\delta_2) = 
\frac{1}{2\pi^2} \int_{-\infty}^\infty \int_{-\infty}^\infty
\exp(-i{\boldmath\mbox{$q\cdot \delta$}}) 
\exp(-\half {\boldmath\mbox{$q\cdot C\cdot q$}}) d^2q
\end{equation}

We denote with $\delta_{ta}$ the amplitude at $z_i$ required for the fluctuation to turn around
at $z=0$ and use the Press-Schechter (1974) prescription for gravitational clustering which 
assumes that any region with $\delta_M \geq \delta_{ta}$ would have turned around by now. 
We further assume that clusters and groups of galaxies are identified with such
regions. Then the probability for two clusters
of masses $M_1,M_2$ to have formed at any time between now and $z_i$ is given by integrating
(2) over all fluctuations $\geq\delta_{ta}$. We denote with $\Delta_{8,i}$ the amplitude
the fluctuation had to have on scale $r_8$ at $z_i$ in order to grow to the observed value of $1/b$ at $z=0$.
A convenient way to characterize $\delta_{ta}$ and
to normalize the density field is by introducing a quantity $Q_{ta}=\frac{\delta_{ta}}{\Delta_{8,i}}$ (Kashlinsky 1991b).
For $\Delta_8=1$ today, $Q_{ta}\simeq 1.65$ almost independently of $\Omega$ or $z_i$. We adopt this value
of $Q_{ta}$ in the calculations in this section. We will show that this set of ``minimal assumptions",
including $b=1$, is justified since 
it gives a fit to the derived spectrum consistent with the APM data.

Expanding in eq.(20) $\exp(-\half{ \boldmath\mbox{$q\cdot C \cdot q$} }) = \sum_{m=0}^\infty
\frac{(-)^m}{m!} q_1^m q_2^m \xi_i^m \exp[-\frac{q_1^2 \Delta^2(M_1)}{2}]\exp[-\frac{q_2^2 \Delta^2(M_2)}{2}]$
we can write 
the probability for two clusters of mass $M_1,M_2$ 
to form at any time between $z_i$ and $z=0$ as:
\begin{equation}
P_{M_1M_2} = \int_{\delta_{ta}}^\infty\int_{\delta_{ta}}^\infty p(\delta_1,\delta_2)
d\delta_1 d\delta_2 = \frac{1}{2\pi^2} \sum_{m=0}^\infty \frac{(-)^m}{m!} 
\left[\frac{\xi_i}{\Delta_i(M_1)\Delta_i(M_2)}\right]^m 
a_m(\zeta_1) a_m(\zeta_2)
\end{equation}
where:
\begin{equation}
a_m(\zeta) = \int_{\zeta}^\infty dx \int_{-\infty}^\infty y^m \exp(-ixy) \exp(-\frac{y^2}{2}) dy
\end{equation}
The quantity $\zeta$ is the number of standard deviations the mass $M$ is
with respect to the ``typical" member of the hierarchy and is given by:
\begin{equation}
\zeta = Q_{ta} \frac{\Delta_8}{\Delta(M)} = Q_{ta} \left[\frac{
\int_0^\infty P(k) W_{TH}(kr_8) k^2 dk}{
\int_0^\infty P(k) W_{TH}(kR(M)) k^2 dk}\right]^\half
\end{equation}
where $R(M)$ is the scale containing mass $M$. The power spectrum in (23) is the original 
\underline{pregalactic} power spectrum of the density field. 

In order to directly evaluate $P_{M_1M_2}$ and the cluster correlation function
we write $\int_{-\infty}^\infty \exp(-ixy-\half y^2) y^m dy = i^m 
\frac{\partial^m}{\partial x^m} \int_{-\infty}^\infty \exp(-ixy-\half y^2) dy =
i^m\sqrt{2\pi} \frac{\partial^m}{\partial x^m} \exp(-\half x^2)$ (Jensen and Szalay 1986).
Then eqs.(21),(22) become:
\begin{equation}
P_{M_1M_2} = \frac{1}{2\pi} \sum_{m=0}^{\infty} \frac{1}{m!} 
\left[\frac{\xi_i}{\Delta_i(M_1)\Delta_i(M_2)}\right]^m 
\left[\frac{\partial^{m-1}\exp(-\frac{\zeta_1^2}{2})}{\partial\zeta_1^{m-1}}\right]
\left[\frac{\partial^{m-1}\exp(-\frac{\zeta_2^2}{2})}{\partial\zeta_2^{m-1}}\right]
\end{equation}
The probability of cluster of mass $M$ to form by now is for Gaussian ensemble given by:
\begin{equation}
P_M=\half {\rm erfc} \left( \frac{\zeta}{\sqrt{2}}\right)
\end{equation}

Eqs.(24),(25) give the cumulative probability of objects to turn around at any time in the past. 
In the gravitational clustering picture $P_{M}$'s include objects that formed at earlier epochs
and by now are incorporated into larger systems. Because clusters and groups form dissipationlessly,
the objects that formed earlier would merge into larger systems and lose their identity as clustering progresses
to larger masses. Since we observe at a fixed epoch ($z=0$), one must translate (24),(25) into
probabilities that object formed (turned-around) today.
Thus the fraction of pairs of clusters that formed today on mass scales $M_1,M_2$ separated by distance $r$ out
of ensemble of density fluctuations given by eq.(19) is $f_{M_1M_2} = 
\frac{\partial^2P_{M_1M_2}}{\partial M_1 \partial M_2}$. Similarly the fraction of clusters in the mass range
$[M,M+dM]$ is $f_M dM=\frac{\partial P_M}{\partial M} dM$  (Press and Schechter 1974). 
From (24),(25) one can show that
\begin{equation}
\frac{ f_{M_1M_2} }{ f_{M_1}f_{M_2} } =\frac{
\frac{ \partial^2P_{M_1M_2} }{ \partial \zeta_1 \partial \zeta_2 } }
{ \frac{ \partial P_{M_1} }{ \partial \zeta_1 }\frac{ \partial P_{M_2} }{ \partial \zeta_2 } }
= \sum_{m=0}^\infty \frac{1}{Q_{ta}^{2m}m!} \left(\frac{\xi_i}{\Delta_8^2}\right)^m 
B_m(\frac{\zeta_1}{\sqrt{2}})B_m(\frac{\zeta_2}{\sqrt{2}})
\end{equation}
where $B_m(x)=\half x^{m-1}H_{m+1}(x)$ (Kashlinsky 1991b) and 
$H_n(x) = (-)^n \exp(x^2)\frac{d^n}{dx^n} \exp(-x^2)$ are the Hermite polynomials.
We will be using the measurements of the cluster correlation function on linear scales, $r>r_8$.
On these scales the ratio on the right-hand-side of (26), $\xi_i/\Delta_8^2$, 
where both the numerator and denominator are evaluated at $z_i$, is roughly
independent of redshift and is approximately equal to the galaxy correlation function today, $\xi(r)$.

Having fixed the fractions of objects that turned-around today on a given scale we can
evaluate the correlation function between clusters of different masses $\xi_{M_1M_2}$. 
By definition, the 2-point correlation function of an ensemble of objects with number
density $n$ is given by the probability to find two objects in small volumes $dV_1, dV_2$
as $d{\cal P}_{12} = n^2(1+\xi)dV_1dV_2$. 
Since the clusters of mass $M_1,M_2$ will make the fraction $f_{M_1M_2}$ of such pairs,
the probability to find them is
$d {\cal P}_{M_1M_2} = f_{M_1M_2}d{\cal P}_{12}$. On the other hand, the mean number density of clusters
of mass $M$ would be $f_M \times n$ and by definition the probability to find two clusters is
$d {\cal P}_{M_1M_2} = f_{M_1}f_{M_2} n^2 (1+\xi_{M_1M_2})dV_1dV_2$. Hence the correlation function
of clusters of mass $M$ is given by:
\begin{equation}
1+ \xi_{M_1M_2} = \frac{ f_{M_1M_2} }{ f_{M_1}f_{M_2} } (1+\xi_i) = \frac{ 
\frac{ \partial^2P_{M_1M_2}}{ \partial \zeta_1 \partial \zeta_2 } }
{ \frac{ \partial P_{M_1} }{ \partial \zeta_1 }\frac{ \partial P_{M_2} }{ \partial \zeta_2 } }
(1+\xi_i)
\end{equation}

Eq.(27) fixes the factor by which the cluster correlation function is amplified over
the underlying correlation function of the hierarchy, $\xi$. 
The amplification is  purely statistical and the discussion does not involve any dynamical effects.
Hence, on linear scales today the amplification factor $\xi_{MM}/\xi_i$ is redshift independent and 
for $r>r_8$ the present-day correlation function for clusters of mass $M$ should be amplified 
over $\xi(r)$ according to:
\begin{equation}
\xi_{MM}(r) = A_M(\frac{\zeta}{\sqrt{2}}) \xi(r)
\end{equation}
where:
\begin{equation}
A_M(x) = \sum_{m=0}^\infty \frac{1}{ Q_{ta}^{2m} m!} \xi^m C_M(x)
\end{equation}
and
\begin{equation}
C_m(x) = \frac{ x^{2m} }{4} \left[ \frac{ H_{m+1}^2(x) }{ x^2 } 
+\frac{ H_{m+2}^2(x) }{ (m+1)Q_{ta}^2}\right]
\end{equation}
In evaluating (28)-(30) we used that the ratio $\xi_i/\Delta_{8,i}$ evaluated at $z_i$ for scales $r>r_8$ 
is $\simeq \xi(r)$, the galaxy 2-point correlation function that is measured today.
As $x\rightarrow 0$ eq.(30) gives $C_0 \rightarrow 1$ and one gets from the first term in (28) that 
at small $x$:
$\xi_{MM}(r) \simeq \xi(r)$. For large masses or $x \gg 1$ the first term in (29) is $C_0 =\zeta^4$ and
the amplification reduces to $\xi_{MM}(r) \simeq Q_{ta}^4 [\Delta_8/\Delta(M)]^4 \xi(r)$ (cf.
Kashlinsky 1987). (The quantity $Q_{ta}$ in this discussion is equivalent to the threshold 
amplitude $b$ of Kaiser 1984).

Eqs.(28)-(30) show that the spectrum of the pregalactic density field can be constrained, and
as we show later determined, by the data on both the slope of the observed
cluster correlation function $\xi_{CC}(r)$ via eq.(28),
and the dependence of the amplitude on the cluster mass via eq.(29). 
Bahcall and Soneira (1983) present the data on both of these for scales $>r_8$ where the present
analysis applies. The slope of the cluster correlation function on very large scales,
$r > 50h^{-1}$Mpc, is poorly determined
from the data (cf. Fig.9 of Bahcall and Soneira 1983). For richness class ${\cal R} \geq1$ Abell clusters,
such as Coma, they approximate $\xi_{\rm CC}(r) \simeq 300 (r/1h^{-1}{\rm Mpc})^{-1.8}$. This 
approximation has large uncertainty at large scales, but for $r <$(30-40)$h^{-1}$Mpc it can be used
as a reasonable approximation to the data. The Zwicky clusters, which are poorer, also exhibit a stronger
correlation amplitude than galaxies (Postman et al 1986), although it is weaker than that of Abell clusters
and is consistent with the amplitude-richness relation proposed in Bahcall and Soneira (1983). Fig.2 in Bahcall
and West (1992) shows the largest compilation of the data on cluster correlation amplitude vs richness.
Thus we use below the data on the cluster correlation function from Bahcall and Soneira (1983) and on
the amplitude-correlation amplitude dependence for clusters of galaxies from Bahcall and West (1992).

In order to compare eqs.(28)-(30) to the data discussed above, we must fix the mass of the clusters
quantitatively. We use Coma, the best studied cluster, as the mass normalization 
point. From Kent and Gunn (1981) we adopt its 
mass to be $M_{\rm Coma} = 1.45\times 10^{15}h^{-1}M_\odot$ (see also
White et al 1993 and references cited therein). The richness of the Coma cluster is adopted from
Bahcall (1981) and Abell et al (1989) to be ${\cal N}_{\rm Coma} = 106$. Assuming 
that both the luminosity function of galaxies 
and the ratio of dark-to-luminous matter are universal for all clusters
would imply that the
cluster richness is proportional to mass. We thus adopt the following relation between cluster
mass and richness:
\begin{equation}
M = M_{\rm Coma} \frac{{\cal N}}{{\cal N}_{\rm Coma}} = 
1.45\times 10^{15} \left(\frac{{\cal N}}{106}\right) h^{-1}M_\odot
\end{equation}
while the comoving scale containing mass $M$ 
that enters in the integrand of denominator of (23) is given by:
\begin{equation}
R(M) = 0.2 \left(\frac{M}{10^{10}h^{-1}\Msun}\right)^\third \Omega^{-\third} h^{-1} {\rm Mpc}
\end{equation}

We can now move to the implications of the cluster correlation data for the spectrum of the pregalactic
density field. For a given $P(k)$ eqs.(3) and (23) specify $\xi(r)$ and $\Delta(M)/\Delta_8$
which in turn uniquely determine both the cluster correlation function vs $r$ and its amplitude vs 
the mass computed from the cluster richness according to eq.(31). For CDM models the value of $\Omega h$ 
is not the only parameter that fixes the
cluster correlation function; the extra dependence on $h$ comes from translating the mass-scales
to linear scales via eq.(32).
Fig.8 plots the cluster correlation in CDM models with the Harrison-Zeldovich power spectrum, 
$n=1$ for various $\Omega$ and $h$. Thin solid lines correspond to the underlying correlation 
function, $\xi(r)$, for given $\Omega,h$. Thick solid lines correspond to the $\xi_{CC} = 300 
(r/1h^{-1}{\rm Mpc})^{-1.8}$ that Bahcall and Soneira find for ${\cal R}\geq 1$ clusters. The numbers
are plotted between 20$h^{-1}$ and 100$h^{-1}$Mpc where the galaxy correlation function is in
the linear regime and thus is given by the pregalactic power spectrum.
Dotted lines correspond to ${\cal N}=5$, dashes to ${\cal N}=10$, dashed-dotted lines to ${\cal 
N}=50$ and dashed-dotted-dotted lines to clusters such as Coma, ${\cal N}=100$. Successful
models should fit both the amplitude and the slope of the measured $\xi_{CC}$, 
while at the same time the underlying
$\xi(r)$ should be in agreement with the APM data constraints in Fig.3. As one can see from the 
figure, the required increase in the amplitude for clusters of richness class ${\cal R}\geq 1$
(${\cal N} \simeq 100$) can be achieved for $\Omega\sim$(0.3-0.4), but such models would not
reproduce the observed slope of $\xi_{CC}(r)$ even on scales $\leq$(30-40)$h^{-1}$Mpc where the
data is more reliable and is definitely not expected to be subject to possible projection effects
(Sutherland 1988). The required slope of $\xi_{CC}(r)$ can be reproduced for $\Omega=0.1$
and $h=0.5$, but then the observed amplitude would be reached for ${\cal N}=10$,
corresponding to poor groups, while clusters such as Coma should have correlation amplitude 
significantly above that observed. Furthermore, such models with $\Omega h\simeq 0.05$ would have
values for $\theta(w=5\times 10^{-3})$ - plotted with asterisks in Fig.3 - that are much bigger
than the values deduced from the APM data. For brevity we do not present the same graph for
$h=1$, but the agreement with the measurements would become even worse for $h >0.5$.

Fig.8 plots the predicted $\xi_{MM}(r)$ for tilted CDM models with $n=0.7$; the line notation is the same
as in Fig.7. The tilted CDM model also does not fit well the data on the cluster correlation 
function or the APM data. E.g. the model can reasonably reproduce the slope and the amplitude
of Abell clusters if $\Omega =0.3, h=0.5$, but then with $\Omega h \simeq0.15$ it would predict
$\theta(w=5\times10^{-3})$ in Fig.3 far above the APM data.

In order to illustrate the dependence of the amplitude on mass or richness for CDM models we computed 
the amplification factor, eq.(29), at $r=25h^{-1}$Mpc. This scale is $\gg r_8$ so it is
reasonable to suppose that the density field there reflects the pregalactic density field. At the
same time, on this scale the underlying $\xi(r)$ is still positive  
in all relevant CDM models.
Fig.9 plots the values of $A(25h^{-1}{\rm Mpc})$ vs the cluster richness for CDM models. Right 
box shows the numbers for $n=1$ and the left box shows them for tilted CDM models. Triangles
correspond to the data from Fig.3 of Bahcall and West (1992). Solid lines correspond to 
$\Omega h=0.1$, dotted to $\Omega h=0.2$, dashes to $\Omega h=0.3$ and the dashed-dotted 
lines correspond to $\Omega h=0.5$. Thin lines of each type are for $h=1$ and thick lines
are for $h=0.5$. One can see that it is difficult to describe the data with one particular
CDM model. In other words, the slope of the pregalactic power spectrum over different range
of scales as probed by Figs.7,8,9 cannot be fitted with one formula given by eq.(2) for any
value of $\Omega, h$ or $n$. Further difficulty would come from constraining the CDM models to
fit the APM data along with the cluster correlation data. As Fig.3 shows, CDM models 
would be consistent with APM data only  if $\Omega h\simeq0.2$ for $n=1$ and $\Omega h\simeq 0.3$ if $n=0.7$.
This would further restrict the models to dotted lines in the left box of the figure and to dashed lines in the
right box.

One can now reverse the problem and instead of fitting various theoretical models to the cluster
correlation amplitude - richness data, one can invert the data to {\it obtain the implied
pregalactic density spectrum from eq.(29) with $x$ given by eq.(23)}. This can be done if one
uses independent measurement of the underlying correlation function, $\xi(r)$, in linear 
regime. Then the latter can be substituted into (29) with $x$ given by (23) 
to give an equation solving which one can
determine, given the measurements of $A(M)$ vs richness for clusters, the
values of $\Delta(M)/\Delta_8$ of the pregalactic density field on mass scale $M$ 
(in turn given by eq.31). First to the choice of scale $r$ on which 
the data on $\xi(r)$ is to be used. Such scale must be $>r_8$ since $\xi(r)$ in eq.(29)
is given by the pregalactic power spectrum at $z_i$. Fig.10 plots $\xi(r)$ 
from the various fits to the APM data. Solid lines are the K92 fits with $k_0^{-1}=35,40,50 h^{-1}$Mpc
from bottom to top. Dotted lines are for the BE93 fit with top and bottom lines corresponding to the
one standard deviation uncertainty. For comparison we also plot with dashed lines the correlation
function according to CDM models (eq.2) with $n=1$ for $\Omega h=0.2$. ($n=0.7$ and $\Omega h=0.3$ line 
would essentially coincide with the thick line over the linear scales; 
hence it is not plotted). All the fits to the APM data
give essentially the same $\xi(r)$ between 10 and 30$h^{-1}$Mpc. 
Thus we choose scale $r=25h^{-1}$Mpc to
substitute the measured $\xi(r)$ into (29). The latter is sufficiently larger
 than $r_8$ to be certain that $\xi(r)$
there reflects the pregalactic density field and, 
at the same time, sufficiently small 
so that the cluster correlation amplitude measurements
are free from the possible uncertainties at larger scales. Using the data on
$r$ anywhere between 15 and 30$h^{-1}$Mpc would
introduce little difference in the results to come. From Fig.10 we adopt 
scale $r=25h^{-1}$Mpc with
the measured correlation function there $\xi(25h^{-1}{\rm Mpc}) =0.07$. The same
number would be given also by a simple power extrapolation of $\xi(r)=(r/r_*)^{-1.7}$ to $r=25h^{-1}$Mpc.

Solid line in Fig.11 plots the amplification factor $A(x)$ at $25h^{-1}$Mpc
vs $x$ according to eq.(29) after adopting $\xi(25h^{-1}{\rm Mpc})=0.07$.
One can see that at small $x$ (small mass scales) $A(x)$ has a very weak
dependence on $x$. Hence for small mass scales the determination of $x$, or $\Delta(M)/\Delta_8 =
\frac{Q_{ta}}{x\sqrt{2}}$, is less accurate since the errors in the data will amplify. On the other hand,
at large $x$ the dependence is quite steep and the determined spectrum will be less sensitive
to observational errors in cluster correlation measurements
for massive clusters. Note that the first (and leading at large $x$ and small $\xi$) 
term in (29) is independent of $\xi(r)$. The two dotted lines show the uncertainty in $A(x)$ introduced by 
assuming e.g. a 25 \% uncertainty in the value of $\xi$, i.e. the lines plotted cover the range of 
$0.05 < \xi(25h^{-1}{\rm Mpc}) < 0.1$. One can see that the values of $\Delta(M)/\Delta_8$
determined in this way depend very weakly on the possible uncertainties in $\xi(r)$.

We now use the data from Fig.3 of Bahcall and West (1992), plotted with triangles in Fig.9,
in order 
to invert eq.(29), with $\xi(25h^{-1}$Mpc)=0.07, to give $\Delta(M)/\Delta_8$ of the pregalactic density
field on scale $M$, or richness ${\cal N}$. Fig.12 plots the results of this inversion. The top 
horizonal axis plots the values of ${\cal N}$ at which $\Delta(M)/\Delta_8$ has been evaluated. The bottom
horizontal axis shows the mass computed according to the normalization given by (31). We emphasize again
that this method gives directly the pregalactic spectrum as it was at $z_i$ independently of the later
gravitational or other effects. The plot in Fig.12 shows a
clearly defined slope of $\Delta(M) \propto M^{-0.275}$ corresponding to the spectral index of
$n \simeq -1.3$. This slope is consistent with the APM implied power spectrum index of $w(\theta)
\propto \theta^{-0.7}$.

Finally, we note that even if the underlying density field is purely gaussian, the distribution
of clusters of galaxies would be non-gaussian (Politzer and Wise 1984, Kashlinsky 1991b). 
Kashlinsky (1991b) analysed the properties of the three-point correlation function of
clusters predicted by this model and found that they compare favorably with the measurements
of the cluster 3-point correlation function by Toth et al (1989) assuming that the pregalactic
power spectrum is close to $n\simeq-1$ on all scales $< 100h^{-1}$Mpc. In principle, one could 
apply similar methods to derive the properties of the pregalactic density field from the 3-point
correlation function of clusters of various richness/mass. The data available at present do
not justify a lengthy addition on this to the paper and we postpone this part of discussion
to a forthcoming paper.

\section{Combining the results from $w(\theta)$ and $\xi_{cc}$: $\Omega$ and the pregalactic
density field}

The agreement of the slope of the spectrum in Fig.12 determined from the cluster correlation amplitude
vs richness data with that deduced from the APM catalog is encouraging and 
shows consistency 
of this approach. Furthermore, the APM dataset fixes the spectrum of the pregalactic density on a given 
linear scale, $r$, while the method outlined in the previous section determines it on a given
mass scale, $M$. The resultant spectra are consistent in slope and 
requiring them to give identical
amplitude at a given linear (or mass) scale could then constrain $\Omega$ according to eq.(32).

Fig.13 shows $\Delta(M)/\Delta_8$ from Fig.12 plotted as function of the linear scale containing
the mass on the lower horizontal axis of Fig.12. The linear scale $r$ in $h^{-1}$Mpc
was computed according to eq.(32). Left box of Fig.13 shows the numbers for $\Omega=1$ (thick
plus signs) and $\Omega=0.1$ (thin plus signs). The open square in the figure shows the value
of 1 for $\Delta(M)/\Delta_8$ at $r_8$; for the pregalactic density field 
this number must be unity {\it by definition}. 
Clearly $\Omega=1$ would be difficult to reconcile with $\Delta(M)/\Delta_8=1$ at $r_8$; 
extrapolating from {\it all} the data points
misses the square by a significant factor. A value of $\Omega<1$ would be required to reproduce the unity
value for $\Delta(M)/\Delta_8$ at $r_8$ as obtained from the cluster correlation amplitude
vs richness dependence.

We now use (32) to determine $\Omega$ by requiring the pregalactic density field 
derived from the cluster correlation function to match that derived from the APM data on the same
scales. The right box of Fig.13 shows the intercomparison. The lines are in the same notation as 
in Fig.10: solid lines correspond to the K92 fit and dotted lines correspond to the BE93 fit. The
plus and square symbols correspond to the same notation as in the left box of the figure. The rhombs
correspond to $\Omega=0.25$ with which the best agreement between the $\Delta(M)/\Delta_8$ 
values derived from the cluster correlation amplitude and APM dataset is achieved. Note that
any deviations from spherical approximation will speed up the collapse (Peebles 1980) 
leading to \underline{smaller} value of $Q_{ta}$.
This will shift the values of $\Delta(M)/\Delta_8$ from the cluster correlation analysis that fit 
APM data best towards even lower values of $\Omega$.

It is interesting to note that the value of $\Omega=0.25$ 
deduced in this analysis is in good agreement
with that required by the dynamics of the Coma cluster. This cluster is by far the best studied
and was also used in the normalization of the mass-richness relation used in Sec.5. The blue mass-to-light
ratio of the Coma cluster is very accurately determined to be 
$M/L_B \simeq 362h$ in agreement with the value of
$M_{\rm Coma}$ used in (31). If one adopts the blue luminosity function of the Schechter form
$\Phi(L) dL = \Phi_* (L/L_*)^{-\alpha} \exp(-L/L_*) d(L/L_*)$ 
and assumes that the entire mass in the Universe is associated
with galaxies and their systems, the total mass-to-light ratio of galactic systems would be uniquely 
related to $\Omega$ via: 
\begin{equation}
\frac{M}{L_B} = \frac{3H_0^2}{8\pi G \Phi_* L_* \Gamma(2-\alpha)} \Omega = 
1330 h \; \; \Omega \; \; \; \; {\rm solar \; units}
\end{equation}
where $\Gamma(x)$ is the gamma function. 
In eq.(33) the numbers were evaluated using the data on the luminosity function from the 
latest APM survey of Loveday et al (1992). Thus dynamics of the Coma cluster implies the same
value of the density parameter as we find from the independent analysis in this section.
(Assuming that the mass-to-light ratio of galaxies varies according to the fundamental plane of
ellipticals will result in $<10\%$ correction, as will be discussed in the next section).

To conclude the sections on the cluster correlation function and the pregalactic density field:
1) We find that CDM models with any $\Omega,h$ cannot fit simultaneously the data on the slope of the
cluster correlation function, the dependence of its amplitude on richness and the APM angular
correlation function.
2) We develop a method to determine the values of $\Delta(M)/\Delta_8$ of the pregalactic density
field directly from the data on the cluster correlation amplitude vs richness. The values determined in this
way are independent of the numerical value of the biasing factor.
3) The slope of the pregalactic density field, $\Delta(M)/\Delta_8$ vs the mass $M$, found in this
way is consistent with that determined from APM.
4) Comparison of the amplitude of $\Delta(M)/\Delta_8$ on a given linear scale $r$ determined
from the cluster correlation data with that 
determined from the APM survey fixes the value of $\Omega$.
5) The value of $\Omega$ determined in this way turns out $\simeq $0.25-0.3; the same 
value is implied by the dynamics of the Coma cluster. 6) If $\Omega=1$ the amplitude of $\Delta(M)/\Delta_8$ 
determined from the cluster correlation data does not pass through unity at 8$h^{-1}$Mpc.

\section{Galaxy formation and high-$z$ objects: constraining the small scale pregalactic
density field}

In the previous sections we discussed constraints on the spectrum of the pregalactic density field from
the present-day data. The smallest scales on which such constraints allow to probe the pregalactic
density field come the cluster correlation amplitude - richness dependence and are around $5h^{-1}$Mpc
(cf. Fig.13). On smaller scales the spectrum of the pregalactic density field is constrained by 
observations of collapsed objects at high-$z$ (Cavalieri and Szalay 1986; Efstathiou and Rees 1988 -
hereafter ER88;
Kashlinsky and Jones 1991; Kashlinsky 1993 - hereafter K93). Since the amplitude of the power
spectrum is fixed by requirement that at $z=0$ it reproduce unity fluctuation in 
galaxy counts over a sphere of
radius $r_8$, for a given spectrum this normalization determines at what $z$ the first objects
of a given mass-scale would typically collapse. E.g. in the CDM models (eq.2) any increase in the amount of
power on large scales would come at the expense of the power on small scales. Consequently, observations of
collapsed objects at high redshifts when used in conjunction with the large-scale structure data can set
strong constraints on CDM models (K93). In this section we discuss what the latest observational data
imply. We also discuss the way the properties of the present day (elliptical) galaxies constrain the 
density fluctuations out of which they grew and collapsed.

The data at high $z$ which we use comes from observations of three types of objects at high redshifts: QSOs,
galaxies and clusters of galaxies:

The latest grism surveys have been successful in finding quasars out to $z\simeq5$ 
(e.g. Schneider et al 1992,1994). There are now
a couple dozen of quasars with $z > 4$ (e.g. Turner 1991 and references cited therein). This highest redshift
comes from an optically selected quasar with $z\simeq 4.9$ (Schneider et al 1991). There are suggestions that 
the true QSO abundance at high redshift may be significantly higher with most quasars hidden by 
dust obscuration (Fall and Pei 1993). On the other hand, analysis of the grism surveys shows 
that the QSO comoving density peaks at $z \sim$(2-3) and drops by a factor $\sim 5$ by redshift $\simeq 4.5$
(Schmidt, Schneider and Gunn 1991). This is also confirmed by data from the radio-loud quasars which
are less likely to be affected by dust (Shaver et al 1996).
Attempts have been made to use the QSO comoving density data at high $z$ to test cosmological models
(ER88; K93; Mahonen et al 1995). In order to do this one has to derive the total collapsed masses
associated with high-redshift quasars and if, as is commonly assumed, QSOs hide an underlying
galaxy, one has to further assume the efficiency with which galactic central nuclei lead to QSO phenomena
(e.g. Begelman and Rees 1978, ER88). The lower bound on the mass comes from the Eddington luminosity
limit on the total quasar power and the upper limit on the efficiency is $<1$. If one assumes that a galaxy
is hidden behind each quasar the total mass that collapsed at its redshift would be $>10^{12}M_\odot$;
same number would be required by reasonable energy conversion efficiencies associated with the central
engines (ER88; Turner 1991). In this case, low-$\Omega$ CDM models would be the most difficult to reconcile
with the QSO abundance at $z\simeq$(4.5-5), although Haenhelt and Rees (1993) argue that the 
required factors may be brought in agreement with CDM requirements at high $z$. While quasars can provide
useful diagnostics of the various models, the uncertainties in their total mass, number densities as well
as in the redshift of their collapse, rather than the redshift at which they are observed, make such constraints
somewhat model dependent.

Clusters of galaxies subtend linear scales comparable to $r_8$ , where the rms density contrast today is unity.
Therefore, existence of collapsed clusters of galaxies at high redshifts would imply a large initial
density contrast on scale comparable to $ r_8$ and would be difficult to reproduce in models with $\Omega=1$
(K93). The data on the existence of clusters of galaxies at high $z$ is only now becoming available, but
it already indicates that galaxies may have assembled into clusters as rich as Coma somewhere before
$z\sim$(1-2). Pascarelle et al (1996) reported a serendipitous discovery of a cluster or group of galaxies
at $z\simeq 2.4$ in the Hubble deep field. Francis et al (1996) report the discovery of a group of red,
and therefore old, galaxies at roughly the same redshift. LeFevre et al (1996) discovered a cluster (or
possibly group) of galaxies at $z\simeq 3.14$. Dickinson (1993) reported observations indicating
presence of rich galaxy clusters with a population of extremely red galaxies at $z=1.2$. Recently,
Jones et al (1997) detected a decrement in the 
temperature of the microwave background of $560\mu$K toward a pair of quasars at $z\simeq 3.8$. Assuming this to 
be due to the Sunyaev-Zeldovich effect they searched for cluster of galaxies in that direction and concluded that the
cluster must be at $z>1$ with the total mass of $\sim10^{15}M_\odot$ (Saunders et al 1997). 
Furthermore, such cluster mass and
redshift are consistent with the quasar pair, observed to have very similar spectra and 
separated by $\simeq 10^{"}$, being
gravitationally lensed images of the same quasar. Deltorn et al (1997) identified
what is probably a massive cluster $(\sim 10^{15}M_\odot)$ at $z\simeq1$ around a high redshift galaxy 3CR184
at the same redshift. The cluster is identified through both the gravitational arc near the central galaxy and
by the excess of galaxies in the redshift distribution around 3CR184.
All this suggests that clusters of galaxies may already have formed at $z\gg 1$. Implications of the
existence of such massive clusters at high redshifts for CDM models were discussed in K93. Fig.2c in K93 shows
that such massive objects must be extremely rare fluctuations, $>$(7-10) standard deviations, 
in the density field of the CDM
models. Most easily they could be explained by assuming low-$\Omega$ \underline{and} the pregalactic 
density field having power in excess of the CDM models at these scales.

Galaxies at high redshifts are, probably, the most useful tool in constraining the power spectrum because in this
case (at least) stellar mass can be fairly reliably estimated from colours 
and stellar populations. Until recently,
only a handful of galaxies, all of them steep spectrum radio sources, have been found at high $z$, the most distant of
which were 0902+34 and 4C41.17 at redshifts of 3.4 and 3.8 (Chambers et al 1990, Lilly 1988). Chambers and Charlot
modelled their $K$-band photometry from Lilly (1988) with a rapid burst of star formation and concluded 
that the redshifts of formation must have been $>4$. Eisenhardt and Dickinson (1992) have re-observed the 0902+34
galaxy at $z=3.8$ in $K$-band and argue for its younger age. Eales et al (1993) have discovered an optically luminous and
massive radio galaxy 6C 1232+39 at $z=3.22$. Lacy et al (1994) found a radio galaxy 8C1435+635 with a record
$z=4.25$. Djorgovski et al (1996) report the discovery of a galaxy responsible for a high-redshift damped
Ly-$\alpha$ system at $z=3.15$; from the velocity field they estimate the dynamical mass of the galaxy
at $>2 \times 10^{11}h^{-1}M_\odot$ within 15$h^{-1}$Kpc.  Hubble deep field observations of
Steidel et al (1996) and Giavalisco et al (1996) find  
population of normal star forming galaxies at $z>3$. Lowenthal et al (1997) have followed with spectroscopic
observations at Keck of the Hubble deep field galaxies and with the 16 confirmed sources concluded that the
comoving density of $z>3$ galaxies is $> 2 \times 10^{-3}h^3$Mpc$^{-3}$ if $\Omega=0.1$ or $> 10^{-2}h^3$Mpc$^{-2}$
if $\Omega=1$. For comparison, the comoving density of $L_*$ galaxies today is $\Phi_* \simeq 1.4\times
10^{-2}h^3$Mpc (Loveday et al 1992).
Trager et al (1997), using the Keck telescope, discovered four high-redshift galaxies: one at $z=3.35$ 
and three at $z\simeq 4$. Using the Ly-$\alpha$ break as redshift indicator Hu and McMahon (1996) reported the
possible discovery of star-forming Ly-$\alpha$ emitting galaxies at $z=4.55$. The above data indicate that galaxies
must have been present already at $z\sim 5$. In an exciting recent discovery Franx et al (1997) identified a
gravitationally lensed arc as galaxy at $z=4.92$ and further find a companion to it with a radial velocity
of only 450 km/sec. Thus the data show that galaxies must already have been present in the Universe at
redshifts beyond $\sim 5$. The existence of such early formation (collapse) of objects on scales
$\sim 10^{12}M_\odot$ is indicative of small-scale power in the pregalactic density field in excess of that given
by the low-$\Omega$ CDM models required by the APM data (K93).

Further indications of when galaxies must have collapsed come 
from recent observations of galaxies at smaller redshifts, but
for which detailed spectroscopy and therefore accurate age determination are available. Hu and Ridgway (1994) 
found galaxy candidates whose colours were consistent with colours of  ellipticals with old stellar populations
shifted to $z\sim 2.5$ and have provided the first evidence that galaxies at high redshifts may already contain old 
stellar populations. A real breakthrough came with observations of the 53W091 galaxy at $z=1.55$ by Dunlop et al 
(1996, hereafter D96). D96 obtained high resolution spectroscopy of 53W091 with the Keck telescope 
that revealed a very old stellar population in the galaxy. By fitting the spectrum with stellar models
they determined its age to be $t_{53W091}=3.5$Gyr with a very small dispersion (0.5Gyr at 90\% confidence
level). This would rule out Einstein-de Sitter Universe for any $H_0>40$km/sec/Mpc and for open or
$\Lambda$-dominated flat models would place formation of 53W091
at high redshifts $z>$(4-5). That this galaxy is unlikely to be either unique or unusual is evidenced by discovery
of more such objects: e.g.  53W069 at $z=1.41$ with age $t_{53W069}=4$Gyr and 0.5Gyr uncertainty (Dey et al
1997). Francis, Woodgate and Danks (1997) report discovery two high redshift galaxies at $z=2.38$ with old stellar
population of $>0.5$Gyr.

Kashlinsky and Jimenez (1997, hereafter KJ97) discussed implications of the 53W091 
data on low-$\Omega$ flat  ($\Omega+\lambda=1$ where $\lambda = \frac{\Lambda}{3H_0^2}$) CDM models. They
pointed out that the redshift of formation of 53W091 required by its age would decrease with decreasing
$\Omega$. But at the same, decreasing $\Omega$ would in such models suppress the power on small scales thereby
delaying
galaxy collapse. KJ97 computed the total stellar mass of 53W091 by fitting the observed flux in all (V,J,H,K) bands
with synthetic galaxy spectra based on the Miller-Scalo initial mass function for stars. The total mass in stars
alone is given in Table 1 of KJ97 and is $\geq 10^{12}M_\odot$ inside the aperture radius subtending
(8-15)$h^{-1}$ Kpc at $z=1.55$. The resultant stellar mass has little dependence on cosmological
parameters or metallicity. KJ97 showed that in the flat $\Lambda$-dominated CDM cosmogonies 53W091 would have
to be greater than $\simeq5$ standard deviations in the pregalactic density field. 

Further observational constraints on $\Omega+\lambda=1$ cosmologies come from the data on high-redshift type Ia
supernovae. The latest measurements include 7 SNIa out to $z\simeq 0.46$ and set limits of $\lambda < 0.35$
at 68\% or $<0.51$ at 95\% confidence levels (Perlmutter et al 1997). Similarly, the data on gravitational lenses 
implies $\lambda < 0.66$ at 95\% confidence level (Kochanek 1996). Together with the KJ97 results this makes flat 
$\Lambda$-dominated cosmogonies less attractive than open Universe cosmogonies. Therefore, in the remainder
of this section we evaluate the limits on the pregalactic density field assuming $\Lambda=0$, but most of the results
can be relatively easily extended to $\Lambda \neq 0$. For $\Omega+\lambda=1$ Universe our main conclusions will not
differ appreciably from the open Universe case.

As discussed above the epoch of galaxy collapse constrains the pregalactic power spectrum on galaxy scales and 
the data on the 53W091 galaxy places strong constraints on 
$\Lambda$-dominated flat Universe CDM models. In what follows we 
generalize the discussion of KJ97 to open Universe. Fig.14 plots the redshift of formation of stellar population
of 53W091 vs $\Omega$. Dotted lines correspond to the best estimate of its age from D96, $t_{53W01}=3.5$Gyr.
Solid and dashed lines correspond to $\pm 0.5$Gyr uncertainty in $t_{53W01}$ at 95\% confidence level. Three
lines of each type correspond to the Hubble constant values of $h=0.5,0.6$ and 0.75. The age of 
53W091 implies large values of the redshift of its formation, $z_{\rm for}\geq 5$, and agrees well with observations
of other high redshift galaxies discussed earlier in this section. 
The numbers for another galaxy, 53W069 (Dey et al 1997), would
be similar, and both galaxies are inconsistent with the $\Omega=1$ Universe for any reasonable value of the
Hubble constant ($h>0.4$).

In order to estimate the likelihood of formation of objects like 53W091 in open CDM models we proceed in the
manner outlined in K93. We start with density field in linear regime at some early epoch $z_i$; in the formalism
outlined in K93 and KJ97 the final results are independent of $z_i$. We denote with $\delta_{col}$ the amplitude which
density fluctuation had to have at $z_i$ in order to collapse at redshift $z$. It can be evaluated from the spherical
model approximation (e.g. Peebles 1980; Gott and Rees 1975; Narayan and White 1988; K93). The time evolution of 
a spherical shell that contained density contrast $\delta_i$ at $z_i$ can be described by its ``Friedman" equation:
$\dot{r}^2 =A/r+C$ where $A=H_0^2\Omega$ and $-C=H_0^2[\third\Omega \delta_i(1+z_i)+\Omega-1]$ and
$r$ is expansion factor of the fluctuation. The
turn-around time is given by $t(\dot{r}=0)=\frac{\pi}{2} A (-C)^{3/2}$ and the collapse time is $t_{col}=2t_{ta}$.
The value of $\delta_{col}$ is then determined by the condition that $t_{col}$ equals the cosmic time that
elapsed between $z_i$ and the redshift $z$ when the fluctuation collapses:
\begin{equation}
\frac{\pi}{H_0} \frac{ \Omega} { [\third \Omega \delta_{col}z_i + \Omega -1]^{3/2} } = 
\int_{z}^{z_i} \frac{dz}{ (1+z)^2 \sqrt{1+\Omega z} }
\end{equation}
In order to normalize the density field at $z_i$ and also eliminate the dependence on $z_i$ we compute also
the amplitude, $\Delta_{8,i}$, which the fluctuation had to have at $z_i$ in order to grow to $1/b$ amplitude at $z=0$.
This can be done by noticing that conservation of mass requires $1+\delta(z) = 
(1+\delta_i)[(r_i/r(z)]^3$. The fluctuation starts with $\Delta_{8,i}$ and must reach density contrast of
$1/b$ at $z=0$. The time it takes is $\int_{r_i}^{r(z=0)} dt/\dot{r}$ with $r(z=0),r_i$ 
related by the above mass-conservation expression. Hence the equation for $\Delta_{8,i}$ is given by:
\begin{equation}
\int_{ \frac{ 1-\Delta_{8,i}/3 }{ 1+z_i } }^{ (1+1/b)^{-\third} } 
\frac{ dr }{ \sqrt{ \Omega r^{-1}+1-\Omega - \frac{5}{3}\Omega\Delta_{8,i}z_i } }
= \int_0^{z_i} \frac{ dz }{ (1+z)^2 \sqrt{1+\Omega z} }
\end{equation}
For $z_i \gg 1$ and $\delta_{col}, \Delta_{8,i}\ll 1$ the ratio $\delta_{col}/\Delta_{8,i}$ determined
from eqs. (34),(35) is independent of $z_i$. We are interested in the pregalactic spectrum of density fluctuations
and in linear regime fluctuations evolve independently of $z$. Hence, in what follows we omit the subscript $i$ in
the ratios $\delta/\Delta_8$ as long as both amplitudes correspond to the pregalactic density field.

Fig.15 plots $\delta_{col}/\Delta_8$ vs $1+z$ at which the fluctuation collapses for $\Lambda=0$. Solid 
lines correspond to $b=1$ and dotted lines to $b=2$, but the solution of eq.(35) scales $\Delta_{8,i} \propto b^{2/3}$.
The three lines of each type correspond to $\Omega=0.1,0.3$ and 1 from bottom to top. One can see that the growth
of density fluctuations slows down after $1+z \simeq \Omega^{-1}$. Fig.15 in conjunction with $z_{\rm for}$ 
shown in Fig.14 allows one
to estimate the density field of 53W091 and the likelihood of its formation in a given cosmological model
given the total mass of the galaxy. (For $\Omega+\lambda=1$ the ratio $\delta_{col}/\Delta_8$ is plotted in 
Fig.1c of K93; in that case the growth of fluctuations freezes at $1+z \simeq \Omega^{-\third}$.)

KJ97 estimated that the stellar mass of 53W091 must be at least $10^{12}M_\odot$ with little dependence on cosmology
or star formation history. Thus in order to estimate the likelihood of 53W091 in the open CDM models we adopt a (conservative)
value for its \underline{total} mass of $5 \times 10^{12}M_\odot$. Fig.16 illustrates the likelihood of 
53W091 forming in the open CDM models. Thick lines in Fig.16 show the value of the rms fluctuation, $\Delta(M)$,
in units of $\Delta_8$ in the open CDM models on the (total) mass scale 
of $5\times 10^{12}M_\odot$. At this mass range the CDM spectra have effective power slope index of $\simeq -2.5$,
so the ratio $\Delta(M)/\Delta_8 \propto M^{-0.1}$ over the mass-scales relevant for 53W091.
The three lines from top to bottom correspond
to $H_0=$75,60 and 50 km/sec/Mpc. Thin lines show the bias-factor-independent values of 
$\delta_{53W091}/(\Delta_8b^{2/3})$ evaluated from eqs.(34),(35) at the redshift plotted in Fig.14.
Notation of the thin lines in Fig.16 is the same as in Fig.14. The values of the bias parameter for such models
are determined by normalization to the COBE DMR data and are $b>1$ (e.g. Stompor et al 1995).
One can see from the figure that similarly to the flat, $\Omega+\lambda=1$, CDM models discussed in KJ97,
the data on 53W091 would require it to be a very rare ($>5$ standard deviations) fluctuation in the density
field of open CDM models.

While the above arguments allow one to rule out or confirm certain cosmological models, they 
show only how rare
the observed object is in a given model and thus do not allow to
estimate the amount of power in the small scale pregalactic density field directly.
Furthermore, estimating the number density of such objects requires assumptions about the gaussianness of the 
pregalactic density field. E.g. at a given number of standard deviations for gaussian models, such as CDM, 
collapsed objects would be less abundant than in models with non-gaussian density field such as
strings (cf. Mahonen et al 1995). A different argument about galaxy formation epoch is required in order to
estimate the small-scale pregalactic density field directly.

We now move to estimating the epoch of galaxy formation from the \underline{present-day} galaxies and 
then estimating the density field from which they formed directly. We assume that galactic haloes formed by
dissipationless gravitational clustering (cf. White and Rees 1978). We assume further that the forming 
halo turns-around at radius $r_{ta}$ with little kinetic energy, i.e. its total energy per unit mass is
$E=-GM/r_{ta}$. The ensuing cold collapse will lead to violent relaxation during which the time-dependent gravitational
potential redistributes energy and particle distribution (Lynden-Bell 1967), while the total mass of the halo 
remains conserved. Thus at the end of the violent relaxation the kinetic energy, $T$, 
and potential energy, $V$, will be related by the virial equilibrium relation: $2T+V=0$. I.e. 
the total energy per unit mass is $E=T+V=-T=-\frac{3}{2}
\sigma^2$ where $\sigma$ is the one-dimensional halo velocity dispersion after virialisation. 
From the above the latter would be 
given by: $\sigma^2=\frac{2}{3} GM/r_{ta}$. We now assume that the halo turns-around at redshift $z_{ta}$
with an over-density $\gamma_{ta}(z_{ta})$ with respect to the average density of the Universe, 
$\frac{3H_0^2}{8\pi G}\Omega(1+z_{ta})^3$,
at that epoch. Therefore for halo that has one-dimensional velocity dispersion $\sigma$ and,
at least initially, had total mass $M$ (which later could have been lost by e.g. tidal stripping) the redshift
of its turn-around is given by:
\begin{equation}
\sigma^2 = \frac{2^{2/3}}{3} (GMH_0)^{2/3} [\Omega \gamma_{ta}(z_{ta})]^\third (1+z_{ta})
\end{equation}
or in terms of numbers:
\begin{equation}
\Omega^\third [\gamma_{ta}(z_{ta})]^\third (1+z_{ta}) = 13.3 \left(\frac{\sigma}{200 {\rm km/sec}}\right)^2
\left(\frac{M}{10^{12}h^{-1}M_\odot}\right)^{-2/3}
\end{equation}
Once the redshift of the halo turn-around has been evaluated one can calculate the redshift of its collapse, 
$z_{col}$, by assuming the collapse to have taken twice the turn-around time (e.g. Gott and Rees 1975).
As expected, at a given $\sigma$ the low mass haloes should have collapsed at higher redshifts.

We turn to computing the turn-around overdensity $\gamma_{ta}(z)=\rho_{ta}/\bar{\rho}(z_{ta})$. The turn-around
time of a fluctuation is given by $t_{ta}= t(\dot{r}=0) = \frac{\pi}{2H_0} \frac{\Omega}{[\frac{5}{3}\Omega 
\delta_i z_i + \Omega -1]^{3/2}}$. On the the other hand, 
the linear scale at the turn-around is
$r_{ta}=r_i \Omega/(\frac{5}{3}\Omega \delta_i z_i +\Omega -1)$, so the mean fluctuation density at the turn-around 
is given by $\rho_{ta}= \frac{3H_0^2}{8\pi G}\frac{(\frac{5}{3}\Omega \delta_i z_i +\Omega -1)^3}{\Omega^2}$. 
Rewriting the right-hand-side of the above expression in terms of cosmic time at turn-around, $t_{ta}$, leads to:
\begin{equation}
\gamma_{ta}(z)=\frac{\rho_{ta}}{\bar{\rho}(z)}=
\left(\frac{\pi}{2}\right)^2 \frac{1}{\Omega(1+z)^3 [H_0t(z;\Omega)]^2}
\end{equation}
For $\Omega=1$ one recovers from eq.(38) a well known result of $\gamma_{ta}(z)=(3\pi/4)^2$. For low values
of $\Omega$ the density contrast at turn around is larger. Fig.17 plots $\gamma_{ta}$ vs $z$ for $\Omega=0.1$
(dashed lines), 0.3 (dotted lines) and 1. At $1+z> \Omega^{-1}$ the values of $\gamma_{ta}$ converge to 
$(3\pi/4)^2$.

Eqs. (37),(38) allow one to estimate the redshift of galaxy collapse, given by $t_{col}=2t_{ta}$, for the halo
of one-dimensional velocity dispersion $\sigma$ and mass $M$. Fig.18 plots the value of $z_{col}$ vs $\Omega$
for galaxy with $\sigma=200$ km/sec. Solid line corresponds to $M=10^{12}h^{-1}M_\odot$, dotted to $M=2\times
10^{12}h^{-1}M_\odot$, and dashed line to $M=4\times 10^{12}h^{-1}M_\odot$. If $\Omega$ is low the redshifts
at which such galaxy should have formed are in general agreement with the data discussed in this section. If, 
however, $\Omega=1$ then massive galaxies would form at redshifts lower than the observations indicate.

In fact, one can narrow down the range of $z_{col}$ if one uses the
observed properties of galaxies in order to compute the total
halo mass and its velocity dispersion. In order to estimate the total halo mass at the time of its formation we proceed as
follows: encouraged by the results that normalization to the observed properties of Coma gave in the previous
section, we assume that this cluster is representative of the global value of the mass-to-light 
ratios associated with galactic systems. In order to relate the initial mass of the halo to the luminosity of the galaxy
it contains, we use the $D_n$-$\sigma$ relation for elliptical galaxies (Dressler et al 1987; Djorgovski and Davis
1987). In obtaining our results we will thus be restricted to elliptical galaxies, but we will assume
that the results are representative of the entire galaxy population. The fundamental plane  relations
of ellipticals are equivalent to the mass-to-light ratio of stellar populations of ellipticals
scaling $\propto L_B^\kappa$. From a detailed study of 37 ellipticals by van der Marel (1991) we adopt $\kappa =0.35$.
Assuming further that the ratio of the dark-to-luminous matter
is constant throughout the Universe and that luminous parts of galaxies formed by gas dissipation inside
formed haloes (White and Rees 1978) would allow to relate the total (initial) mass of formed haloes 
to the blue luminosity of the present day galaxies. (By ``initial"
we mean the mass at the time of the halo collapse; clustering processes would have later stripped parts, or all,
of the halo material to form a continuous dark matter distribution of clusters and groups of galaxies.) Thus the mass
of the halo at the time of its collapse is assumed to be given by:
\begin{equation}
M_{\rm Halo} = \left(\frac{M}{L_B}\right)_{\rm Coma} \frac{ \Gamma(2-\alpha) }{ \Gamma(2+\kappa-\alpha) } 
\left(\frac{L_B}{L_*}\right)^{1+\kappa}
\simeq 1.1 \left(\frac{M}{L_B}\right)_{\rm Coma} \left(\frac{L_B}{L_*}\right)^{1.35}
\end{equation}
The $\Gamma$-function factors in (39) come from requiring that $(M/L)_{Coma} =
\int \Phi(L) M(L) dL/\int \Phi(L) LdL$ with $\Phi(L) = L_*^{-1} \Phi_* (L/L_*)^{-\alpha} \exp(-L/L_*)$

Observationally it is known that inner parts of elliptical galaxies are dominated by both stellar
and halo potentials (e.g. Rix et al 1997) which would be consistent with ellipticals forming by dissipational collapse
(Kashlinsky 1982). Thus central velocity dispersions of elliptical galaxies would not provide a good measurement of
the halo potential well. The latter can be determined from dynamics in the outer parts of elliptical galaxies,
where the only available measurements today come from X-ray data from hot gas coronae. The best measurements
of temperature electron density profiles around ellipticals come from ASCA (Matsumoto et al 1997, Matsushita
1997) and ROSAT (Davis and White 1996) satellites. Since the cooling time of the gas is of the order of $H_0^{-1}$
the gas is likely to be hydrostatic. Then its temperature will be related to the halo velocity dispersion 
via the equation of hydrostatic equilibrium which for isothermal halo profile reads: 
$\frac{d(k_{\rm Boltzman}n_{gas}T)}{dr} = -\mu m_p \frac{\sigma^2}{r}$,
where $\mu$ is molecular weight and $m_p$ is the proton mass (e.g. Awaki et al 1994). 
Temperature profiles (Matsushita 1997) indicate that the gas in outer parts can be assumed 
isothermal and that the gas has density profile $n_{gas} \propto r^{-\beta_{gas}}$
with $\beta_{gas} \simeq$(2.5-3). Then the halo velocity dispersion is related to the measured X-ray temperature
via:
\begin{equation}
\sigma = \left(\frac{k_{\rm Boltzman} T}{\mu \beta_{gas} m_p}\right)^\half
\end{equation}
Eqs. (37)-(40) along with the numbers in Fig.15 allow now to determine the density fluctuation spectrum given
photometric ($L_B$) and X-ray ($T$) data on galaxies.

We have computed the values of $\delta(M)/\Delta_8$ using eqs(37)-(40) for a sample of galaxies
for which good X-ray and photometric data are available. The halo velocity dispersion was computed according
to eq.(40) from the ROSAT (Davis and White 1996) and ASCA (Matsumoto et al 1997) observations. The data
on the blue absolute luminosities were taken from Faber et al (1989) who determined the absolute luminosities
using the fundamental plane relations. Only galaxies which appear in both the X-ray and Faber et al catalog were
used. The ROSAT determined X-ray temperatures and gas profiles for galaxies have larger uncertainties than ASCA;
hence of the ROSAT data listed in Table 1 of Davis and White (1996) we kept only those galaxies that
have quoted uncertainties in $T$ of less than 25\%. The ASCA data available are listed in Table 3 of Matsumoto
et al (1997). Note that the atomic modelling of the spectra alone can introduce \underline{systematic}
uncertainties of $\sim 20\%$ (Matsushita 1997). The final dataset for which we found the accurate X-ray
temperature measurements and photometry included some 20 galaxies.

Fig.19 plots the values of $\delta(M)/\Delta_8$ vs the initial halo mass computed from eqs.(37)-(40)
assuming a ``universal" hot gas profile of $\beta_{gas}=2.5$ and
$\mu =0.6$; the numbers in eq.(40) are not very sensitive to
the value of $\beta$ observed to lie between 2 and 3.
The left box shows the results for $\Omega=0.3$, as required by the Coma cluster dynamics. The right
box assumes the initial dark haloes to have the total mass-to-light ratio like Coma, but the total $\Omega$
to be 1. Note
that this prescription gives the actual amplitude of the density fluctuation, rather than its rms value
$\Delta(M)$, and hence should have larger scatter. Additional scatter in Fig.19 is introduced by the uncertainties
in the current X-ray data and modelling and a further $\simeq$ 15\% uncertainty in the individual distances 
from using the fundamental plane relations. The slope in Fig.19 is quite obvious and is in good agreement with
the slope determined for large scales. The scatter is larger though, but is expected given the above 
observational uncertainties, the simplifying assumptions that were used in this section and the fact that the
actual $\delta(M)$ rather than the rms, $\Delta(M)$ is shown. 
On the other hand, the X-ray temperatures do not correlate
with the central velocity dispersion, and so the Faber-Jackson blue luminosity - central velocity
dispersion correlation is unlikely to reproduce the trend in Fig.19. It is likely that the numbers in Fig.19
reflect the slope and amplitude of the pregalactic density field on the corresponding mass scales.

\section{Reconstructed density field and cosmological paradigms}

The methods outlined in sections 3-7 allowed us to reconstruct pregalactic density field from various
astronomical data on scales from $\sim 1$ to 100$h^{-1}$Mpc. Fig.20 summarizes the results in
terms of the quantity:
\begin{equation}
\frac{\Delta(M)}{\Delta_8} = 
\left[\frac{\int_0^\infty P(k) W_{TH}(kr) k^2 dk}{\int_0^\infty P(k) W_{TH}(kr_8) k^2 dk}\right]^\half
\end{equation}
with $r$ being the linear scale containing mass $M$ according to eq.(32). 
Left box corresponds to $\Omega =0.25$, the value suggested by dynamics of the Coma cluster and galaxy
luminosity function. Right box plots the reconstructed field for $\Omega=1$, implied by inflationary
prejudices.

The open box in the figure shows 
the ``normalization" point where the above quantity is unity by definition and irrespective of the bias
factor. The three dotted lines show the BE93 fit to the APM data within one standard deviation 
uncertainty and the three lines
correspond to the less accurate, but earlier, K92 fits to the data. The lines are drawn following the
peculiar velocities analysis in sec.4 suggesting the constancy of the bias parameter, $b$. They are
plotted on scales $>r_8$ where the APM data are less affected by non-linear gravitational effects. On scales
$>100h^{-1}$Mpc the APM data probably are not useful for reliably probing the pregalactic density field.
The rhombs show the values for $\Delta(M)/\Delta_8$ reconstructed in Secs.5,6 from the data on the cluster 
correlation amplitude vs richness relation and normalizing the cluster richness-mass relation to Coma. 
This reconstruction allows to determine the pregalactic density field on
scales from $\simeq 5$ to $\simeq 20h^{-1}$Mpc. The two methods should give the same results where
they reconstruct the density field on the same scales. For $\Omega=0.3$ the density
fields reconstructed from the two independent methods coincide. If $\Omega=1$ the discrepancy between
the two amplitudes is significant, although the slopes coincide. Thus we conclude that 
$\Omega\simeq$(0.25-0.3) is in better agreement with the data and the pregalactic density field
on scales $5 h^{-1}{\rm Mpc} < r <100h^{-1}$ is likely to be close to that in the left box of Fig.20.

The plus signs in Fig.20 show the density field reconstructed in Sec.7 by using the fundamental plane
relations and X-ray data for measuring halo velocity dispersion. The total halo masses at the time of their
collapse/formation were normalized to reproduce the observed blue mass-to-light ratio of Coma. Plotted is
the value of $\delta(M)$, not the rms $\Delta(M)$, in units of $\Delta_8$; the numbers plotted depend on 
the amplitude of the bias parameter $\propto b^{2/3}$. Both the slope and amplitudes of the density field
on galactic scales are consistent 
with extrapolation from the density field reconstructed on larger scales
assuming a simple power law for $P(k)$ with the slope of $\sim -1$. The slight
excess in the amplitude ($\sim 30\%$) of the plus signs should not be regarded as worrying since what is
plotted is the actual $\delta(M)$, not its rms value. Furthermore, the galaxies used in the fundamental plane
and X-ray studies are found predominantly in rich clusters and 
may indeed have formed out of higher peaks of the density field.
Also the X-ray temperatures determined from ASCA and ROSAT observations for galactic
X-ray emission may have systematic errors of $\sim 20\%$ due to uncertainties in the atomic physics involved
(Mushotzky, private communication). Such systematic errors may be sufficient to bring the amplitudes
of the plus signs by the necessary $\simeq 30\%$. In addition,  many of the assumptions used here may be too
simplifying: e.g assuming the same slope of the gas profile, $\beta_{gas}$, or there may be deviations from 
the isothermal density profile, $\rho_{halo}\propto r^{-2}$ assumed here for all haloes. Given these uncertainties
we find that the consistency in the density field plotted in Fig.20 from the various datasets and methods
is quite reasonable.

Fig.21 juxtaposes the reconstructed density field with the predictions of the CDM models for the 
Harrison-Zeldovich  (right box) and tilted spectra. Notation of the data points is the same as in Fig.19.
Only the data points for $\Omega=0.3$ (left box in Fig.19) are shown because of the discrepancy between the
cluster correlation data and the APM methods for $\Omega=1$. CDM models are shown with dashed lines: thick
dashed lines correspond to $\Omega h=0.5$ and thin lines to $\Omega h=0.2$ for $n=1$ and to 0.3 for
$n=0.7$. One can see that the models that fit the large-scale part of the pregalactic spectrum,
miss the points on small scales by a significant factor. That is,
for any value of the excess power, $\Omega h$, or $n$ one would have to assume a
significant scale dependence in the bias parameter $b$ in order to fit the density field over the entire range of 
scales. 

The most economical conclusion to draw from the points in Fig.20 would be to assume that 1) the Universe
is open with $\Omega\simeq$0.25-0.3, the value which is also suggested by the dynamics of the Coma cluster; and 2) the pregalactic
power spectrum has constant power index of $\sim -1$ all the way from $\sim 30 h^{-1}$Mpc to $\sim
1h^{-1}$Mpc. For primordial adiabatic fluctuations this could be achieved only if the smaller wavelengths
which enter the horizon in the radiation dominated era have more power when compared with the scale-free
primordial spectrum, $P_i(k) \propto k^n$. Computations show that this may be done with strings 
(Albrecht and Stebbins 1992) although the models are still not definite enough to predict a unique
power spectrum. Another possibility could be the primeval-baryon-isocurvature (PBI) model
where the shape of the spectrum of the density field could be further evolving after recombination due
to reionisation and other effects (Peebles 1987). Or one can assume that the primordial power 
spectrum in the CDM models was not scale free and contained 
extra power on small scales. However, the latter
may lack theoretical motivation if, as we have argued on the basis of cluster correlation and galaxy 
data the Universe is open, and CDM model predictions are tied to and/or motivated by inflationary scenario.

Fig.20 plots the pregalactic density field in units of $\Delta_8$ and thus independently of the latter's value. In
order to assign a numerical value to $\Delta(M)$ at some 
early epoch, say the epoch of recombination, we have to compute
the value of $\Delta_8$ at that time. Assuming purely gravitational evolution, i.e. ignoring pressure forces,
Thompson drag and other effects which can become important in some models, would give at $z=1000$ the
values of $\Delta_8 = (0.55, 1.2, 2.8) \times 10^{-3} b^{2/3}$ for $\Omega = (1,0.3,0.1)$ respectively. 
The value of $b$ can be determined from normalization 
to the microwave background anisotropy measurements. We omit this since it would require assumptions about the
nature of the density field, dark matter, reionisation history, etc.

\section{Conclusions}

The results of the paper can be summarized as follows:

1) Our analysis of the APM survey results on $w(\theta)$ in six 0.5 magnitude wide slices shows that the
various evolutionary corrections are not significant out to the limits of the catalog at $b_J \simeq 20.5$. 
This allows to make
accurate fits of the various spectra to the APM data. Of these, the BE93 fit for the power spectrum of
the density field traced by galaxies gives the best approximation to all six slices. If light traces mass,
this fixes the pregalactic density field on scales $10h^{-1}{\rm Mpc} < r <100h^{-1}$Mpc. 
We then choose the value of the
angular scale at which the angular correlation function is $\simeq 3.3$ times the systematic error from the
plate gradients in all six slices to test the requirements the data sets on CDM models. We find that  in order
to fit the APM data, CDM models would require $\Omega h \simeq 0.2$ if $n=1$ or $\Omega h\simeq 0.3$ if $n=0.7$.

2) The velocity field implied by the galaxy correlation function determined from
the APM data is consistent with the Great Attractor peculiar velocities and the bias parameter which
is constant over the range of scales probed by both datasets, $10h^{-1}{\rm Mpc} < r < 60 h^{-1}$Mpc.
This suggests that pregalactic density field is proportional to the power spectrum of galaxy clustering
determined from the APM on scales $>r_8$.
Furthermore, if $\Omega^{0.6}/b=1$
the amplitude of the dot velocity correlation function at zero lag would be significantly
larger, $\simeq (1300 {\rm km/sec})^2$, than the values indicated locally
and by the central velocity dispersions of typical galaxy systems. This suggests that the velocity data
are in agreement with open Universe. Furthermore, we show that the quantity constructed from the
the velocity correlation tensor, $\sqrt{\Sigma(r)- \Pi(r)}$, can be used to determine $\Omega^{0.6}/b$.
Its amplitude, predicted by both the APM and and the (local) CfA data on $\xi(r)$, is 
$\simeq400 \frac{\Omega^{0.6}}{b}$km/sec on scales $<40h^{-1}$Mpc and should be detectable in the 
current velocity surveys if $\Omega^{0.6}/b=1$.

3) We calculate in detail the amplification in the cluster correlation function for clusters
of various masses due to gravitational clustering assuming only that clusters of galaxies should be
identified with regions with turn-around time less than the age 
of the Universe. We use the data on both
the slope of the cluster correlation function for Abell 
clusters of richness ${\cal R} \geq 1$, and on the 
dependence of the cluster correlation amplitude on the cluster richness/mass. Cluster masses for given
${\cal R}$ are normalized to Coma. We find that, for any value of $\Omega h$, 
CDM models cannot simultaneously fit data on the cluster correlation 
slope and amplitude, its amplitude-richness dependence and 
the APM catalog data on $w(\theta)$. We show that by using the data on the underlying galaxy correlation function, $\xi(r)$ at some fixed $r$
(we chose $r=25h^{-1}$Mpc), enables one to invert the amplitude-richness data to obtain the values
of the rms fluctuations, $\Delta(M)/\Delta_8$, for the {\it pregalactic density field} on cluster
mass-scales. Applying the
method to the data we obtain the pregalactic density field whose slope is in excellent agreement with
the APM field on the same scales. Requiring then that the two amplitudes 
coincide fixes $\Omega$ to be 0.3 independently
of the bias factor; the same value that is implied by the dynamics of the Coma cluster.

4) We then use the data on objects at high redshift to constrain the small-scale part of the pregalactic
density field. We argue that data on the existence of clusters of galaxies at redshifts $\geq$(1-2), 
observations of galaxies
at $z\sim 5$ and the recent determination of galaxy ages from Keck observations for a number of high-$z$ galaxies
imply more power on small scales than predicted in CDM models normalized to the large-scale galaxy correlation
data. We then reconstruct the pregalactic density field out of which modern-day (elliptical) galaxies have
formed. To do this we use the data on blue absolute luminosities of Faber et al (1989)
and assume the fundamental plane relations 
to determine the initial halo mass (normalized further to reproduce the mass-to-light ratio of the 
Coma cluster) and the X-ray data from ROSAT and ASCA satellites to determine the halo velocity dispersion.
Despite the simplifying assumptions we recover a density field which is
in good agreement with the simple extrapolation
of the reconstructed density field from larger scales.

5) At the end of these steps we reconstruct the pregalactic density field over linear scales
encompassing two orders of magnitude, $1h^{-1}{\rm Mpc} < r <100h^{-1}$Mpc. The recovered field 
is difficult to fit with CDM models and constant bias parameter over this range of scales. 
We argue that the most economical explanation of the density field on all scales would be to assume
that the results in Fig.20 represent the pregalactic density field with constant bias parameter $b\simeq 1$
and $\Omega\simeq 0.3$ as required by dynamics of galaxy systems and the consistency between the X-ray
observations of rich clusters and Big-Bang nucleosynthesis (White et al 1993)

\section{Acknowledgements}
I am grateful to people who provided the data used in the paper: to Sandy Faber for providing the absolute photometric
data for galaxies, Kyoko Matshushita for permitting me to use the results from her thesis on ASCA observations of galaxies,
Carlton Baugh for the data on the BE93 power spectrum, Mike Vogeley for the CfA data used for peculiar velocity
calculations and Will Sutherland for the APM data binned in six narrow magnitude slices. Particular thanks go to Richard
Mushotzky for several valuable discussions on X-ray observations of galaxies and to Raul Jimenez on the age
determinations of high redshift galaxies.
\newpage
\section{References}
Abell, G., Corwin, H.G. and Ollowin, R.P. 1989,Ap.J.Suppl,{ 70},1.\\
Albrecht, A. and Stebbins, A. 1992,Phys.Rev.Lett.,{ 68},2121.\\
Awaki, H. et al 1994,PASJ,{ 46},L65.\\
Bahcall, N. 1981,Ap.J.,{247},787.\\
Bahcall, N. and Soneira, R. 1983,Ap.J.,{270},20.\\
Bahcall, N. and West, M. 1992,Ap.J.,{392},419.\\
Bardeen, J., Bond, J.R., Kaiser, N. and Szalay, A. 1986,Ap.J.,{304},15.\\
Baugh, C. and Efstathiou, G. 1993,MNRAS,{265},145. (BE93)\\
Baugh, C. and Efstathiou, G. 1994,MNRAS,{267},323.\\
Begelman, M. and Rees, M.J. 1978,MNRAS,{185},847.\\
Bennett, C. et al 1996, Ap.J.,{464},L1.\\
Bertschinger, E. and Dekel, A. 1989,Ap.J.,{336},L5.\\
Bertschinger, E. et al 1989,Ap.J.,{364},370.\\
Blumenthal, G. et al 1984,Nature,{311},517.\\
Broadhurst,T., Ellis, R. and Glazebrook, K. 1992,Nature,{355},55.\\
Bruzual, G. 1983,Ap.J.,{273},105.\\
Bond, J.R. and Efstathiou,G. 1985,Ap.J.,{285},L45.\\
Chambers, K.C. and Charlot, S. 1990,Ap.J.,{348},L1.\\
Chambers, K.C. et al 1990,Ap.J.,{263},21.\\
Cavalieri, A. and Szalay, A. 1986,Ap.J.,{311},589.\\
Cen, R. et al 1992,Ap.J.,{399},L11.\\
Clutton-Brock, M. and Peebles, P.J.E., 1981,A.J.,{86},1115.\\
Collins, C.A., Nichol, R.C. and Lumdsen, S.L. 1992,MNRAS,{254},295.\\
Da Costa, L.N. et al 1994,Ap.J.,{437},L1.\\
Davis, D.S. and White, R.E. 1996,Ap.J.,{470},L35.\\
Davis, M. and Peebles, P.J.E. 1983,Ap.J.,{267},465.\\
Davis, M. et al 1985,{292},371.\\
Dekel, A. et al 1993,Ap.J.,{412},1.\\
Deltorn, J.-M., et al 1997,Ap.J.,{483},L21.\\
Dey, A. et al 1997,preprint.\\
Dickinson, M. 1993,BAAS,{182},6703.\\
Djorgovski, G. and Davis, M. 1987,Ap.J.,{313},59.\\
Djorgovski, G. et al 1996,Nature,{382},234.\\
Dressler, A. et al 1987a,Ap.J.,{313},42.\\
Dressler, A. et al 1987b,Ap.J.{313},L37.\\
Dunlop, J. et al 1996,Nature,{381},581. (D96).\\
Eales, S.A. et al 1993,Ap.J.,{409},578.\\
Efstathiou, G. and Rees, M.J. 1988,MNRAS,{230},5P. (ER88)\\
Efstathiou, G. et al 1990,Nature,{348},705.\\
Eisenhardt, P. and Dickinson, M. 1992,Ap.J.,{399},L47.\\
Faber, S. et al 1989,Ap.J.Suppl.,{69},763.\\
Fall, S.M. and Pei, Y.C 1993,Ap.J.,{402},479.\\
Francis, P.J. et al 1996,Ap.J.,{457},490.\\
Francis, P.J., Woodgate, B. and Danks, A. 1997,Ap.J.,{482},L25.\\
Franx, M. et al 1997,astro-ph/9704090.\\
Geller, M. and Huchra, J., 1989,Science,{246},897.\\
Giavalisco, M., Steidel, C.S. and Macchetto, F.D. 1996,Ap.J.,{470},189.\\
Gorski, K. 1988,Ap.J.,{332},L7.\\
Gorski, K. et al 1989,Ap.J.,{344},1.\\
Gorski, K. et al 1996,Ap.J.,{464},L11.\\
Gott, R. and Rees, M.J. 1975,Astron. Astrophys.,{45},365.\\
Groth, E. and Peebles, P.J.E. 1977,Ap.J.,{217},385.\\
Groth, E., Juszkiewicz, R. and Ostriker, J. 1989,{346},558.\\
Haenhelt, M. and Rees, M.J. 1993,MNRAS,{263},168.\\
Hamilton, A. et al 1991,Ap.J.,{374},L1.\\
Hu, E.M. and Ridgway, S.E. 1994,A.J,{107},1303.\\
Hu, E.M. and McMahon, R.G. 1996,Nature,{382},231.\\
Jensen, L. and Szalay, A. 1986,Ap.J.,{305},L5.\\
Jones, M.E. et al 1997,Ap.J.,{479},L1.\\
Juszkiewicz, R. and Yahil, A. 1989,Ap.J.,{346},L49.\\
Kaiser, N. 1983,Ap.J.,{273},L17.\\
Kaiser, N. 1984,Ap.J.,{284},L9.\\
Kaiser, N. 1987,MNRAS,{227},1.\\
Kashlinsky, A. 1982,MNRAS,{200},585.\\
Kashlinsky, A. 1987,Ap.J.,{317},19.\\
Kashlinsky, A. 1991a,Ap.J.,{383},L1.\\
Kashlinsky, A. 1991b,Ap.J.,{376},L5.\\
Kashlinsky, A. 1992a,Ap.J.,{399},L1. (K92)\\
Kashlinsky, A. 1992b,Ap.J.,{386},L37. \\
Kashlinsky, A. 1993,Ap.J.,{406},L1. (K93)\\
Kashlinsky, A. 1994, in ``Evolution of the Universe and its observational quest",p. 181.,ed. K.Sato,
Universal Academy Press, Tokyo, Japan. \\
Kashlinsky, A. and Jimenez, R. 1997,Ap.J.,{474},L81. (KJ97).\\
Kashlinsky, A. and Jones, B.J.T. 1991,Nature,{349},753.\\
Kashlinsky, A., Tkachev, I. and Frieman, J. 1994,Phys.Rev.Lett.,{73},1582.\\
Kent, S. and Gunn, J. 1981,A.J.,{87},945.\\
Kochanek, C. 1996,Ap.J.,{466},638.\\
Lacy, M. et al 1994,MNRAS,{271},504.\\
Lahav, O. et al 1991,MNRAS,{251},128.\\
LeFevre, O. et al 1996,Ap.J.,{471},L11.\\
Lilly, S. 1988,Ap.J.,{333},161.\\
Limber, D.N. 1953,Ap.J.,{117},134.\\
Loveday, S. 1992,Ap.J,{390},338.\\
Lowenthal, J.D. et al 1997,astro-ph/9612239.\\
Lucy, L.B. 1974,A.J.,{79},745.\\
Lynden-Bell, D. 1967,MNRAS,{136},101.\\
Maddox, S. et al 1990,MNRAS,{242},43p. (MESL)\\
Mahonen, P., Hara, T. and Miyoshi, S.J. 1995,Ap.J.,{454},L81.\\
van der Marel 1991,MNRAS,{253},710.\\
Mathewson, D., Ford, V.L. and Buchorn, M. 1992,Ap.J.,{389},L5.\\
Matsumoto, H. et al 1997,Ap.J.,{482},133.\\
Matsushita, K. 1997,Ph.D. Thesis,University of Tokyo.\\
Melott, A. 1992,Ap.J.,{393},L45.\\
Narayan, R. and White, S.D.M. 1988,MNRAS,{231},97P.\\
Pagel, B. 1997,Cambridge Universe Press.\\
Pascarelle, S.M. et al 1996,Ap.J.,{456},L21.\\
Peacock, J. 1991,MNRAS,{253},1p.\\
Peacock, J. and Dodds, S.J. 1994,MNRAS,{267},1020.\\
Peebles, P.J.E. 1982,Ap.J.,{263},L1.\\
Peebles, P.J.E. 1980,``Large-scale structure of the Universe",Princeton Univ. Press.\\
Peebles, P.J.E. 1987,Nature,{327},210.\\
Peebles, P.J.E. 1988, in ``Large scale motions in the Universe: a Vatican study week",
eds. V.C. Rubin and G.V. Coyne, Princeton Univ Press, p.457.\\
Peebles, P.J.E. 1989,Ap.J.,{344}L53.\\
Peebles, P.J.E. 1990,Ap.J.{362},1.\\
Perlmutter, S. et al 1997,Ap.J.,in press. astro-ph/9608192.\\
Picard, A. 1991,Ap.J.,{368},L7.\\
Politzer, H.D. and Wise, M.B. 1984,Ap.J.,{285},L1.\\
Postman, M. and Lauer, T.R. 1995,Ap.J.,{440},28.\\
Press, W. and Schechter, P. 1974,Ap.J.,{187},425.\\
Rix, H.-W.  et al 1997,astro-ph/9702126.\\
Saunders, W. et al 1991,Nature,{349},32.\\
Saunders, R. et al 1997,Ap.J.,{479},L5.\\
Schechter, P. 1976,Ap.J.,{203},297.\\
Schneider, D. Schmidt, M. and Gunn, J. 1991,A.J.,{102},837.\\
Schneider, D. et al 1992,PASP,{104},678.\\
Schneider, D. et al 1994,A.J.,{107},880.\\
Schmidt, M., Schneider, D. and Gunn, J. 1991,in``The space distribution of quasars",ed. Crampton,D.,ASP 
Vol.21.\\
Shaya, E., Peebles, P.J.E. and Tully, B. 1995,Ap.J.,{454},15.\\
Shaver, P.A. et al 1996,Nature,{384},439.\\
Smoot, G. et al 1992,Ap.J.,{396},L1.\\
Strauss, M. and Willick, J.A. 1995,Phys.Reports,{261},271.\\
Steidel, C.S. et al 1996,Ap.J.,{462},L17.\\
Stompor, R., Gorski, K.M. and Banday, A.J., 1995,MNRAS,{277},1225.\\
Sutherland, W. 1988,MNRAS,{234},159.\\
Toth, G., Hollosi, J. and Szalay, A. 1989,Ap.J.,{344},75.\\
Trager, S.C. et al 1997,astro-ph/9703062.\\
Turner, E. 1991,A.J.,{101},5.\\
Vittorio, N., Juszkiewicz, R. and Davis, M. 1987,Nature,{323},132.\\
Vittorio, N. and Silk, J. 1985,Ap.J.,{285},L39.\\
Vogeley, M. et al 1992,Ap.J.,{391},L5.\\
Walker, T.P. et al 1991,{376},51.\\
White, S.D.M. and Rees, M.J. 1978,MNRAS,{183},341.\\
White, S.D.M. et al 1993,Nature,{366},429.\\
Willick, J. 1990,Ap.J.,{351},L5.\\
Yoshii, Y. and Takahara, F. 1988,Ap.J.,{326},1.\\
\newpage
{\bf Figure captions}

Fig.1: Selection function for the six 0.5 magnitude wide slices of the APM data. The lines are
computed using the $K$-correction as $\delta m =Kz$ with $K=3$, but the dependence on $K$ is weak for
$K\leq 4$. Solid lines correspond to $\Omega=0.1$ and dotted to $\Omega=1$. The six curves of each type
correspond to the magnitude range of each of the APM slices shown on top of the boxes in Fig.2.

Fig.2: Open triangles show the APM data in the narrow magnitude slices with the magnitude ranges shown on
top of the boxes. Thick dashed lines are for $\Omega=1$ and a power law $\xi(r)$, which should give a good fit to the
data at small angular scales. Thick dotted lines, that mostly merge with the thick dashed lines,
is the power law fit for $\Omega=0.1$. The upper thick lines of each type are for the clustering 
pattern stable
in physical coordinates, i.e. $\xi(r;z) = (r/r_*)^{-1.7} (1+z)^{-1.7}$. Lower thick lines correspond to
clustering pattern which is stable in comoving coordinates: $\xi(r;z) = (r/r_*)^{-1.7} (1+z)^{-3}$. Three solid
lines correspond to the best BE93 fit to the APM power spectrum and $\pm$ one standard deviation uncertainty.
For the BE93 lines the power spectrum was taken to evolve $\propto (1+z)^{-3}$. The numbers are shown 
$K=3$.

Fig.3: The angular scale, $\theta(w)$, at which the projected angular correlation reaches a given value, $w$, is
plotted vs the mean blue magnitude of the APM six slices. Left boxes correspond to $w=2\times10^{-3}$ 
which is just above the
systematic error, $w_{pp}\simeq 1.5\times10^{-3}$. The right boxes are for $w=5\times 10^{-3}$. Thick plus signs
correspond to the APM data plotted in Fig.2. The data is compared with CDM predictions: upper boxes show
the values of $\theta(w)$ for $n=1$ 
and lower ones for the tilted CDM model with $n=0.7$. $\times$ signs correspond to $\Omega h=0.5$, 
open squares to $\Omega h=0.3$, triangles to $\Omega h=0.2$, rhombs to $\Omega h=0.1$ and asterisks to
$\Omega h=0.05$. Only CDM models which match the observed values would be 
consistent with APM data.

Fig.4: The values of the rms density contrast, $\Delta(M)$, in units of $\Delta_8$ are plotted for the various fits
to the APM data. Dotted lines show the BE93 fit with the one standard deviation uncertainty. Solid lines
correspond to the K92 fit with $k_0=35,40,50 h^{-1}$Mpc from bottom up. Thick dashed line corresponds to CDM
models with $\Omega h=0.2$, $n=1$ and $\Omega h=0.3$, $n=0.7$.

Fig.5: The values of the ``typical velocity" profile derived from eq.(11)
using the data on the galaxy correlation function, 
$\xi(r)$, is plotted for $\Omega^{0.6}/b=1$. Dashed line corresponds to the velocity field from the CfA data on
$\xi(r)$. Solid lines correspond to $\xi(r)$ given by the BE93 fit to the APM data with one standard deviation 
uncertainty.

Fig.6: The values of the transverse minus parallel velocity correlations according to the data on $\xi(r)$.
Notation is the same as in Fig.5. 

Fig.7: Cluster-cluster correlation function for CDM models with $n=1$ and various $\Omega, h$ on scales
$20h^{-1}{\rm Mpc} < r < 100h^{-1}{\rm Mpc}$. Thin solid lines show the underlying correlation
function in the CDM model specified on top of each box. 
Dotted lines shows the amplified correlation function for clusters of richness
${\cal N}= 5$, dashes to ${\cal N}= 10$, dashed-dotted lines to ${\cal N}= 50$, and the dashed-double-dotted lines
to ${\cal N}= 100$. For comparison, the richness class ${\cal R}\geq 1$ clusters, such as Coma, have ${\cal N}= 100$.
Thick solid line corresponds to the approximation $\xi_{CC}=300 (r/1h^{-1}{\rm Mpc})^{-1.8}$ which Bahcall and 
Soneira (1983) find for ${\cal R}\geq 1$ clusters. 

Fig.8: Same as Fig.7 only for tilted CDM models with $n=0.7$. Same notation as in Fig.7.

Fig.9: Amplification of the correlation function for clusters $A= \xi_{MM}(r)/\xi(r)$ evaluated for CDM models
at $r=25h^{-1}$Mpc is plotted vs cluster richness. Triangles correspond to the data points from Fig.3 of 
Bahcall and West (1992). Solid lines are for CDM models with $\Omega h=0.1$, dotted for $\Omega h=0.2$,
dashed for $\Omega h=0.3$, and the dashed-dotted lines correspond to $\Omega h=0.5$. Thin lines of each type are
for $h=1$ and thick lines are for $h=0.5$.

Fig.10: Correlation function $\xi(r)$ according to the various fits to APM is plotted vs $r$. Dotted line correspond
to CDM model with $n=1$ and $\Omega h=0.2$ or $n=0.7$ and $\Omega h=0.3$ (the two practically coincide). Dotted
lines correspond to the BE93 fit with one standard deviation uncertainty. Three solid lines correspond to the
K92 fit with the transition scale $k_0^{-1}=35,40,50h^{-1}$Mpc. 

Fig.11: The amplification factor $A(x)$ is plotted according to eq.(29). Solid line corresponds to the underlying
correlation $\xi=0.07$ and the two dotted lines show the results of, say, 25\% uncertainty in $\xi$.

Fig.12: The spectrum of the pregalactic density field, $\Delta(M)/\Delta_8$, obtained from the data on the cluster
correlation amplitude - richness dependence. The upper horizontal axis shows the richness corresponding to the
particular value of $\Delta(M)/\Delta_8$. The lower horizontal axis shows the values of $M$ computed from 
normalization to Coma (eq.31).

Fig.13: Plots $\Delta(M)/\Delta_8$  from Fig.12 vs linear scale (eq.32). Left box: thick plus signs correspond to
$\Omega=1$, thin plus signs to $\Omega=0.1$. Open box corresponds to $\Delta(M)/\Delta_8=1$ at $r=8h^{-1}$Mpc; the
numbers miss this point if $\Omega=1$. Right box: dotted lines correspond to the BE93 fit to the APM
data with one standard deviation uncertainty. Solid lines are for K92 fit with transition scale $k_0^{-1}=35,
40,50h^{-1}$Mpc. Rhombs correspond to the data from Fig.12 with $\Omega\simeq0.25$ when it would coincide with the
independently deduced lines from the APM data. Other signs are in the same notation as in Fig.12.

Fig.14: The redshift of formation of 53W091 according to the age estimate $t_{age}=3.5\pm 0.5$ Gyr
of D96 is plotted vs $\Omega$ for $\Lambda=0$. Dotted lines correspond to $t_{age}=3.5$Gyr, solid to 
$t_{age}=3$Gyr and dashes to $t_{age}=4$Gyr. Three lines of each type corresponds to $h=0.5,0.6,0.75$ from bottom
up.

Fig.15: Values of $\delta_{col}/\Delta_8$ at early epoch, $z_i$, vs redshift of collapse $z$ are plotted
for $\Lambda=0$. Solid lines correspond to $\Delta_8$ evolving to 1 today, or $b=1$; dotted lines are for
$b=2$. Three lines of each type correspond to $\Omega=1,0.3,0.1$ from up down.

Fig.16: The values of $\delta_{col}/\Delta_8$ implied by the redshift of collapse of 53W091 plotted in Fig.14
are shown vs $\Omega$ for zero cosmological constant Universe. Three thick solid lines show the predictions
of this ratio in CDM models with the value of $\Omega$ shown on the horizontal axis. The total mass of 53W091 there
was taken to be $5\times 10^{12}M_\odot$ (cf.KJ97). The three thick lines correspond to $h=0.5,0.6,0.75$ from bottom
up.

Fig.17: The overdensity of the fluctuation at turn-around plotted vs the turn-around redshift. Solid line is
for $\Omega=1$, dotted for $\Omega=0.33$ and the dashed line is for $\Omega=0.1$.

Fig.18: The redshift of collapse (formation) of galaxy with the halo dispersion $\sigma=200$km/sec plotted vs
$\Omega$ according to eq.(37). The total (initial halo) mass is taken to be $10^{12}h^{-1}M_\odot$ (solid line), 
$2\times 10^{12}h^{-1}M_\odot$ (dotted line) and $4\times 10^{12}h^{-1}M_\odot$ (dashed line).

Fig.19: Density fluctuations $\delta(M)/\Delta_8$ computed according to eqs.(37)-(40) vs the mass scale they
contain. The numbers were computed normalizing the total mass-to-light ratio of galaxies to Coma at
$M/L=362h$. Left box assumes $\Omega=0.3$, corresponding to the value implied by Coma dynamics, and 
the right box corresponds to $\Omega=1$.

Fig.20: Pregalactic density field, $\Delta(M)/\Delta_8$ reconstructed over the entire range of $1h^{-1}{\rm Mpc} 
< r < 100h^{-1}$Mpc. Left box corresponds to $\Omega=0.3$, right box to $\Omega=1$. Open square corresponds to
the value of 1 at $r_8$. The lines from the APM data are plotted only for linear
scales $r>r_8$. Solid lines are the K92 fits to the APM data with the transition scale $k_0^{-1}=
35,40,50h^{-1}$ from bottom up. The three dotted lines correspond to the BE93 fit with one standard deviation
uncertainty.  Rhombs show the values obtained from inverting the data on
cluster correlation amplitude vs richness
according to eq.(29). The rhombs coincide with the APM data only if $\Omega \simeq 0.3$. Plus signs show the
values reconstructed according to eqs.(37)-(40)
from the data on the fundamental plane of galaxies and the ROSAT and ASCA measurements of
the halo velocity dispersion.

Fig.21: The results from the left box of Fig.20 are juxtaposed with CDM models. In the left box
the dashed lines show the CDM density field with $n=1$ and $\Omega h=0.5$ (thick dashed line) and $\Omega h=0.2$.
The right box shows the tilted CDM models with $n=0.7$ and $\Omega h=0.5$ (thick dashed line) and
$\Omega h=0.3$. Other symbols have the same notation as in Fig.20.

\newpage
Table 1. Data from Bertschinger et al (1990)\\
\begin{tabular}
{c c c c}
 & $r=0$ & $r=40h^{-1}$Mpc & $r=60h^{-1}$Mpc \\
\hline
$V(r)$ (km/sec) & $457 \pm 61$ & $388 \pm 67$ & $327 \pm 82$\\
$(L,B)$ & $(156 \pm 7; -19 \pm 12)$ & $(177 \pm 9; -15 \pm 17)$ & $(194 \pm 13; 5\pm 26)$\\
\hline
\end{tabular}
\newpage
Table 2. $\sqrt{\nu(0)}$ in km/sec computed according to eq.(12) for the various
fits to $\xi(r)$.\\
\begin{tabular}
{c c c c c c c c c}
$r_f$ & $\frac{\Omega^{0.6}}{b}$ & $\xi(r)$=$(\frac{r}{r_*})^{-1.7}$ & & K92 & & BE93 & & CfA \\
 & & 40Mpc/h & 60Mpc/h & 40Mpc/h & 60Mpc/h & 40Mpc/h & 60Mpc/h & 40Mpc/h \\
\hline
0 & 1.00 & 1184 & 1230 & 1180 & 1220 & 1212 & 1239 & 1300 \\
0 & 0.50 & 678 & 681 & 674 & 669 & 682 & 693 & 732 \\
0 & 0.33 & 538 & 515 & 537 & 512 & 545 & 517 & 567 \\
0 & 0.25 & 478 & 443 & 478 & 442 & 483 & 445 & 497 \\
\hline
\hline
12 & 1.00 & 613 & 691 & 604 & 673 & 608 & 657 & \\
12 & 0.50 & 455 & 449 & 452 & 442 & 453 & 435 & \\
12 & 0.33 & 419 & 387 & 418 & 384 & 418 & 380 & \\
12 & 0.25 & 406 & 363 & 405 & 361 & 405 & 359 & \\
\hline
\end{tabular}
\newpage
\clearpage
\begin{figure}
\centering
\leavevmode
\epsfxsize=1.0
\columnwidth
\epsfbox{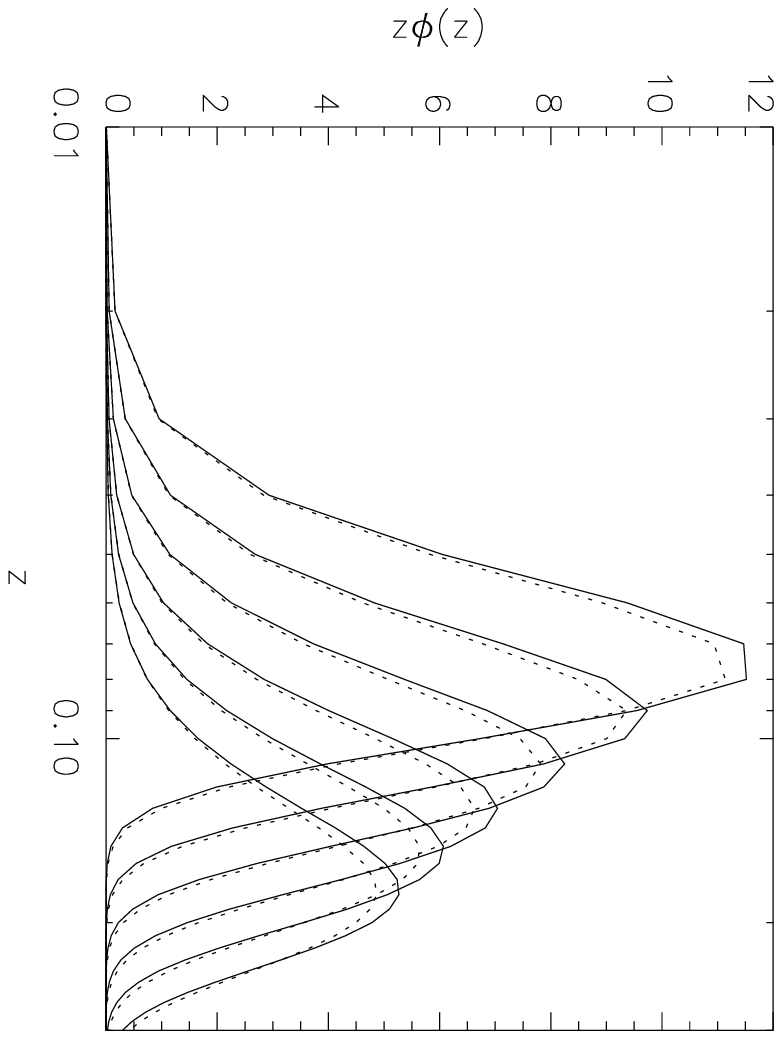}
\caption[]{ }
\end{figure}

\clearpage
\begin{figure}
\centering
\leavevmode
\epsfxsize=1.0
\columnwidth
\epsfbox{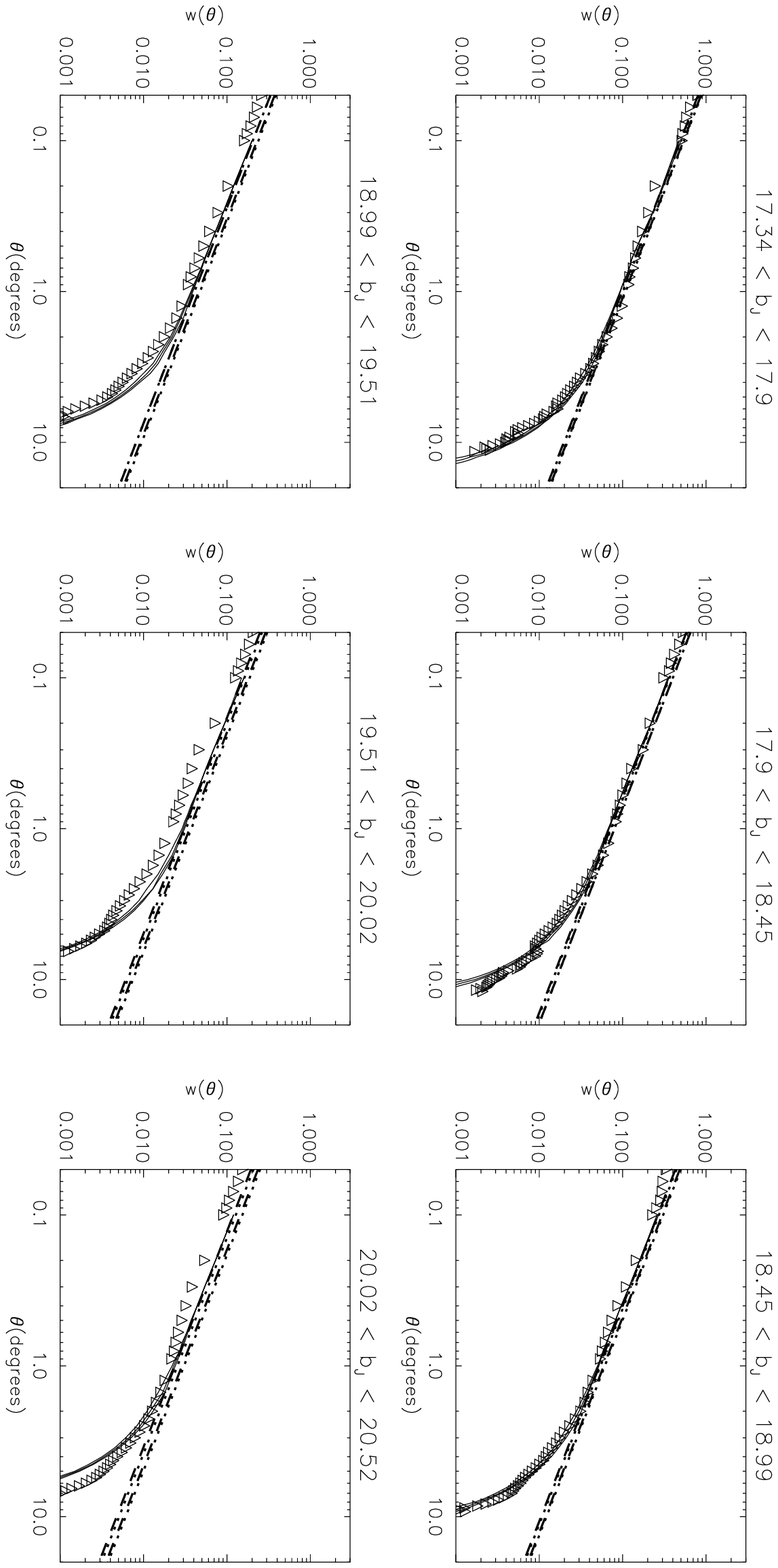}
\caption[]{ }
\end{figure}

\clearpage
\begin{figure}
\centering
\leavevmode
\epsfxsize=1.0
\columnwidth
\epsfbox{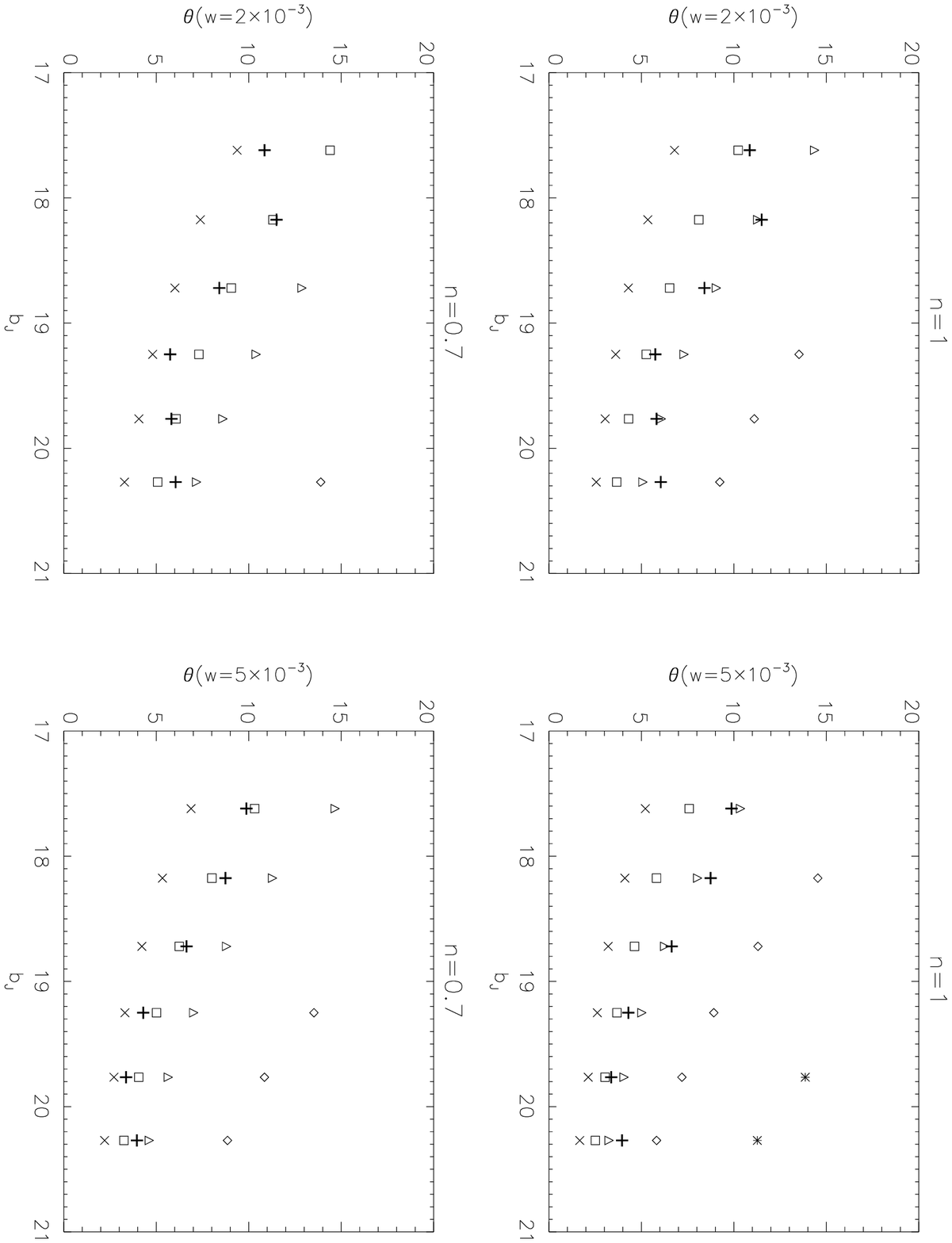}
\caption[]{ }
\end{figure}

\clearpage
\begin{figure}
\centering
\leavevmode
\epsfxsize=1.0
\columnwidth
\epsfbox{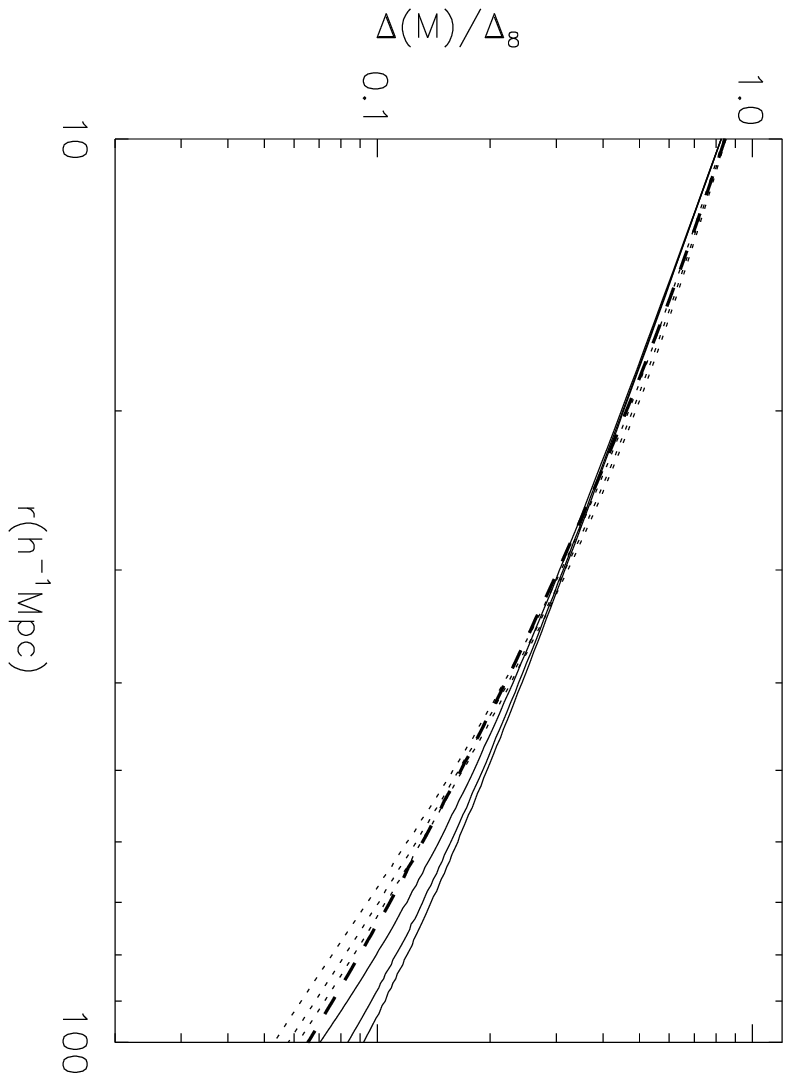}
\caption[]{ }
\end{figure}

\clearpage
\begin{figure}
\centering
\leavevmode
\epsfxsize=1.0
\columnwidth
\epsfbox{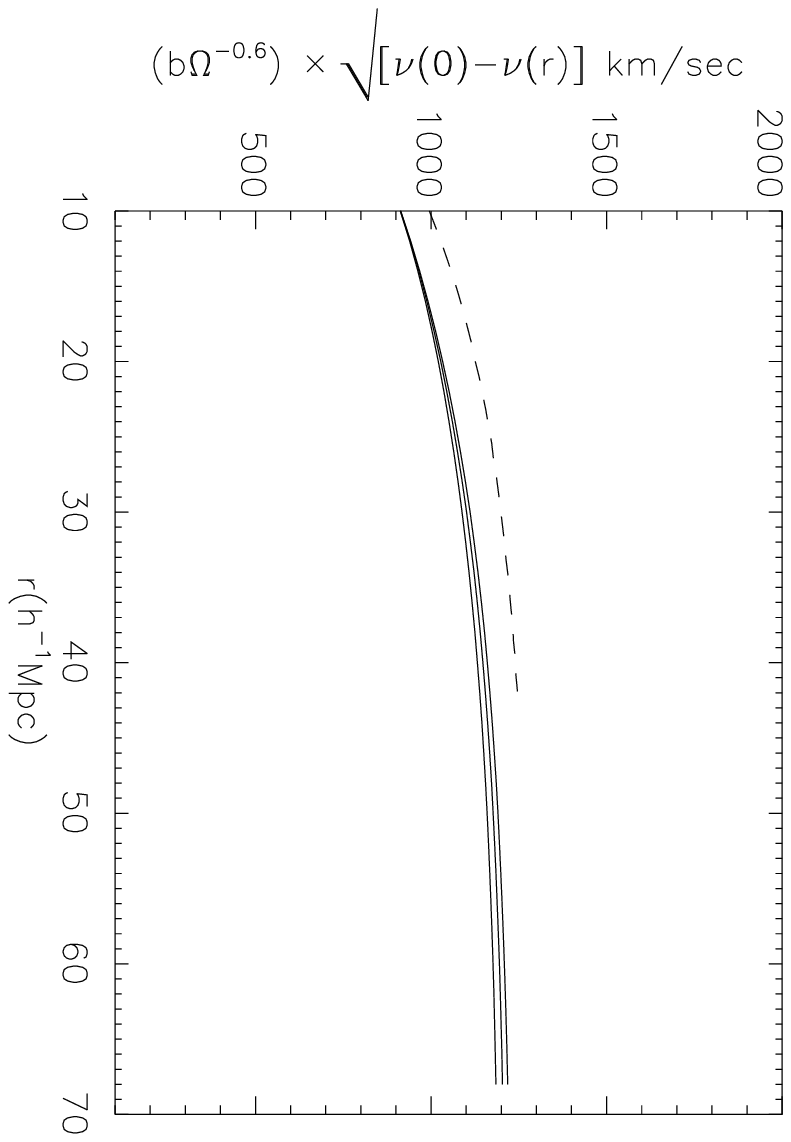}
\caption[]{ }
\end{figure}

\clearpage
\begin{figure}
\centering
\leavevmode
\epsfxsize=1.0
\columnwidth
\epsfbox{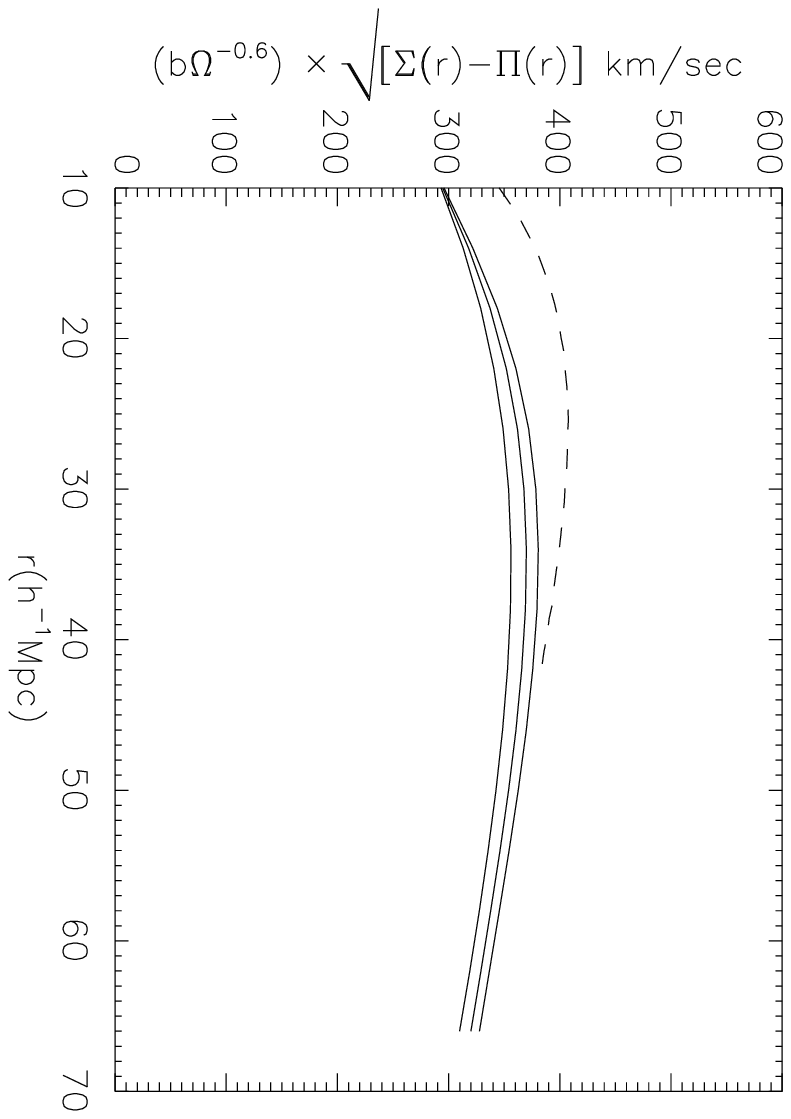}
\caption[]{ }
\end{figure}

\clearpage
\begin{figure}
\centering
\leavevmode
\epsfxsize=1.0
\columnwidth
\epsfbox{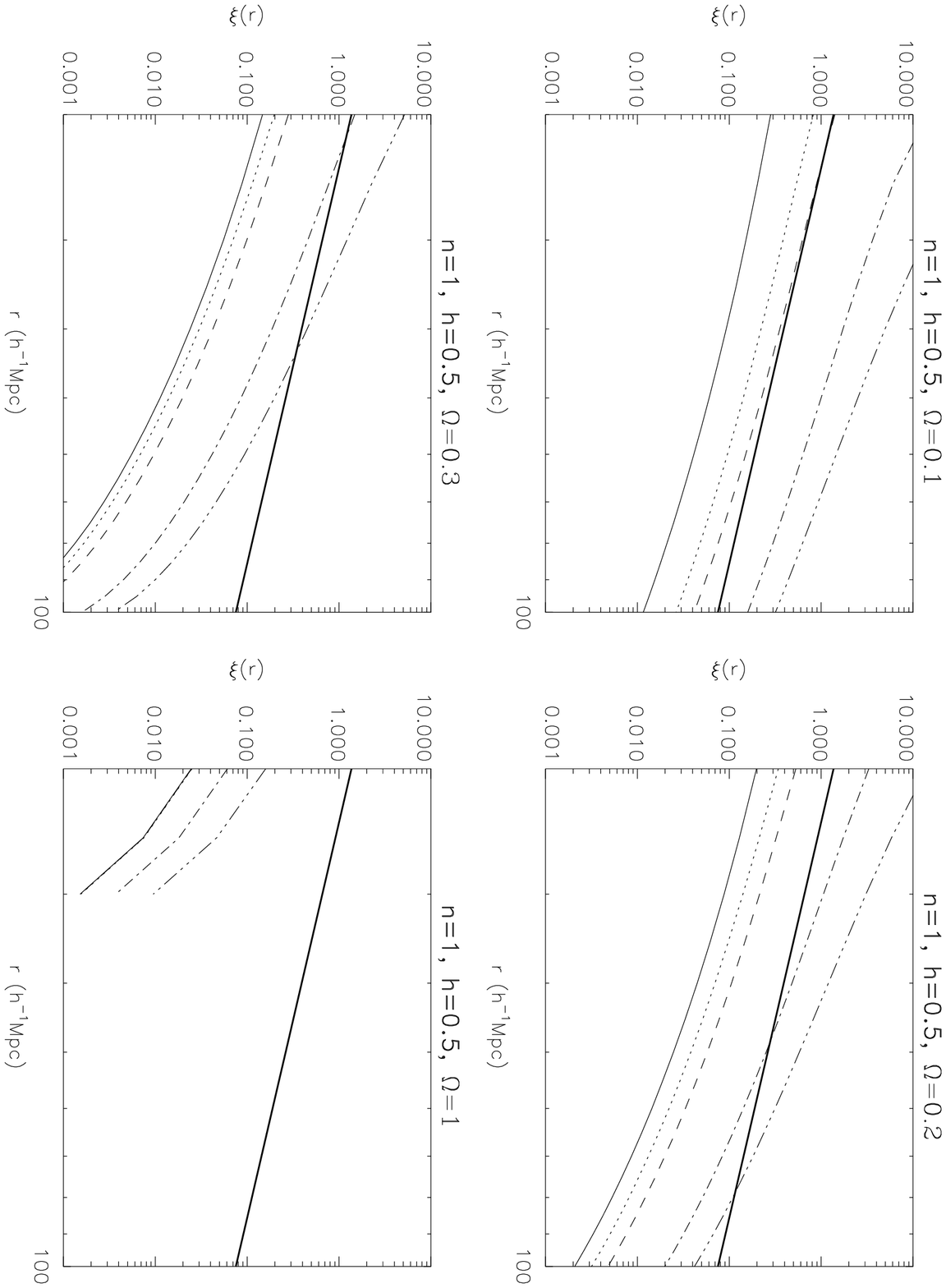}
\caption[]{ }
\end{figure}

\clearpage
\begin{figure}
\centering
\leavevmode
\epsfxsize=1.0
\columnwidth
\epsfbox{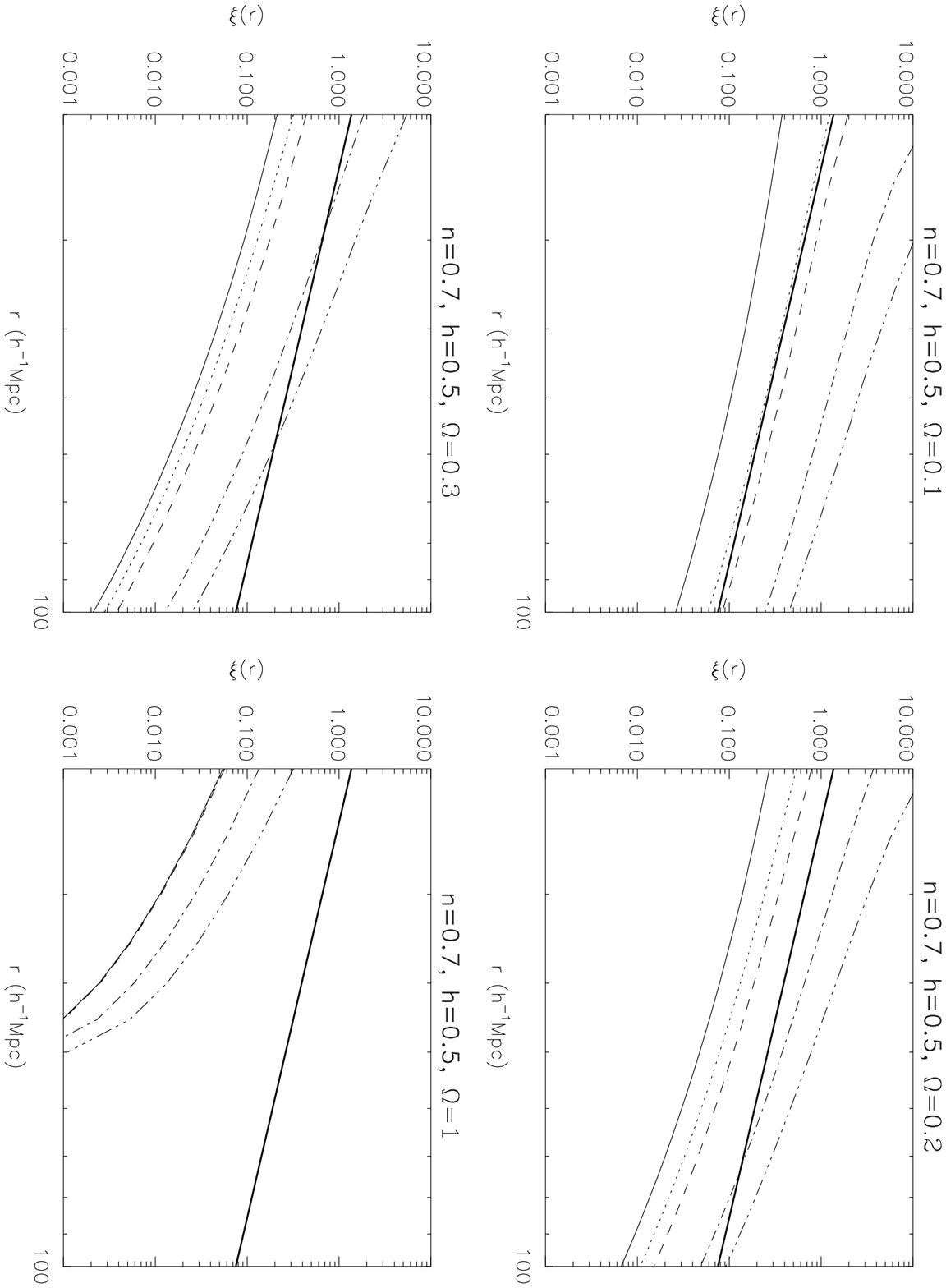}
\caption[]{ }
\end{figure}

\clearpage
\begin{figure}
\centering
\leavevmode
\epsfxsize=1.0
\columnwidth
\epsfbox{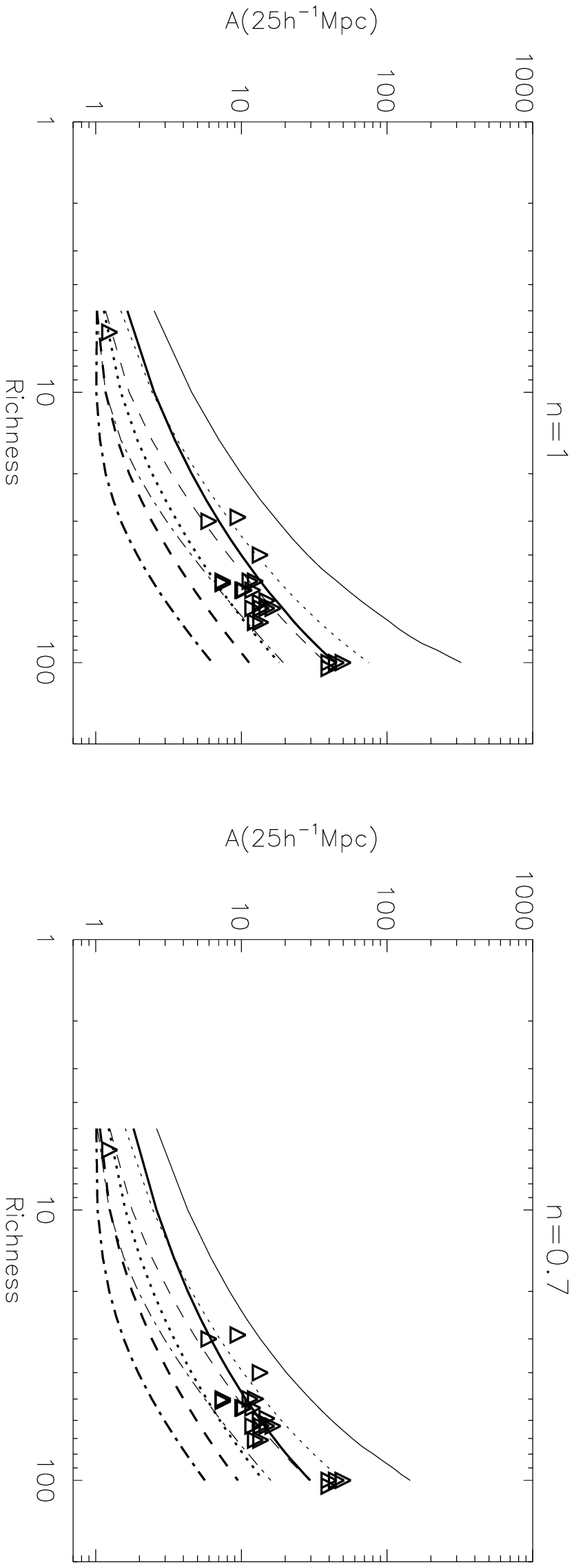}
\caption[]{ }
\end{figure}

\clearpage
\begin{figure}
\centering
\leavevmode
\epsfxsize=1.0
\columnwidth
\epsfbox{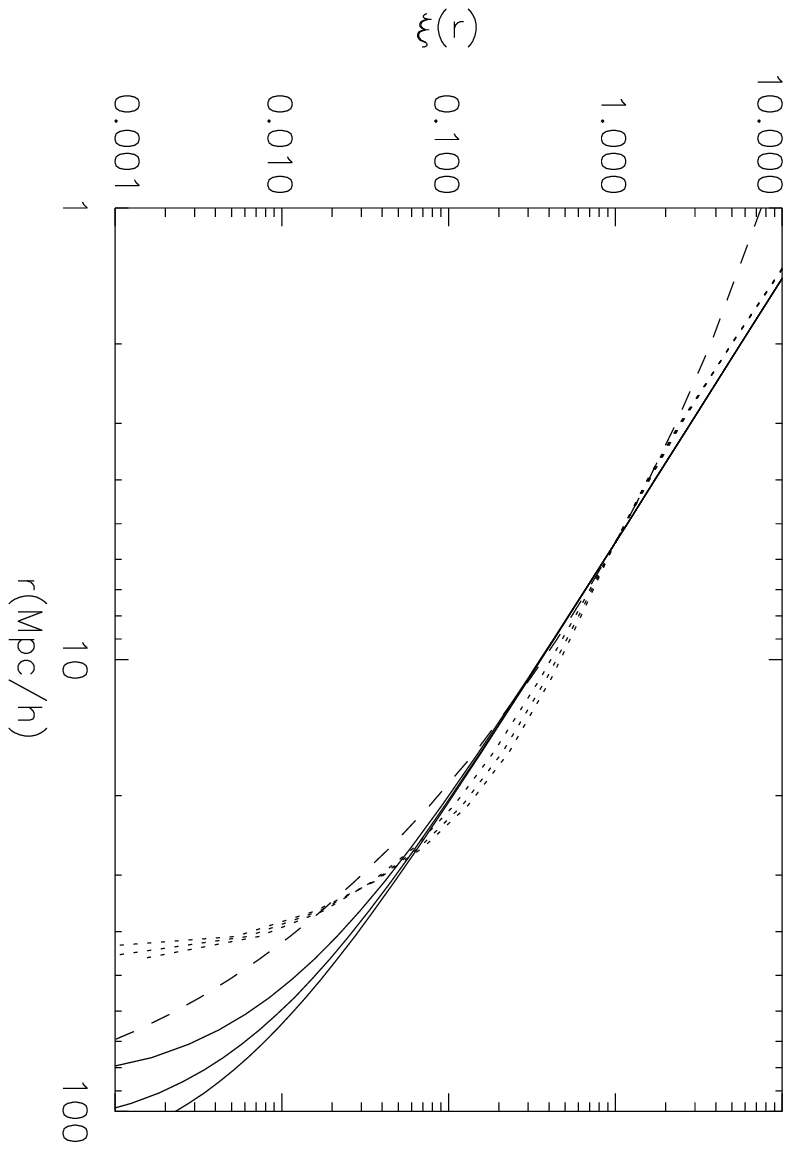}
\caption[]{ }
\end{figure}

\clearpage
\begin{figure}
\centering
\leavevmode
\epsfxsize=1.0
\columnwidth
\epsfbox{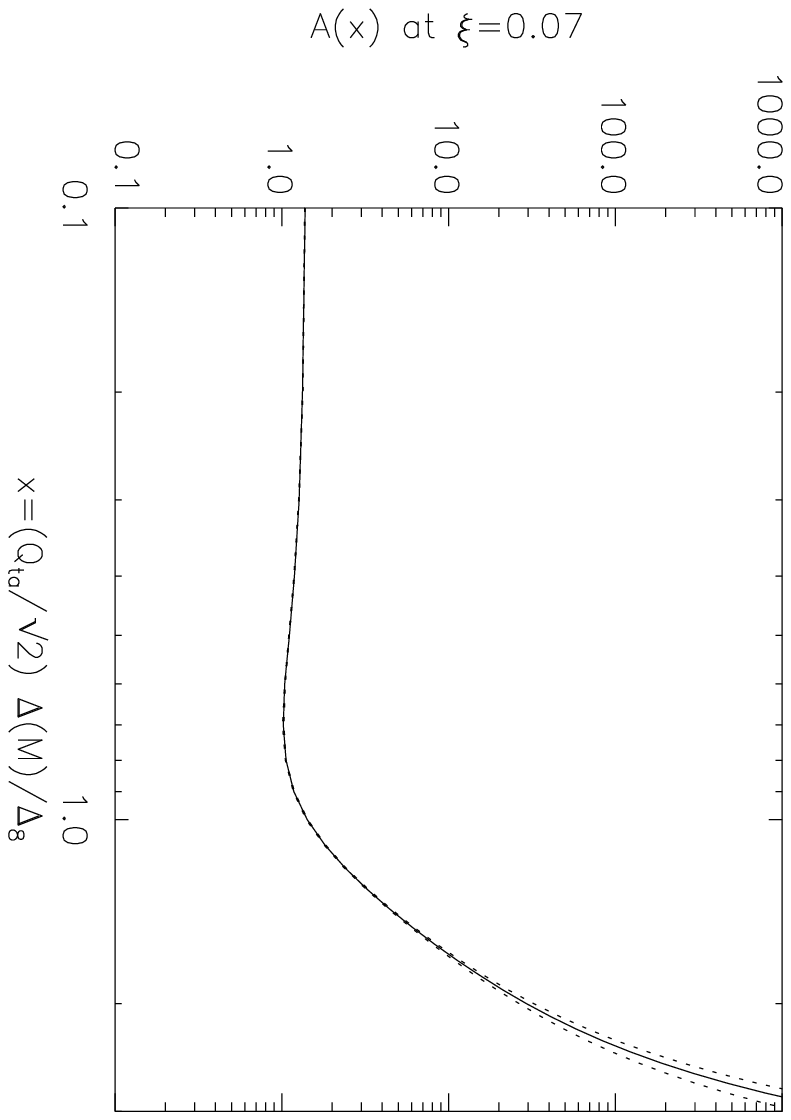}
\caption[]{ }
\end{figure}

\clearpage
\begin{figure}
\centering
\leavevmode
\epsfxsize=1.0
\columnwidth
\epsfbox{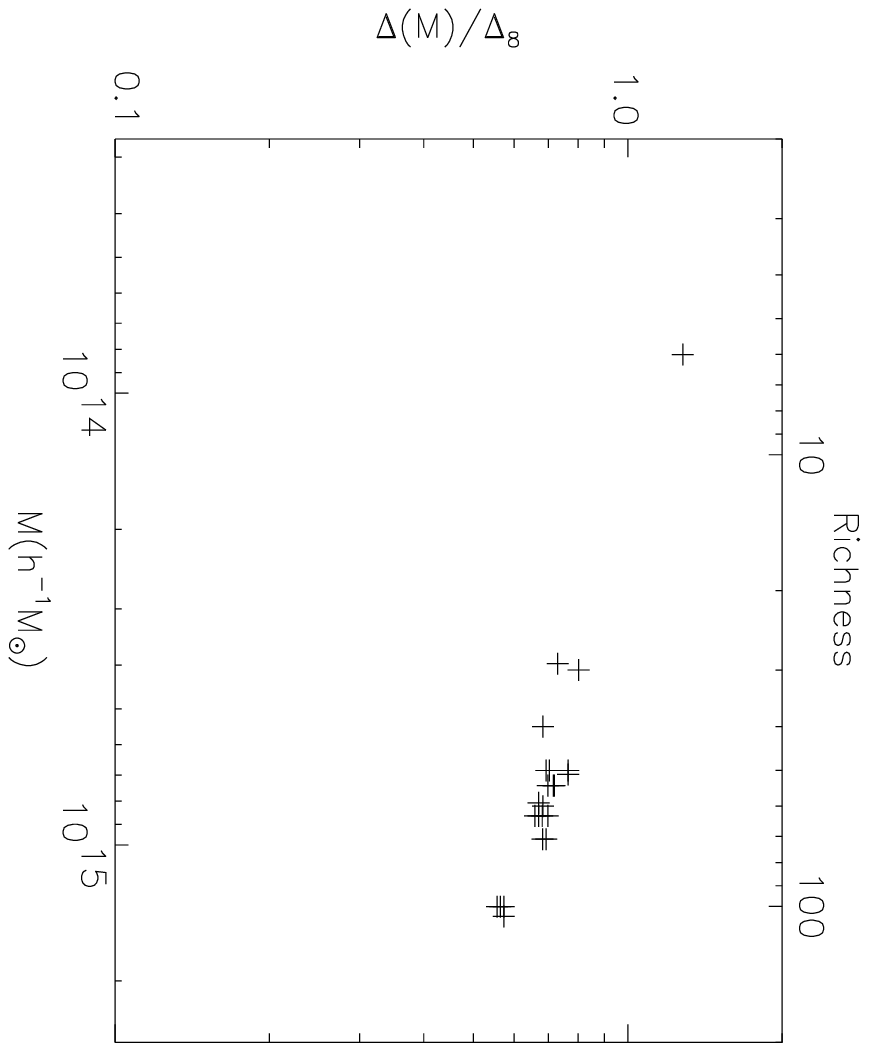}
\caption[]{ }
\end{figure}

\clearpage
\begin{figure}
\centering
\leavevmode
\epsfxsize=1.0
\columnwidth
\epsfbox{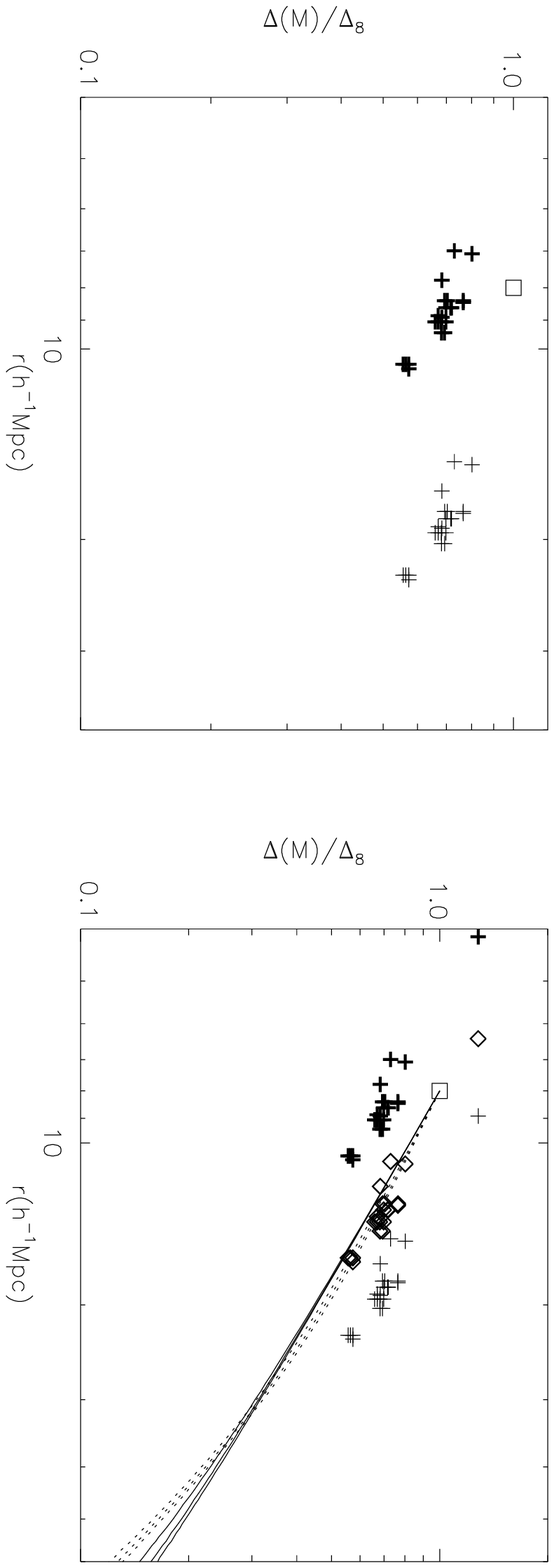}
\caption[]{ }
\end{figure}

\clearpage
\begin{figure}
\centering
\leavevmode
\epsfxsize=1.0
\columnwidth
\epsfbox{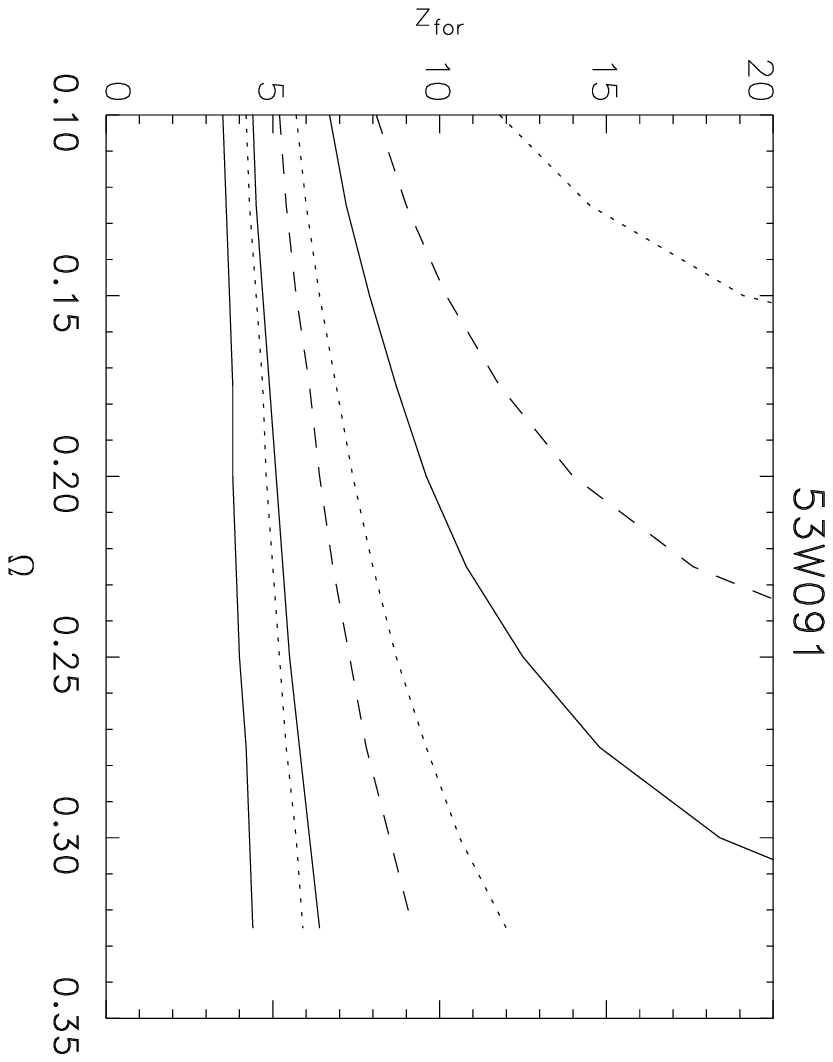}
\caption[]{ }
\end{figure}

\clearpage
\begin{figure}
\centering
\leavevmode
\epsfxsize=1.0
\columnwidth
\epsfbox{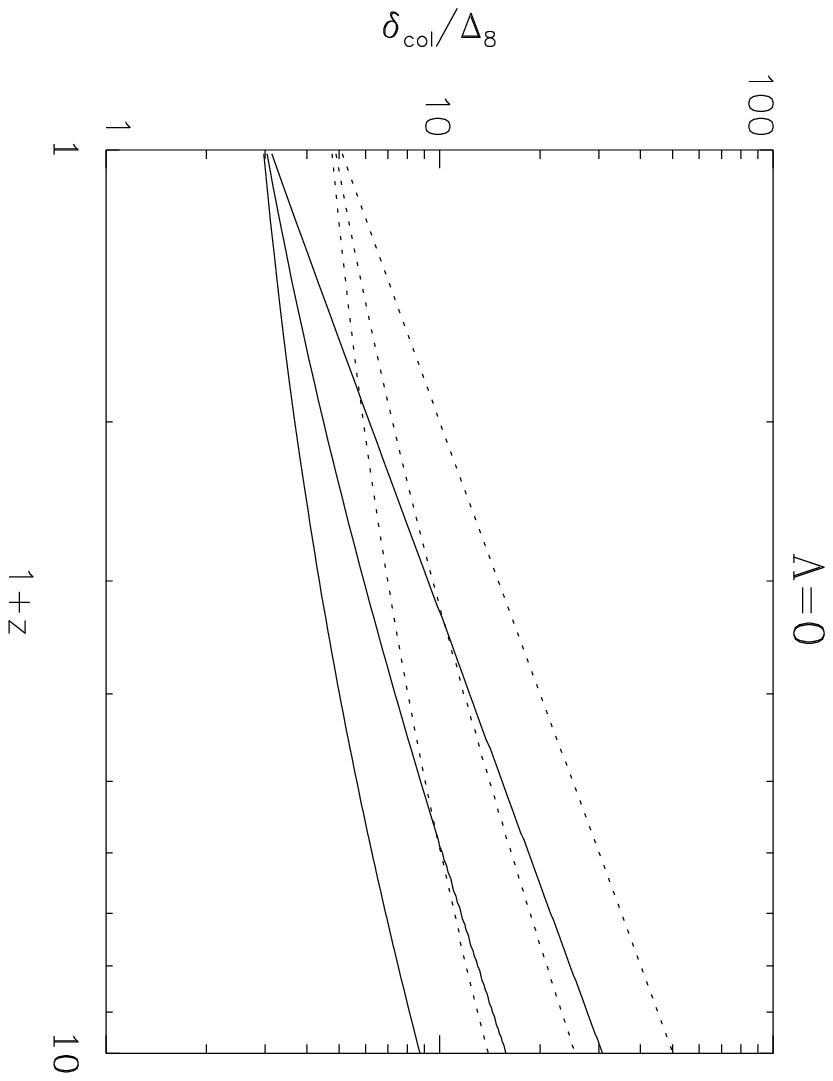}
\caption[]{ }
\end{figure}

\clearpage
\begin{figure}
\centering
\leavevmode
\epsfxsize=1.0
\columnwidth
\epsfbox{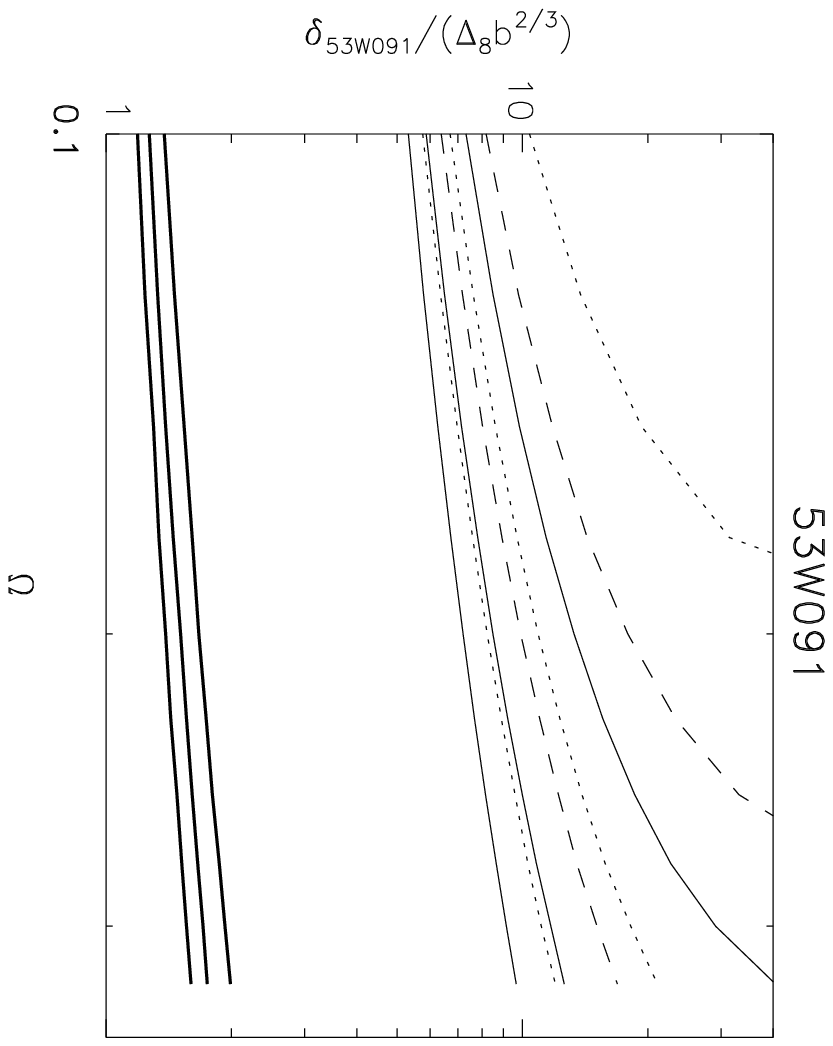}
\caption[]{ }
\end{figure}

\clearpage
\begin{figure}
\centering
\leavevmode
\epsfxsize=1.0
\columnwidth
\epsfbox{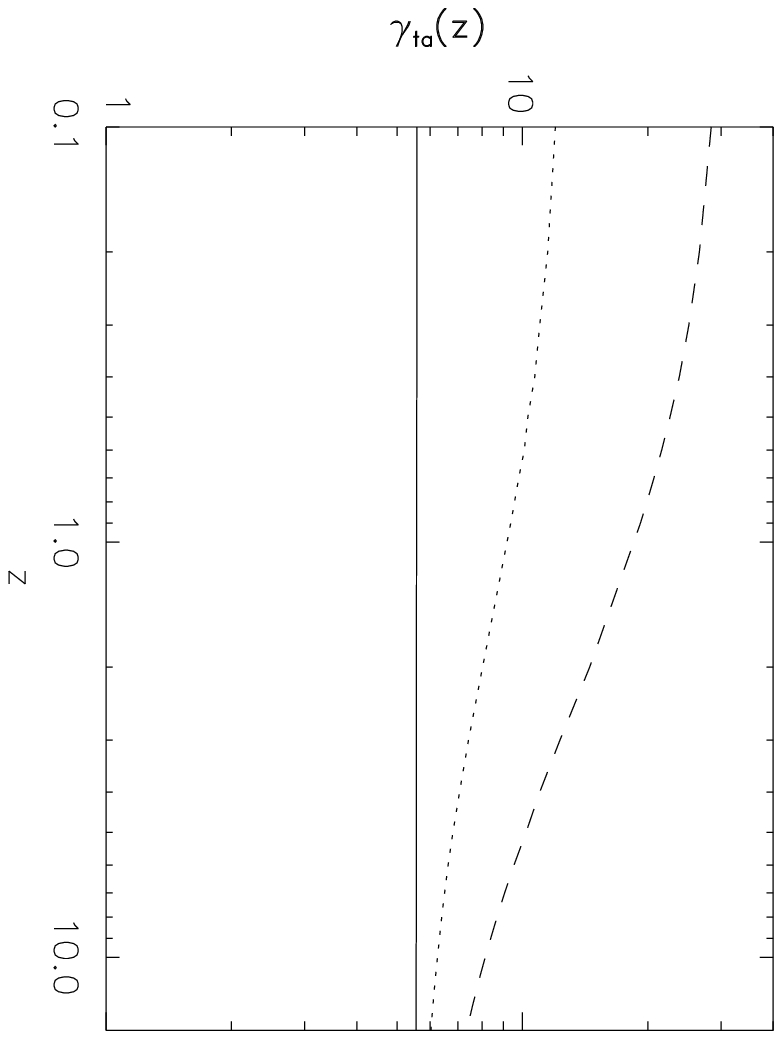}
\caption[]{ }
\end{figure}

\clearpage
\begin{figure}
\centering
\leavevmode
\epsfxsize=1.0
\columnwidth
\epsfbox{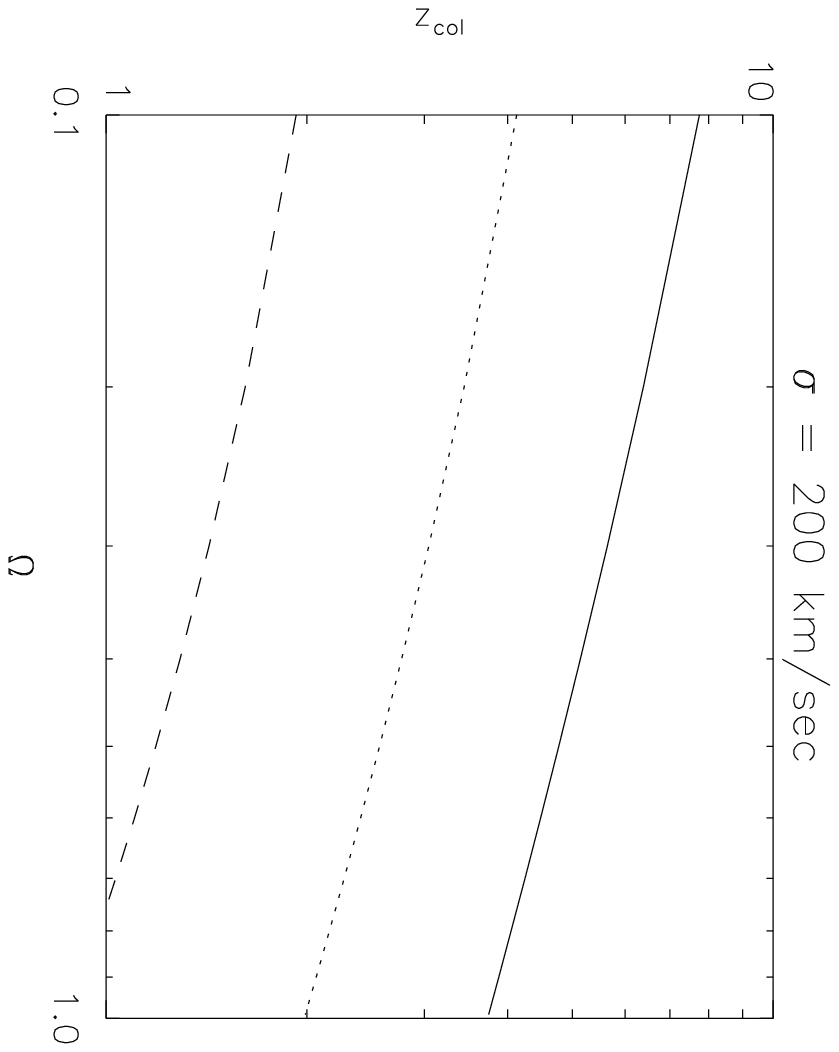}
\caption[]{ }
\end{figure}

\clearpage
\begin{figure}
\centering
\leavevmode
\epsfxsize=1.0
\columnwidth
\epsfbox{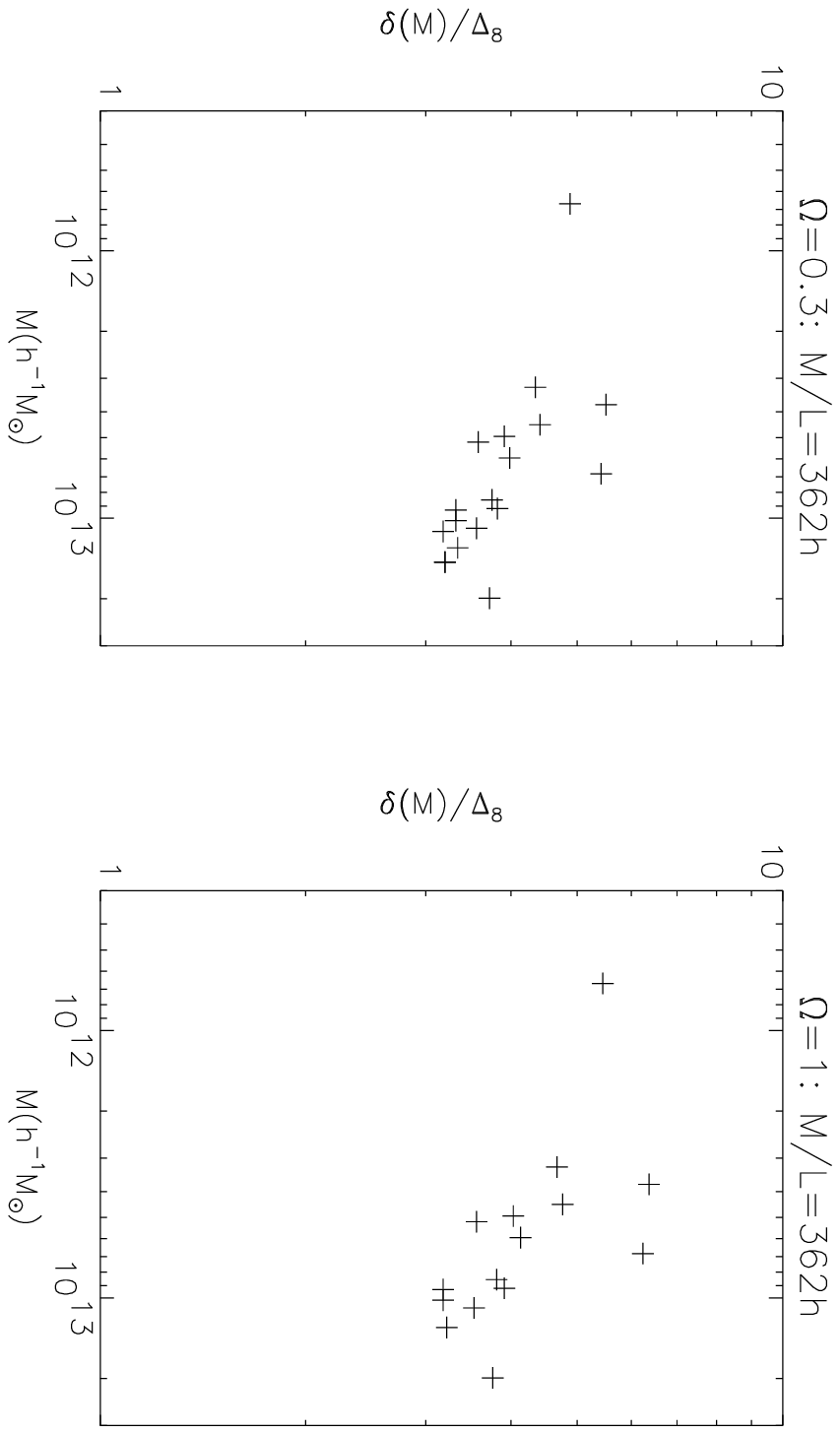}
\caption[]{ }
\end{figure}

\clearpage
\begin{figure}
\centering
\leavevmode
\epsfxsize=1.0
\columnwidth
\epsfbox{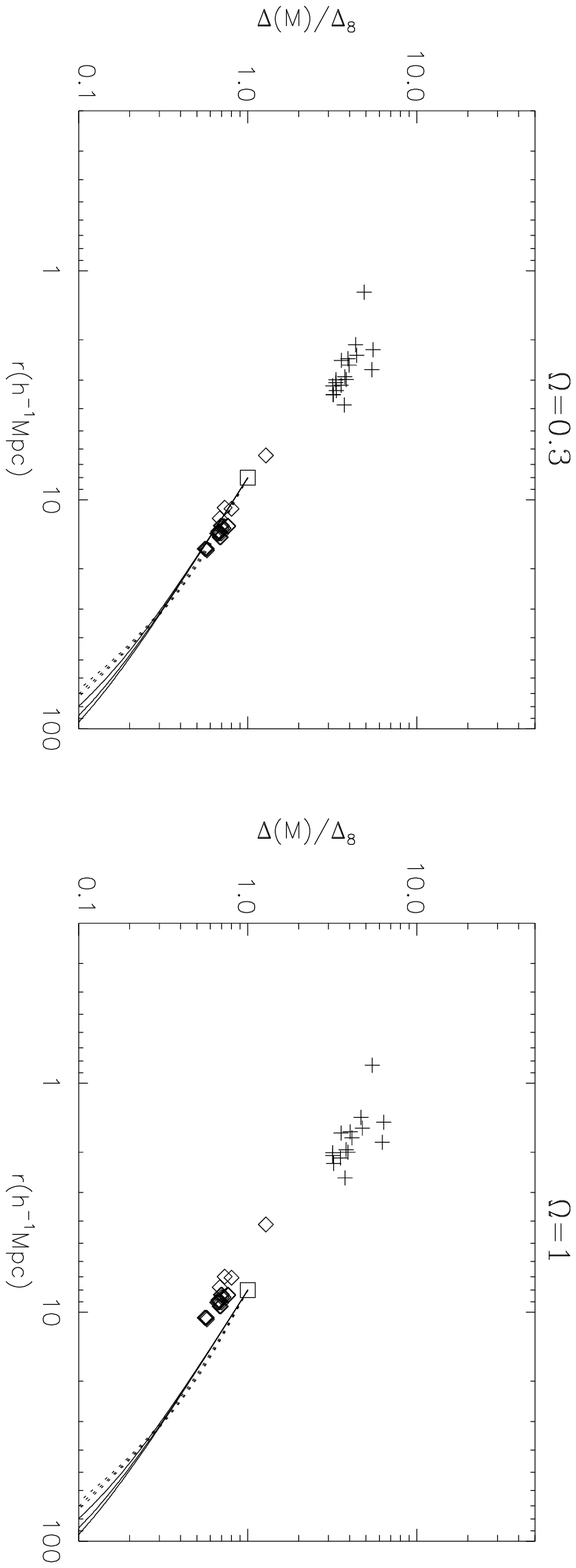}
\caption[]{ }
\end{figure}

\clearpage
\begin{figure}
\centering
\leavevmode
\epsfxsize=1.0
\columnwidth
\epsfbox{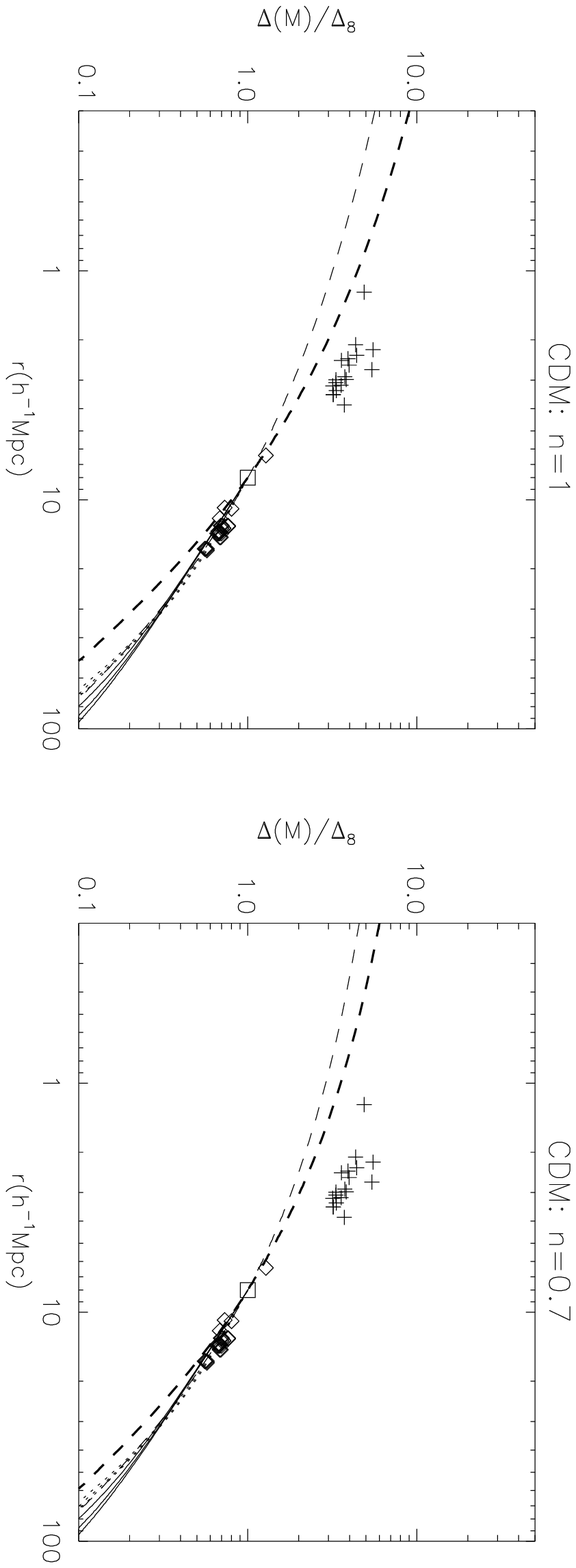}
\caption[]{ }
\end{figure}

\end{document}